\documentclass[12pt]{article}
\usepackage{fancyhdr}
\usepackage{setspace}
\usepackage{amsmath}
\usepackage{graphicx, graphics}
\usepackage{amssymb}
\usepackage{amsfonts}
\usepackage{lscape}
\usepackage{natbib}
\usepackage{float}
\usepackage{multirow}
\usepackage{caption}
\usepackage{ulem}
\usepackage{rotating}
\usepackage{setspace}
\usepackage{enumitem}
\usepackage[margin=1.0in]{geometry}

\usepackage{todonotes}
\def\citetpos#1{\citeauthor{#1}'s (\citeyear{#1})}

\usepackage{appendix}

\usepackage{hyperref}
\hypersetup{
    colorlinks=true,        
    linkcolor=red,          
    citecolor=blue,         
    filecolor=magenta,      
    urlcolor=cyan           
}
\usepackage[all]{hypcap}

\usepackage{subcaption}
\usepackage{color}
\definecolor{Blue}{rgb}{0.3,0.3,0.9}
\definecolor{Red}{rgb}{0.9,0.3,0.3}
\definecolor{crimson}{cmyk}{0.00, 0.83, 0.71, 0.35}
\definecolor{maroon}{rgb}{.6,.1,.20}

\newtheorem{theorem}{Theorem}

\newtheorem{definition}{Definition}

\newtheorem{proposition}{Proposition}

\usepackage{natbib}
\defcitealias{AscariMavroeidis}{\scshape AM}

\begin{document}
\title{Coherence without Rationality at the Zero Lower Bound\thanks{We thank Seppo Honkapohja, Martin Ellison, Klaus Adam, Alexandre Kohlhas, and James Moberly for insightful comments. We also thank seminar participants at the Bank of Finland and De Nederlandsche Bank and participants at the CEPR European Summer Symposium in International Macroeconomic (ESSIM) 2022 conference, the 2022 Barcelona Summer Forum Workshop on Expectations in Dynamic Macroeconomic Models, the Dynare 2022 Conference, 2022 Bank of Finland-CEPR conference, and ASSA 2023. The views expressed in this paper are those of the authors and not necessarily those of the Bank of Finland or De Nederlandsche Bank.}}

\author{Guido Ascari\thanks{Corresponding author: Department of Economics, University of Pavia, Via San Felice 5, 27100 Italy. E-mail address: guido.ascari@unipv.it.}\\\textit{University of Pavia} \\ \textit{De Nederlandsche Bank}
\and Sophocles Mavroeidis\thanks{Department of Economics, University of Oxford, Manor Road, OX1 3UQ, UK. E-mail address: sophocles.mavroeidis@economics.ox.ac.uk.}\\\textit{University of Oxford} \\\textit{and INET Oxford}
\and Nigel McClung\thanks{Email: nigel.mcclung@bof.fi.} \\ \textit{Bank of Finland}
}

		\maketitle

		\begin{abstract}

Standard rational expectations models with an occasionally binding zero lower bound constraint either admit no solutions (incoherence) or multiple solutions (incompleteness). This paper shows that deviations from full-information rational expectations mitigate concerns about incoherence and incompleteness. Models with no rational expectations equilibria admit self-confirming equilibria involving the use of simple mis-specified forecasting models. Completeness and coherence are restored if expectations are adaptive or if agents are less forward-looking due to some information or behavioral friction. In the case of incompleteness, the E-stability criterion selects an equilibrium. 

			\vspace{15pt}
			
			\noindent\textit{Keywords}: 	incompleteness, incoherence, expectations, zero lower bound\\
			\textit{JEL classification}: C62: E4: E52
		\end{abstract}
\baselineskip=17pt
\thispagestyle{empty}

\newpage

    \textit{The irrationality of a thing is no argument against its existence, rather a condition of it.} \\  \indent Friedrich Nietzsche, ``Human, All Too Human: A Book for Free Spirits'', 1878. 

\section{Introduction}
In the last 15 years since the Great Financial Crisis, central banks in Western economies had to face the problem of a zero (or effective) lower bound (ZLB) on the nominal interest rate. This spurred a very large and important literature on the topic. At least from the seminal contribution by \cite{BenhabibSG}, it is well-known that rational expectations (RE) models with a ZLB on the nominal interest rate generally admit multiple equilibria and also multiple steady states. However, the stochastic element in the ZLB literature is often very stylized with one single (often discount factor) shock that occurs only once and has either a stochastic or a known duration.

More recently, \citet[henceforth AM]{AscariMavroeidis} highlight an even more serious concern regarding this type of model when stochastic shocks hit the economy, a standard assumption in macroeconomic models.
They show that in models featuring a ZLB constraint, a stochastic environment and RE, equilibrium existence is not generic, i.e., the model is incoherent, and when these model do admit an equilibrium, they generally admit more equilibria than previously acknowledged, i.e., the model is incomplete.\footnote{Following \citetalias{AscariMavroeidis} we will use the terms incoherence and incompleteness to mean the non-existence of equilibria and the multiplicity of equilibria, respectively. Hence, a model is coherent if it admits at least one equilibrium, and complete if the equilibrium is unique.} Specifically, \citetalias{AscariMavroeidis} derive conditions for existence of a rational expectations equilibrium (REE), and for existence and uniqueness of a minimum state variable (MSV) equilibrium for dynamic forward-looking models with occasionally binding constraints. These conditions are difficult to interpret. Therefore, \citetalias{AscariMavroeidis} highlight a different and more fundamental problem in models with occasionally binding constraints and standard stochastic shocks than the ones already noted in the literature in this class of models, such as the  indeterminacy of REE equilibria in linear models and/or multiplicity of steady states. Section \ref{sec: coherence} reviews the \citetalias{AscariMavroeidis} results in more detail.

Given that a model without an equilibrium cannot be of any use, this paper  points to a possible route to tackle the incoherence problem: 
abandoning the full-information RE assumption. We show that the problem of incoherence and incompleteness hinges on the assumption that agents have RE. Non-existence of REE is by itself a compelling and novel reason to investigate the possibility of non-rational equilibria. Indeed, one of the main results from this paper is that a standard New Keynesian model with the ZLB constraint can fail to yield a REE and still admit other types of self-confirming equilibria. To illustrate this point, we consider two distinct equilibrium concepts which have been associated with different types of deviations from full-information RE. 

First, we investigate one of the most studied deviations from RE, that is, adaptive learning as typified by \cite{evans2001learning}. Adaptive learning agents have imperfect knowledge about the economy's structure, but learn to forecast macroeconomic variables by recursively estimating the parameters of a subjective forecasting model using simple statistical tools like least squares. A classic question examined in adaptive learning applications is whether agents eventually learn to forecast rationally, and hence whether the learning economy converges to a REE. However, given that we are interested in cases where a REE does not exist, we assume that agents learn by recursively estimating
forecasting models that are mis-specified and under-parameterized relative to the forecasting models that agents would have in a REE. Under this assumption, we derive analytically conditions for the economy to settle on a self-confirming equilibrium in which agents make optimal forecasts within their class of forecasting rule. This form of self-confirming equilibrium, which is distinct from REE, is often labelled \textit{restricted perceptions equilibrium (RPE)} in the learning literature (e.g. see \cite{evans2001learning} or \cite{branchRPEsurvey2}). Importantly, we prove that a RPE can exist when the RE model is incoherent and hence no REE exists. 

Second, we consider bounded rationality as a possible deviation from RE.  Boundedly rational agents are less forward-looking than rational agents, for instance because they are myopic \`{a} la \cite{gabaix2020}, have imperfect common knowledge as in \cite{angeletos2018forward}, or have finite planning horizons similar to \cite{WoodfordXie2020}. In this setting, too, a unique \textit{bounded rationality equilibrium (BRE)} may exist, even if a REE does not. Hence, both adaptive learning and bounded rationality might alleviate, under certain conditions, the coherence problem of the standard NK model with a ZLB constraint. Finally, we also investigate the implications of combining the two deviations from rationality.

The derivation of an adaptive learning RPE and BRE in an incoherent REE framework is the central contribution of the paper. In this respect, some remarks are noteworthy. 

First, adaptive learning can ensure completeness and coherence all by itself. Specifically, we prove that a unique \textit{temporary} equilibrium always exists in our model with a ZLB constraint and adaptive learning agents, provided that agents do \textit{not} observe current endogenous variables before market clearing takes place---a very common assumption in the learning literature. 

Second, a RPE emerges as a self-confirming equilibrium, even if the underlying model does not admit a REE. The learning literature has typically focused on the question of whether a REE can be learnable, because the underlying model admits a REE solution. Here, instead, we investigate whether adaptive learning can generate self-confirming equilibria even when a REE does not exist. When agents do \textit{not} observe current endogenous variables, expectations are predetermined, and a temporary equilibrium always exists, but it is not necessarily self-confirming. To the best of our knowledge, our finding that self-confirming adaptive learning equilibria exist when there is no REE is a novel and intriguing addition to the literature. 

Third, and related to the previous point, whenever the NK model does not admit a REE, it is impossible for agents to form self-confirming beliefs about the dynamics of inflation and output (i.e., as implied by a standard MSV in our simple model). The economy can easily diverge into a deflationary spiral if agents attempt to learn these dynamics using simple statistical techniques. 
Hence, while it is a curse to be smart, it is a blessing to be simple-minded, because the non-rationality of agents' beliefs can save the economy from spiralling out of control and lead it to a coherent and complete self-confirming RPE. 

Fourth, the source of the problem of rational incoherence can be intuitively explained in terms of income and substitution effects, following \cite{Bilbiie}. A similar intuition is behind the so-called ``forward guidance puzzle'' and its proposed solutions that hinge on weakening agents' forward-lookingness \citep[e.g.,][]{del2012forward, mckay2016power, angeletos2018forward,gabaix2020, WoodfordXie2020, EGP2021}. Hence, we show that weakening the `rationality' of agents kills several birds with one stone, because it simultaneously solves different problems highlighted by the literature (forward-guidance puzzle, belief-driven liquidity traps, existence of an equilibrium) that share the same mechanism as a common source.

Fifth, a basic takeaway from the existence analysis is that the baseline NK model with RE is incoherent \textit{when negative shocks are sufficiently large in magnitude or sufficiently persistent}, but can still admit RPE or BRE. A fundamentals-driven RE liquidity trap must, therefore, be relatively short-lived compared to the duration of actual liquidity trap events experienced by Japan, the Euro Area and the U.S., because persistent shocks would make the RE model incoherent. This is not true for the RPE or BRE, where a liquidity trap can be highly persistent. In this sense, one could argue that a RPE or a BRE could explain why the economy did not blow up after a large shock such as the Great Financial Crisis.

Finally, a second contribution of the paper concerns the stability properties of these equilibria under learning, that is, the issue of whether RPE and REE can emerge from a process of learning. Following the adaptive learning literature, we employ the expectational stability or ``E-stability" criterion to select an equilibrium that may arise through an economy-wide adaptive learning process in which agents recursively update the parameters of their subjective forecasting models using simple statistical techniques such as least squares. We find there is a unique E-stable RPE when a RPE exists. Similarly, only one MSV REE can be E-stable. 

After a brief literature review, the paper proceeds as follows. Section \ref{sec: model} introduces a simple model of the ZLB that nests our different assumptions about expectations formation as special cases. Section \ref{sec: coherence} illustrates the problem of rational incoherence and the possibility of irrational coherence. Section \ref{sec: learning} shows how adaptive learning resolves incompleteness issues, and also discusses the plausibility of the RPE concept. Section \ref{sec: conclusion} concludes. The proofs of all the Propositions can be found in the \hyperref[sec: Appendix]{Appendix}.

\subsection{Literature review}
This paper contributes to an already large literature about deviations from RE and the ZLB. Earlier work on adaptive learning at the ZLB studied monetary and fiscal policies that can prevent an economy with learning agents from getting stuck in a liquidity trap \citep{EvansGH2008, BenhabibEvansHonkapohja, EHMstag},\footnote{See also \cite{EM2018JMCB} for a related discussion on interest rate pegs and adaptive learning.} unconventional policies such as forward guidance \citep{ColeJMCB, EGP2021}, ``make-up" strategies such as price level targeting \citep{HM2020} or average inflation targeting \citep{HMcC2021}. \cite{ChristianoEichenbaumJohannsen} show that the E-stability criterion selects one of multiple equilibria of a model with a transitory demand shock that can drive the economy into a liquidity trap. This finding is closely related to our result about E-stability of REE in the case of incompleteness. However, their model assumes that the economy returns to a steady state after the shock dissipates, whereas our framework allows for multiple, recurring liquidity trap episodes, consistent with the recurrence of ZLB events in the U.S.~and elsewhere. Thus, we extend insights from \cite{ChristianoEichenbaumJohannsen} to models with recurring demand shocks. More generally, the above mentioned papers do not consider existence and stability of equilibria of models with recurring, fundamentals-driven liquidity traps. 

A significant strand of the adaptive learning literature focuses on self-confirming ``misspecification equilibria" that can emerge if agents recursively learn to forecast using a misspecified forecasting rule. In a misspecification equilibrium, agents do not understand the true equilibrium law of motion for economic variables, but observable macroeconomic outcomes nonetheless confirm their subjective beliefs about specific statistical properties of the economy. RPE is a special case of misspecification equilibrium involving a ``simple" under-parameterized forecasting model that omits some variables which affect the macroeconomic dynamics. In a RPE, agents forecast optimally within their class of forecasting rules in the sense that forecast errors are orthogonal to their forecasting model. The properties of RPE and misspecification equilibria, as well as their emergence through adaptive learning, has been explored in \cite{branchRPEsurvey}, \cite{branchRPEsurvey2}, \cite{evans2001learning}, \cite{MarcetSargent1989JPE}, \cite{EHSargent93}, \cite{BranchEvansJET06}, \cite{BranchEvansEL}, \cite{BullardEvansHonkapohja}, \cite{EvansMcGoughJET2020} and \cite{BruceDavidGeorge}, \cite{HommesSorger}, \cite{HommesZhu}, \cite{BranchGasteiger}, among many others. Empirical support for RPE and related misspecification equilibria comes from experiments involving monetary sticky price economies \citep{Adam07} and analysis of survey and macroeconomic data involving estimation of New Keynesian frameworks \citep{HommesNKBLE}.\footnote{See also \cite{SlobodyanWouters}, \cite{Ormeno}, \cite{BeShears}, \cite{Assenza}, and \cite{BranchGasteiger} for additional empirical support for small misspecified forecasting rules.} 
 
A number of earlier works, including \cite{angeletos2018forward}, \cite{gabaix2020} and \cite{WoodfordXie2020}, study BRE and issues related to the ZLB. Among other things, these papers show that deviations from RE that make agents less forward-looking than rational agents can resolve the so-called NK paradoxes of the ZLB, such as the prediction that forward guidance announcements can have arbitrarily large effects on the economy (``forward guidance puzzle"). Importantly, contributions to this literature typically treat the ZLB regime as arising from a transitory shock, usually with a known duration, after which time the economy returns to steady state forever. Models employing shocks with known duration are not susceptible to the issues of equilibrium existence and multiplicity that we study here. Our contribution, therefore, is to embed bounded rationality into models with recurring stochastic shocks, and to show that these deviations from RE resolve the problem of incoherence and incompleteness identified by \citetalias{AscariMavroeidis}. 

Finally, \cite{MertensRavn}, \cite{NakataSchmidt2019, NakataSchmidt2020}, and \cite{Bilbiie}, among others, study conditions for the existence of both fundamentals-driven and confidence-driven liquidity trap equilibria, which are caused by fundamental shocks to the economy and non-fundamental (sunspot) shocks, respectively.\footnote{Additionally, \cite{BianchiMelosiRottner} study implications of fundamentals-driven liquidity traps in a nonlinear New Keynesian model.}  One takeaway from these papers is that the fundamentals-driven liquidity trap equilibrium is unlikely to exist if shocks are too persistent, but sunspot equilibria can feature very persistent liquidity traps. However, to our knowledge, confidence-driven liquidity trap equilibria have only been derived in coherent models (i.e. models that admit at least one MSV solution). An incoherent model fails to admit confidence-driven liquidity trap equilibria, and tight restrictions on the support of \textit{fundamental} shocks are necessary for existence of both MSV and confidence-driven liquidity trap equilibria.

\section{Model and expectations formation mechanisms} \label{sec: model}
We employ a model that nests the simple New Keynesian model as well as reflects the reduced-form of the alternative bounded rationality models explored by  \cite{gabaix2020}, \cite{angeletos2018forward}, \cite{WoodfordXie2020}:
\begin{eqnarray}
x_t&=& M \hat{E}_{t}x_{t+1}- \sigma (i_t -N \hat{E}_{t}\pi_{t+1})+ \epsilon_t,  \label{eq:IS}\\
\pi_t&=&\lambda x_t+M_f \beta \hat{E}_{t} \pi_{t+1},\label{eq:PC}\\
i_t&=&\max\{\psi \pi_t,-\mu\} \label{eq:MP},
\end{eqnarray}
where $x_t$ is the output gap, $i_t$ the nominal interest rate and $\pi_t$ is the inflation rate. If $M=N=M_f=1$, the model nests the simple three-equation New Keynesian model of \cite{Woodford} where (\ref{eq:IS}) is the Euler equation, (\ref{eq:PC}) is the NK Phillips Curve and (\ref{eq:MP}) the monetary policy rule, described by the simplest Taylor rule but with a ZLB constraint. The model is log-linearized around the zero inflation steady state and $0<\beta<1$, $0<\sigma, \lambda,\mu$, and $\psi>1$ (i.e. the ``Taylor principle" holds). Bounded rationality implies, instead, $0<M,N,M_f\le1.$ Note that $\hat E$ denotes (possibly non-rational) expectations and $\hat E=E$ denotes model-consistent (rational) expectations.

We follow earlier work, including \cite{Eggertsson}, \cite{NakataSchmidt2019}, \cite{ChristianoEichenbaumJohannsen}, and \citetalias{AscariMavroeidis}, and assume that the demand shock, $\epsilon_t$, follows a two-state Markov process with transition matrix:
\begin{eqnarray*}
K:=\begin{pmatrix} p & 1-p\\ 1-q & q
\end{pmatrix},
\end{eqnarray*}
\noindent with $0<p=Pr(\epsilon_t=\epsilon_1|\epsilon_{t-1}=\epsilon_1)\le1$, $0<q=Pr(\epsilon_t=\epsilon_2|\epsilon_{t-1}=\epsilon_2)\le1$.
If we assume $q=1$ and $\epsilon_2=0$, similar to \cite{Eggertsson} or \cite{ChristianoEichenbaumJohannsen}, then we have a model in which a transitory shock, $\epsilon_t=\epsilon_1\neq0$, displaces the economy from steady state, but the economy eventually returns to the absorbing steady state of the model when $\epsilon_t=\epsilon_2=0$. In the standard RE version of the model there are two non-stochastic steady states: one with zero inflation, and one with zero nominal interest rates. However, equilibrium inflation and output in the temporary state ($\epsilon_t=\epsilon_1$) depend on whether agents have full-information RE or whether they are boundedly rational in some way.

We consider three models of expectations formation. First, agents have full-information RE in the special case of the model with no discounting in the Euler equation and Phillips curve (\ref{eq:IS})-(\ref{eq:MP}) and model-consistent expectations.
\begin{definition} \label{def: RE}
Agents have \textbf{full-information rational expectations (RE)} if and only if $\hat E=E$ and $M=M_f=N=1$ in the NK model given by Equations (\ref{eq:IS})-(\ref{eq:MP}).
\end{definition}
A REE, defined in Section \ref{sec: coherence}, is a solution of the model (\ref{eq:IS})-(\ref{eq:MP}) obtained under these assumptions. In keeping with the literature, we treat full-information RE as the benchmark model of expectations formation, against which we compare ZLB dynamics under alternative expectations formation mechanisms. Particular attention is paid to the possibility that agents do not have full knowledge about the structure of the economy, and consequently expectations can be model-inconsistent (i.e., $\hat E\neq E $). The adaptive learning literature in particular studies agents with imperfect knowledge who learn to forecast the law of motion for aggregate variables using standard statistical tools like least squares. In this setting, imperfect knowledge can imply model-inconsistent expectations, but the focus of a large swath of this literature is whether agents can form self-confirming beliefs, either by learning a REE, or some non-rational, self-confirming equilibrium if their subjective forecasting models are mis-specified with respect to the rational forecasting models. 

\begin{definition} \label{def: IK}
Agents have \textbf{imperfect knowledge} if and only if $\hat E\neq E$; $M=M_f=N=1$ in the NK model given by Equations (\ref{eq:IS})-(\ref{eq:MP}).
\end{definition}

Definition \ref{def: IK} follows the ``Euler equation approach" to imperfect knowledge, which treats the Euler equation form of the first-order conditions of agents' optimization problem under RE, (\ref{eq:IS})-(\ref{eq:PC}), as agents' subjective decision rules under imperfect knowledge. 
The alternative is the so-called ``infinite horizon approach'' of \cite{Preston2004} according to which optimizing learning agents with imperfect knowledge learn to forecast the path of interest rates, output and inflation.\footnote{See \cite{BullardEusepi} for comparison of Euler equation learning and infinite horizon learning.} Therefore, our definition of imperfect knowledge involves both non-rational beliefs and sub-optimal decision-making, in keeping with a large literature on imperfect knowledge and learning. Our main conclusion that imperfect knowledge can lead to coherence when the model is rationally incoherent continues to hold under infinite-horizon learning.\footnote{For brevity, we give those results in Appendix \ref{oappesec: RPE IH}.}

We can deviate from RE without relaxing the assumption that agents have full knowledge about the structure of their economic environment. For instance, \cite{gabaix2020} derives a model in which households and firms are relatively myopic due to cognitive limitations. In this setting, myopia implies a change in the model structure in the form of discounting in the aggregate demand curve (\ref{eq:IS}) (i.e., $M<1$) and additional discounting in the Phillips curve (\ref{eq:PC}) (i.e. $M_f<1$). However, nothing in \citetpos{gabaix2020}  model prevents agents from having full knowledge about the world they inhabit, and therefore nothing prevents these boundedly rational agents from having model-consistent expectations. Hence, \citetpos{gabaix2020} behavioral model shows how we can deviate from full-information RE without sacrificing the assumption that agents have perfect knowledge. Bounded rationality models by \cite{angeletos2018forward} and \cite{WoodfordXie2020} may also lead to reduced-form structural models with additional discounting in the structural equations. If $M,M_f$ or $N$ is less than one, we say that agents are boundedly rational.
\begin{definition} \label{def: BR}
Agents are said to be \textbf{boundedly rational} if and only if $\hat E=E$ and \\$min\{M,M_f,N\}<1.$
\end{definition}

\section{Coherence: existence of an equilibrium}\label{sec: coherence}

To put the whole paper into context, it is worth clarifying the main contributions of \citetalias{AscariMavroeidis}.
While the stochastic element in the literature on the ZLB is often very stylized, featuring one single (often discount factor) shock that occurs only once and has either a stochastic or a known duration, \citetalias{AscariMavroeidis} consider the general problem of the conditions for existence and uniqueness of equilibria in dynamic forward-looking models with RE when some variables are subject to occasionally binding constraints, like in the ZLB case, and when recurrent stochastic shocks hit the economy, a standard assumption in macroeconomic models. \citetalias{AscariMavroeidis} propose to use a method based on  \cite{GourierouxLaffontMonfort1980} that studied this problem in the context of simultaneous equations models with endogenous regime switching, and derived conditions for existence and uniqueness of solutions,
which \cite{GourierouxLaffontMonfort1980} label as coherency conditions.
The problem of existence of equilibria, i.e., coherence, in more standard stochastic environments commonly used in macroeconomic models is obviously fundamental and a first-order concern for this literature.\footnote{Even though there is a large and expanding literature on solution algorithms for such models, \citep[see e.g.,][]{FernandezGordonGuerronRubio2015,GuerrieriIacoviello2015, gustetal2017AER,  AruobaCubaBordaSchorfheide2018, AruobaCubaBordaHigaFloresSchorfheideVillalvazo2021,eggertsson2021toolkit}, there are no general conditions for existence of equilibria for this class of models, as say, the Blanchard-Kahn conditions for standard linear dynamic RE models. Moreover, NK models with a ZLB are often presented as (log)linear approximations around an equilibrium of some originally nonlinear model, whose existence needs to be checked as an obvious precondition of the analysis. A number of theoretical papers provide sufficient conditions for existence of MSV equilibria in NK models \cite[see][]{egg2011nberma, Bonevaetal2016JME, Armenter2018, ChristianoEichenbaumJohannsen,Nakata2018,Nakataschmidt2019JME}, while \citetalias{AscariMavroeidis} provide both necessary and sufficient conditions that can be applied more generally.} 

There are two main takeaways from \citetalias{AscariMavroeidis}. First, the question of coherence  is a nontrivial problem in models with a ZLB constraint and \citetalias{AscariMavroeidis} were only able to provide some general results for a limited class of models. A typical New Keynesian (NK) model with a ZLB constraint is not generically coherent both when the Taylor rule is active and when monetary policy is optimal under discretion. The restrictions on the support of the shocks that are needed to restore an equilibrium are difficult to interpret because they are asymmetric and because they depend both on the structural parameters and on the past values of the state variables. \citetalias{AscariMavroeidis} show that the assumption of orthogonality of structural shocks is incompatible with coherence, because if a model admits multiple shocks, their support restrictions cannot be independent from each other.
Second, imposing the (somewhat awkward) support restrictions needed to guarantee existence of a solution causes another serious problem: multiplicity of MSV solutions, i.e., incompleteness.\footnote{In \citetalias{AscariMavroeidis}, an MSV equilibrium is defined as usually intended, that is, as a function of the state variables of the model. However, an incoherent model could in principle admit other types of equilibria, but, to the best of our knowledge, no work in the literature, including \citetalias{AscariMavroeidis}, has found them. We use the terminology MSV and REE interchangeably in the case of incoherence.} \citetalias{AscariMavroeidis} show the existence of many MSV solutions, possibly up to $2^k$ MSV equilibria, where $k$ is the number of (discrete) states that the exogenous variables can take, for example, using a $k$-state approximation of an AR(1) process. 
While the literature on the ZLB has recognized the possibility of multiple steady states and/or multiple equilibria, and of sunspots solutions due either to indeterminacy or to belief-driven fluctuations between the two steady states, this is a novel source of multiplicity, that concerns `fundamental' solutions, i.e., MSV ones. This is particularly relevant because numerical solution algorithms usually search for a solution of this type. The multiplicity of MSV solutions arises from the interaction between RE and the non-linear nature of the problem, as we will show below. Our paper investigates whether relaxing the full-information RE assumption could alleviate the problems highlighted by \citetalias{AscariMavroeidis} by breaking this interaction.

\subsection{Rationality without Coherence} \label{sec: ratnocoh}

We start by assuming full-information RE to illustrate the problem of incoherence. For simplicity, we focus on MSV REE, but some of the insights from our paper can be extended to study non-fundamental ``sunspot" equilibria which feature extraneous volatility. Since our model, (\ref{eq:IS})-(\ref{eq:MP}), is a purely forward looking model with a two-state discrete-valued exogenous shock, the MSV REE law of motion for $Y_t=(x_t,\pi_t)'$ will assume the form $Y_t=\textbf{Y}_{j}$ where $Y_t=\textbf{Y}_{1}$ if $\epsilon_t=\epsilon_1$ and $Y_t=\textbf{Y}_{2}$ otherwise.

\begin{definition} \label{def: REE}
\textit{ \textbf{Rational expectations equilibrium (REE)}.} 
$\textbf{Y}=(\textbf{Y}_1',\textbf{Y}_2')'$\textit{ is a rational expectations equilibrium if and only if} $\textbf{Y}_j$ \textit{solves (\ref{eq:IS})-(\ref{eq:MP}) given} $\hat E_t (Y_{t+1}|\epsilon_t=\epsilon_j)=Pr(\epsilon_{t+1}=\epsilon_1|\epsilon_t=\epsilon_{j})\textbf{Y}_1+Pr(\epsilon_{t+1}=\epsilon_2|\epsilon_t=\epsilon_{j})\textbf{Y}_2,$  for $j=1,2$.
\end{definition}

There are up to four MSV REE of (\ref{eq:IS})-(\ref{eq:MP}). First, there is a possible solution in which interest rates are always positive (``PP" solution). Then, there is a potential solution with binding ZLB if and only if $\epsilon_t=\epsilon_1$, which we refer to as the ``ZP" solution. Analogously, there could be a ``PZ" solution with binding ZLB if and only if  $\epsilon_t=\epsilon_2$. Finally, it is possible that the ZLB is always binding (``ZZ" solution). We add a superscript $i$ to $\textbf{Y}$ to distinguish between the REE (i.e. $\textbf{Y}^i$ where $i=PP,ZP,PZ,ZZ$). 
Following \citetalias{AscariMavroeidis}, if at least one of the four possible REE exist then the model is coherent.

\begin{proposition} \label{prop1}
Consider (\ref{eq:IS})-(\ref{eq:MP}) and suppose $M=M_f=N=1$, $\epsilon_2\ge0$. A rational expectations equilibrium (REE) exists if and only if $\epsilon_1 \ge \bar{\epsilon}_{REE}$, where $\bar{\epsilon}_{REE}$ is a constant that depends on the model's parameters, defined in Equation \eqref{eq: cutoff REE} in Appendix \ref{appe sec: prop1}.
\end{proposition}
Proposition \ref{prop1} generalizes Proposition 5 of \citetalias{AscariMavroeidis} to the case with $q<1$. It establishes that under the conventional assumption that the Taylor rule (\ref{eq:MP}) satisfies the Taylor Principle and recurrent demand shocks, we need to restrict the magnitude of the shocks, $\epsilon_t$, to get a REE. For a solution to exist, $\epsilon_1$ cannot be too negative (i.e. the shock cannot be too ``big", in absolute value). The lower bound on $\epsilon_1$, denoted as $\bar{\epsilon}_{REE}$, is increasing in $p$ for standard parameters, which means that a model with more persistent shocks requires tighter restrictions on the magnitude of the shocks for an equilibrium to exist. This explains why fundamentals-driven liquidity trap cannot be persistent in a REE. A ``big" shock is needed to take the economy into a liquidity trap, but then, for a REE to exist, it cannot be persistent. 
Thus, the model is not generically coherent; solutions only exist for special calibrations of the shock process and solutions do not exist if the shocks are too persistent (i.e. $p$ is very high) or if the shock is big ($\epsilon_1$ is very low). 

\paragraph{Intuition from a special case.} While Proposition \ref{prop1} deals with the case with $q<1$, the assumption that the high demand state is absorbing ($q=1$) and equal to zero ($\epsilon_2=0$) is helpful for intuition.\footnote{The assumption $q=1$ is standard in the literature \citep[e.g.,][]{Eggertsson, ChristianoEichenbaumJohannsen, Bilbiie}. To explain the intuition, we borrow heavily from \citetalias{AscariMavroeidis} and \cite{Bilbiie}.} Under this assumption, the economy under full-information RE either returns to the steady state with zero inflation (i.e. $\pi_t=x_t=i_t=0$) or the steady state with zero interest rates (i.e. $i_t=-\mu$, $\pi_t=-\mu<0$ $x_t=-\mu(1-\beta)/\lambda<0$). The ``temporary state" value of output when $\epsilon_t=\epsilon_1 <0$ (assuming for brevity that we go back to the zero-inflation steady state) is given by:
\begin{eqnarray}
x_t&=&\nu(p)E_tx_{t+1}-\sigma \max\{\frac{\psi \lambda}{1-\beta p}x_t,-\mu\}+\epsilon_1, \label{simple}\\
\nu(p)&:=& \left(1+\frac{\lambda \sigma}{1-\beta p}\right)>1, 
\end{eqnarray}
which we obtain by substituting the Phillips curve and Taylor rule into (\ref{eq:IS}). From (\ref{simple}), it is apparent that for any $p$, sufficiently low values of $\epsilon_1$ preclude unconstrained interest rates. Thus, for a sufficiently negative demand shock, output will be given by:
\begin{eqnarray}
x_t=\frac{1}{1-p\nu(p)}(\sigma \mu+\epsilon_1),
\end{eqnarray}
if a solution of the model exists at all. However, if the negative demand shock is sufficiently persistent, so that $p\nu(p)>1$, then $x_t$ and therefore temporary inflation, $\pi_t=\frac{\lambda}{1-\beta p}x_t$ are decreasing in $\epsilon_1.$ This implies that sufficiently large $\epsilon_1$ will increase $x_t$ and $\pi_t,$ precluding existence of a solution in which the ZLB binds. Therefore, for a solution to exist we need to either restrict $p$ to be small enough to ensure $p\nu(p)<1$, which in turn implies a solution for any $\epsilon_1$, or, alternatively, we need to restrict $\epsilon_1$ to be close to zero.

\begin{figure}[htp]
\caption{Incoherence and Income vs. Substitution}
\centering
    \begin{subfigure}{3.2in}
        \includegraphics[width=\linewidth]{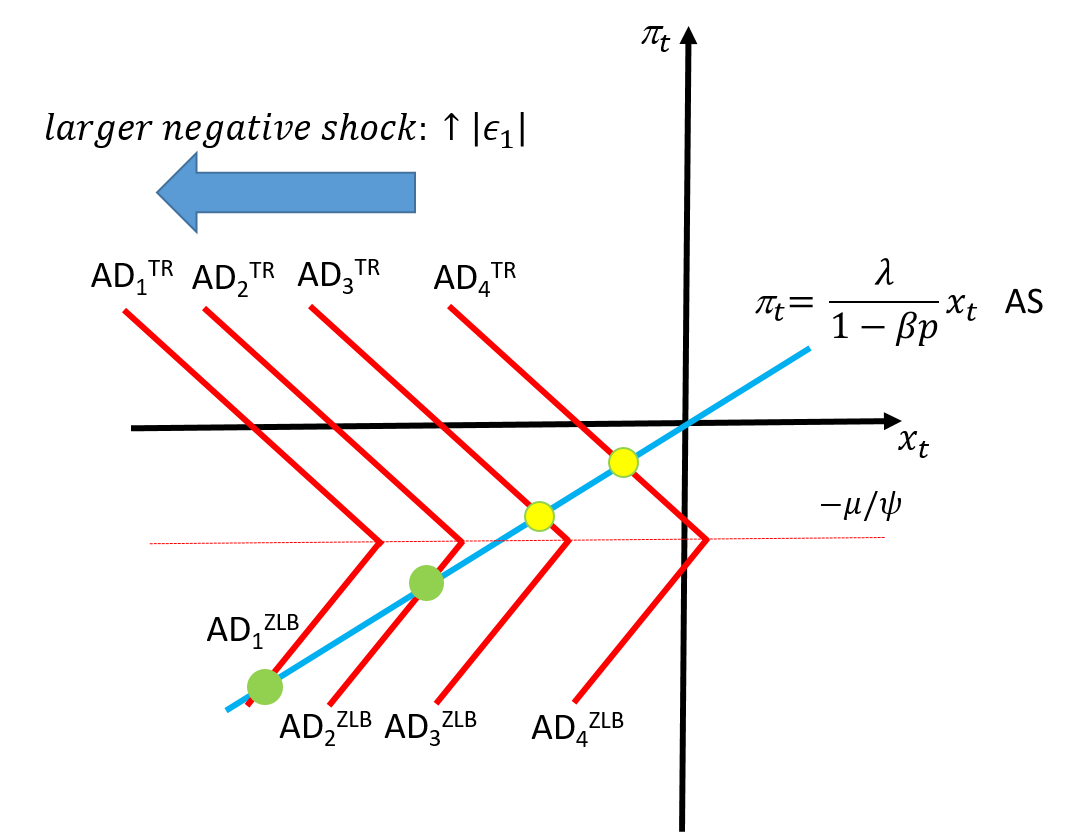}
        \caption{$p\nu(p)<1$}
        \label{fig: pvless1}
    \end{subfigure}
    \begin{subfigure}{3.2in}
        \includegraphics[width=\linewidth]{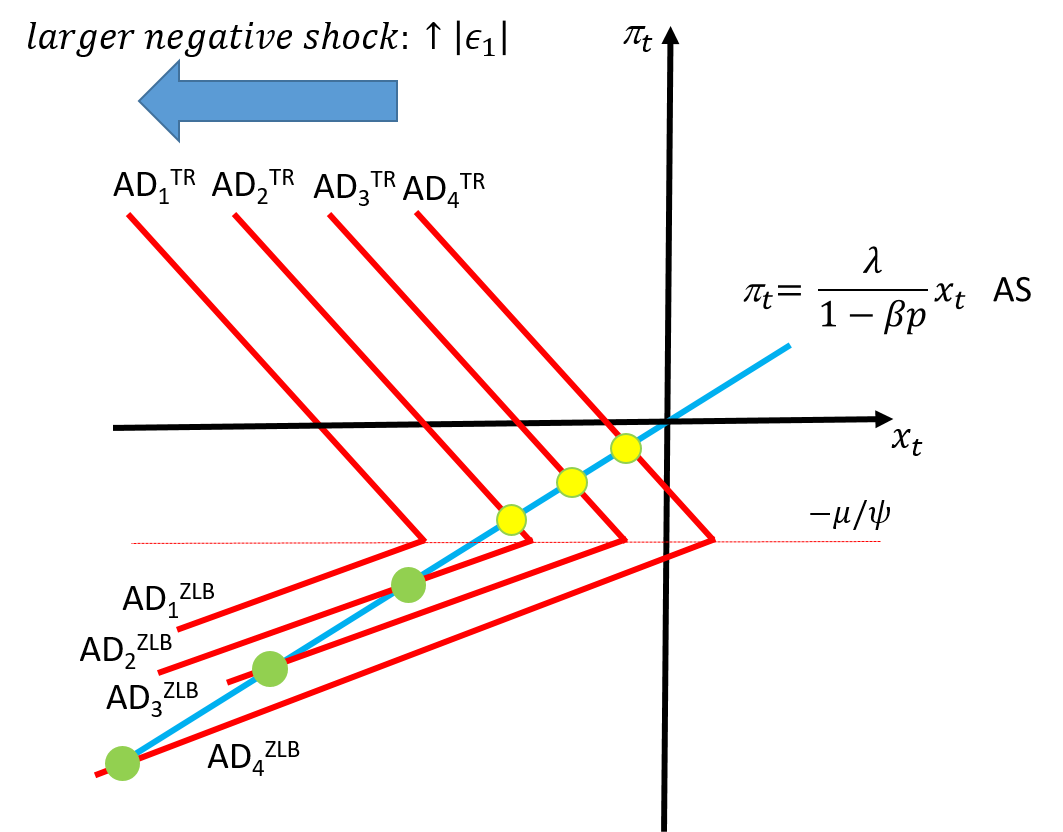}
        \caption{$p\nu(p)>1$}
        \label{fig: pvmore1}
    \end{subfigure}
 \caption*{\footnotesize
        Note: ``$AS$" (``$AD$") stands for aggregate supply (demand) curve; ``$ZLB$'' stands for zero-lower-bound regime; ``$TR$'' stands for Taylor rule. The ``$AD$" is piecewise linear depending on whether the ZLB is binding ($AD^{ZLB}$) or slack ($AD^{TR}$).  Yellow (green) dots indicate equilibria with a positive (zero) interest rate.}
    
\label{fig: inutition}
\end{figure}

Figure \ref{fig: pvless1} illustrates the determination of demand for the case $p\nu(p)<1$. It can be seen that a solution exists for any $\epsilon_1$. Figure \ref{fig: pvmore1} illustrates equilibrium determination when $p \nu(p)>1$. It is apparent that two solutions exist if $\epsilon_1$ is small, but no solution if $\epsilon_1$ is large in magnitude. In this case, the model is generally incoherent, while, if we impose support restrictions, i.e., $\epsilon_1>\bar{\epsilon}_{REE}$, the model is incomplete. The issue of incompleteness will be tackled in Section \ref{sec: learning}.\footnote{In fact two or four solutions exist in the two cases, respectively, depending on whether one assumes the economy returns to the zero-inflation steady state---as in Figures \ref{fig: pvless1} and \ref{fig: pvmore1}---or one assumes the economy goes to the permanent liquidity trap steady state---not depicted in Figures \ref{fig: pvless1} and \ref{fig: pvmore1}. Moreover, the figures express visually the way the condition $p\nu(p)\lesseqqgtr 1$ relates to the relative slope of the $AS$ and the $AD$ curve under the ZLB. See \citetalias{AscariMavroeidis}.}

How should we interpret this restriction on $p$ and $\epsilon_1$? Following \cite{Bilbiie}, there are two effects of the demand shock, $\epsilon_1$, when interest rates are pegged at zero. First, a larger demand shock (i.e., a more negative value of $\epsilon_1$) raises real interest rates given a fixed nominal rate, inducing households to save more. This intertemporal substitution effect should put downward pressure on inflation and output.  At the same time, $\nu(p)>1$ implies strong income effects at the ZLB;  current income, $x_t$, responds by \textit{more} than proportionally to an increase in expected future output, $E_tx_{t+1}$. For high values of $p$, an exogenous increase in real interest rates (via lower $\epsilon_1$) raises demand and inflation through this income effect. In the case where $p \nu(p)>1$, the income effect dominates the substitution effect, and the negative demand shock has the counter-intuitive effect of raising inflation at the ZLB, while lowering inflation away from the ZLB (see the green and yellow dots respectively in Figure \ref{fig: pvmore1}) . In this scenario, we need to make sure that $\epsilon_1$ is not \textit{too} negative. On the other hand, if $p \nu(p)<1$ then intertemporal substitution effects dominate and a larger negative shock (more negative $\epsilon_1$) pushes down inflation and output, which in turn ensures that a solution with a binding ZLB always exists. 

In sum, we can discuss the problem of incoherence in our model in terms of income and substitution effects. RE implies that agents are entirely forward-looking, which in turn allows for a scenario where income effects dominate substitution effects. Tight restrictions on the persistence parameter, $p$, are necessary to avoid such cases, while restrictions on $\epsilon_1$ are essential to ensure equilibrium existence when income effects are strong. Much of the rest of this paper investigates whether deviations from RE can ensure that these substitution effects dominate income effects when $p \nu(p)>1$, thus opening up the possibility that non-rational solutions exist even when rational solutions do not.

\subsection{Coherence without Rationality}\label{sec: cohnorat}
We now turn to the question of what happens if no REE exists. Specifically, we investigate the possible existence of non-rational equilibria. First, we look at the case of imperfect knowledge as in Definition \ref{def: IK}. Agents with imperfect knowledge are assumed to recursively estimate simple subjective forecasting models in the spirit of the adaptive learning literature. We assess existence of \textit{temporary equilibria} when agents are learning. Then, we ask if there exists an adaptive learning process that could generate an equilibrium where agents expectations are confirmed. We show that a self-confirming RPE may emerge as the outcome of an adaptive learning process where agents use an under-parameterized forecasting rule and attempt to forecast period-ahead inflation and output using their estimates of the long-run average of both variables. Second, bounded rationality does not need to imply imperfect knowledge, and so it is important to consider what happens when agents are boundedly rational as in Definition \ref{def: BR}. It turns out that bounded rationality in the form of discounting ($M,M_f,N<1$) can imply an even more complete resolution of the problem of incoherence than RPE, provided that the discount factors are exogenously given and do not depend on the magnitude of the shock.

\subsubsection{Restricted Perceptions}\label{sec: RPE}
The model (\ref{eq:IS})-(\ref{eq:MP}) has a single state
variable, $\epsilon_{t}$, which follows a regime-switching process.
Consequently, the REE law of motion for output and inflation is a
regime-switching intercept---see Definition \ref{def: REE}. Rational agents
are assumed to know the functional form of the REE solution. However, agents
without RE could fail to grasp the structure of the REE, particularly so in
the case of incoherence when no such equilibrium exists. 
Consequently, they might try to forecast inflation and output using an
under-parameterized forecasting model which omits the state variable,
$\epsilon_{t}$. Agents with these restricted perceptions instead try to
forecast the \textit{unconditional} mean of output and inflation:
\begin{equation}
\hat E_{t} Y_{t+j}=Y^{e}_{t} = Y^{e}_{t-1}+t^{-1}\left( Y_{t-k}-Y^{e}%
_{t-1}\right) ,\label{RPEPLM}%
\end{equation}
where $Y^{e}_{t}$ is the agents' most recent least squares estimate of the
unconditional mean of $Y=(x,\pi)^{\prime}$ using all data available from
$t=0,\hdots,t-k$ where $k=0$ if agents have current information and $k=1$ if
agents have lagged information and only observe endogenous variables after
markets clear. We assume a decreasing gain parameter equal to $t^{-1}$, but
more generally the gain parameter could be a small constant, $g_{y}\in(0,1]$
for $y=x,\pi$ (``constant-gain learning"), or a mix of constant-gain and
decreasing-gain learning as in \cite{MarcetNicolini}.

If we substitute (\ref{RPEPLM}) into the model (\ref{eq:IS})-(\ref{eq:MP})
with $M=M_{f}=N=1$ then we have the following result.

\begin{proposition}
\label{prop: CC temp RPE} The model (\ref{eq:IS})-(\ref{eq:MP}) with
$M=M_{f}=N=1$ and expectations formed according to (\ref{RPEPLM}) with $k=1$
is coherent and complete for all $\sigma,\lambda,\psi>0$.
\end{proposition}

Coherence and completeness means in this context that the model admits a ``temporary equilibrium'', that is, it  has a unique solution for the endogenous variables $Y_{t}$  for any given $p,q,\epsilon_1,\epsilon_2$, provided that $Y_t$ is not observed contemporaneously (i.e. $k=1$). We consider this to be an inherently significant finding. From a theoretical perspective, it shows that relying on the lagged information assumption, commonly employed in the adaptive learning literature, suffices to solve the coherence problem in a NK model with a ZLB constraint.\footnote{If $k=0$ then a temporary equilibrium can fail to exist for small values of $t$ with decreasing-gain, or sufficiently large constant gain parameters. Therefore, under contemporaneous information we need to restrict the magnitude of the gain parameter to get a solution. \cite{EM2018JMCB} document that constant-gain learning models with contemporaneous information can lead to unreasonable predictions when interest rates are pegged. Proposition \ref{prop: CC temp RPE} is a complementary result that favors the lagged information assumption.} Intuitively, learning implies that expectations are predetermined, and this simplifies the task of computing the market clearing equilibrium allocation relative to the nontrivial fixed point problem needed to solve for the REE. From an empirical perspective, inflation has been  mostly low but stable during and after the Great Recession, contrary to the prediction of deflationary spirals in an RE model. This proposition could provide a possible account of this period, so that inflation is actually determined by a temporary equilibrium, where agents update their beliefs based on an under-parameterized forecast rule as data becomes available with a lag.

Though a temporary equilibrium for the economy always exists, learning agents do not have expectations that are necessarily consistent with the data they observe. An equilibrium, instead, is a \textit{self-confirming equilibrium} if the learning agents' subjective inflation and output forecasts coincide with the true unconditional means of inflation and output, that is if:
\begin{eqnarray*}
\hat E_t Y_{t+j}=E(Y)=\bar q\mathbf{\hat Y}_2+(1-\bar q)\mathbf{\hat Y}_1,
\end{eqnarray*}
where $ Y=(x,y)'$, $\mathbf{\hat Y}_j$ is $Y_t$ when $\epsilon_t=\epsilon_j$ and $\bar q=Pr(\epsilon_t=\epsilon_2)=(1-p)/(2-p-q)$. If the agents form conditional forecasts using the unconditional mean of inflation and output (i.e. if $\hat E_t Y_{t+j}=E(Y)$) then agents' beliefs about the long-run averages of inflation and output are true and self-confirming only if $\mathbf{\hat Y}_j$ solves (\ref{eq:IS})-(\ref{eq:MP}) given $\hat E_tY_{t+j}=E(Y)=\bar q\mathbf{\hat Y}_2+(1-\bar q)\mathbf{\hat Y}_1$ and $\epsilon_t=\epsilon_j$ for $j=1,2$.

\begin{definition} \label{def: RPE}
\textit{\textbf{Restricted perceptions equilibrium (RPE)}.} $\mathbf{\hat{ Y}}=(\mathbf{\hat Y}_1',\mathbf{\hat Y}_2')'$ \textit{ is a restricted perceptions equilibrium if and only if (i)} $\mathbf{\hat Y}_j$ \textit{solves (\ref{eq:IS})-(\ref{eq:MP}) given} $E_tY_{t+1}=\bar{\mathbf{ Y}}:=\bar q \mathbf{\hat Y}_2+(1-\bar q) \mathbf{\hat Y}_1$ \textit{and } $\epsilon_t=\epsilon_j$ \textit{for $j=1,2$; and (ii)} $E(Y_t)=\bar{\mathbf{Y}}$.\footnote{See \citet[sec. 3.6 and 13.1]{evans2001learning}, \cite{branchRPEsurvey} and \cite{branchRPEsurvey2} for a thorough discussion of the RPE concept.}
\end{definition}

There are four possible RPE of (\ref{eq:IS})-(\ref{eq:MP}) indexed by $i=PP,ZP,PZ,ZZ$, which are analogous to the REE discussed earlier. In a RPE, agents have ``restricted perceptions'' in the sense that they omit key fundamental state variables from their forecasting models, that is, they use an under-parameterized forecast rule. In our simple model, $\epsilon_t$ is the only state variable. Consequently, the natural under-parameterized forecast rule for this model omits $\epsilon_t$ as (\ref{RPEPLM}) does.  This RPE concept also makes the analysis tractable, leading to the following useful result.

		\begin{figure}[htp!]
			\caption{Restricted Perceptions Equilibrium\label{fig:rpe}}
			\centering
					\includegraphics[scale=.5]{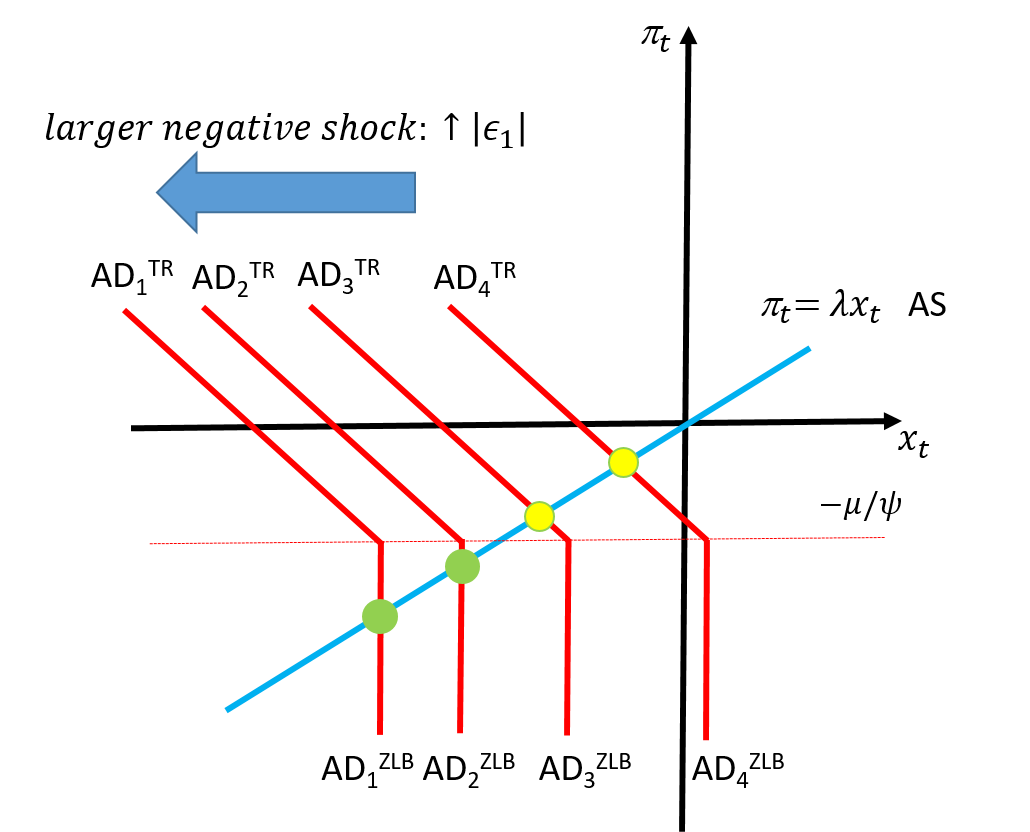}
      \caption*{\footnotesize
        Note: ``$AS$" (``$AD$") stands for aggregate supply (demand) curve; ``$ZLB$'' stands for zero-lower-bound regime; ``$TR$'' stands for Taylor rule. The ``$AD$" is piecewise linear depending on whether the ZLB is binding ($AD^{ZLB}$) or slack ($AD^{TR}$). Yellow (green) dots indicate equilibria with a positive (zero) interest rate.}
		\end{figure}

    \begin{proposition} \label{prop: RPE existence}
Consider (\ref{eq:IS})-(\ref{eq:MP}) and suppose $M=M_f=N=1$, $\epsilon_2\ge0$. Then:
\begin{enumerate}
    \item[i.] A restricted perceptions equilibrium (RPE) exists if and only if $\epsilon_1\ge\bar{\epsilon}_{RPE}$, where $\bar{\epsilon}_{RPE}$ depends on the model's parameters, see Equation \eqref{eq: RPE cutoff} in Appendix \ref{appe sec: prop3}, and satisfies $\bar{\epsilon}_{RPE}=-\infty$ if $q=1$.
    \item[ii.] $\bar{\epsilon}_{REE}\ge\bar{\epsilon}_{RPE}$ if and only if $p+q\ge1$.
\end{enumerate}
    \end{proposition}
Proposition \ref{prop: RPE existence} is one of the main results of this paper. It tells us that models with persistent shocks (i.e. $p+q>1$) admit non-rational equilibria but \textit{not} rational equilibria if $\epsilon_1\in[\bar{\epsilon}_{RPE},\bar{\epsilon}_{REE})$.\footnote{We note that $Corr(\epsilon_t\epsilon_{t-1})=\left(E(\epsilon_t\epsilon_{t-1})-[E(\epsilon_t)]^2\right)/(E(\epsilon_t^2)-[E(\epsilon_t)]^2)=p+q-1$. If $p+q=1$, then there is no distinction between the REE and RPE because $\epsilon_t$ is i.i.d. } Thus we can gain traction in an otherwise incoherent model of the ZLB by assuming restricted perceptions.

As in the case of REE, it is useful to study RPE when $q=1$ and $\epsilon_2=0$ to develop intuition, see Figure \ref{fig:rpe}. In this case, we have $\bar q=1$ and so the RPE forecast is simply equal to one of the two non-stochastic steady states of the model. Substituting the forecast consistent with the economy reverting to the zero inflation steady state into the model---so $\hat{E}_{t} x_{t+1}=\hat{E}_{t} \pi_{t+1}=0$ in (\ref{eq:IS})-(\ref{eq:MP})---and solving for equilibrium output in the temporary state with $\epsilon_t=\epsilon_1$ gives: $x_t=\sigma\mu+\epsilon_1$, assuming the ZLB binds. Thus, effectively the perceived $p$ is equal to zero and the slope of the aggregate demand curve becomes vertical in the temporary state under a ZLB. It follows that a RPE exists for \textit{any} $p$ and $\epsilon_1$. No support restrictions for the shock distribution are needed. Restricted perceptions ensures that the income effects of raising real rates do not dominate the substitution effects, and thus equilibrium is ensured for any values of $p$ and $\epsilon_1$, in accordance with Proposition \ref{prop: RPE existence}.

\subsubsection{Bounded Rationality} \label{sec: BRE}
Assuming bounded rationality in the form of discounting ($M,M_f,N<1$) yields the following proposition that illustrates how deviations from RE ameliorate incoherence concerns, as in Proposition \ref{prop: RPE existence}.
\begin{proposition}\label{prop:BR coherence}
Consider (\ref{eq:IS})-(\ref{eq:MP}) and suppose $min\{M, M_f, N\}<1$ and $\epsilon_2\ge0$. Then:
\begin{enumerate}
    \item[i.] A bounded-rationality equilibrium (BRE) exists if and only if $\epsilon_1 \ge \bar{\epsilon}_{BR}$, for some constant $\bar{\epsilon}_{BR}$ that depends on the model's parameters (see Equation \eqref{eq: BRE cutoff} in Appendix \ref{appe sec: prop4}). 
    \item[ii.] If $(M-1)(1-M_f\beta)+\lambda\sigma N<0$ then $\bar{\epsilon}_{BR}=-\infty$.
\end{enumerate}
\end{proposition}

Again, we can understand the coherence result in terms of the income and substitution effects of shocks that raises real interest rates at the ZLB. Assume $q=1$ and $\epsilon_2=0$. The BRE value of output in the temporary state binding ZLB is given by:
\begin{eqnarray}
x_t&=&\nu^{BR}(p)E_tx_{t+1}-\sigma \max\{\frac{\psi \lambda}{1-M_f \beta p}x_t,-\mu\}+\epsilon_1,\\
\nu^{BR}(p)&:=& \left(M+N\frac{\lambda \sigma}{1-\beta M_f p}\right). \nonumber
\end{eqnarray}
In this bounded rationality model, output at the ZLB is, therefore, given by
\begin{eqnarray}
x_t=\frac{1}{1-p\nu^{BR}(p)}(\sigma \mu+\epsilon_1).
\end{eqnarray}
Clearly, substitution effects dominate income effects if and only if $p\nu^{BR}(p)<1$, similar to the RE case. However, unlike the RE case, we have $\nu^{BR}(p)<1$ for any $p$ if and only if
\[(M-1)(1-M_f\beta)+\lambda\sigma N<0,\]
which is the condition in Proposition \ref{prop:BR coherence}. Therefore, myopia can ensure that substitution effects dominate income effects for any $p$ (i.e., implying existence of a MSV solution for any $p$ and $\epsilon_1$).

Not only does $(M-1)(1-M_f\beta)+\lambda\sigma N<0$ ensure coherence in the case of bounded rationality, it also ensures existence of a unique BRE (``completeness"), as formalized in the following proposition. 

\begin{proposition} \label{prop: BR completeness}
Consider the model given by (\ref{eq:IS})-(\ref{eq:MP}) and assume $\psi>1$. A unique bounded rationality equilibrium (BRE) exists for any $p,q,\epsilon_1$ and $\epsilon_2\ge 0$ if and only if $(M-1)(1-M_f\beta)+\lambda\sigma N<0$. Further, there exist $\epsilon^{PP,BR}$ and $\epsilon^{ZP,BR}$ such that $\epsilon^{PP,BR}>\epsilon^{ZP,BR}$ and
\begin{enumerate}
    \item[i.] The PP solution is the unique BRE if and only $\epsilon_1>\epsilon^{PP,BR}$.
    \item[ii.] The ZP solution is the unique BRE if and only if $\epsilon^{PP,BR}\ge \epsilon_1>\epsilon^{ZP,BR}$.
    \item[iii.] The ZZ solution is the unique BRE if and only if $\epsilon_1\le \epsilon^{ZP,BR}$.
\end{enumerate}
\end{proposition}

Although the condition $(M-1)(1-M_f\beta)+\lambda\sigma N<0$ completely mitigates concerns about incoherence and incompleteness, it requires a rather high degree of discounting in the Euler and Phillips curve equations. As it turns out, the condition is satisfied by Gabaix's preferred calibration: $M=0.85$, $M_f=0.8$, $N=1$, $\beta=0.99$, $\lambda=0.11$, $\sigma=0.2$. For that calibration, we have:
 \begin{eqnarray*}
 (M-1)(1-M_f\beta)+\lambda\sigma N=-0.0092<0.
 \end{eqnarray*}
 On the other hand, it is not satisfied for the calibration in \cite{mckay2016discounting}: $M=0.97$, $M_f=N=1$, $\beta=0.99$, $\lambda=0.02$, $\sigma=0.375$. That calibration yields:
  \begin{eqnarray*}
 (M-1)(1-M_f\beta)+\lambda\sigma N=0.0072>0.
 \end{eqnarray*}
Thus bounded rationality offers a full solution of the problems of incoherence and incompleteness for some, but not all, calibrations featured in the literature. 

\subsubsection{BRE, RPE and Coherence}

Bounded rationality and imperfect knowledge constitute two distinct departures from RE that are widely discussed in the literature, and they both mitigate concerns about coherence. In this regard, several points are worth considering.

First, bounded rationality might seem to provide a more robust resolution to the problem relative to imperfect knowledge, as coherence can be ensured for any assumption about $p$, $q$ and $\epsilon_t$ if $M,M_f,N$ are sufficiently small. 
However, this need not be the case 
if prices are relatively flexible or if agents choose their discount factors optimally as in \cite{Moberly}. 

\begin{figure}[htp]
 \caption{Region of Coherence of the REE, RPE, and of the BRE}
 \centering
    \begin{subfigure}{2in}
        \includegraphics[scale=.35]{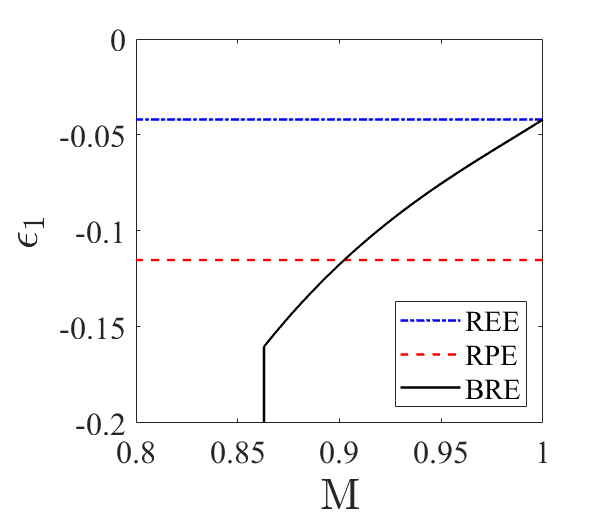}
		\caption*{(a)}
        \label{fig: BR coherence0.98}
    \end{subfigure}
     \begin{subfigure}{2in}
       \includegraphics[scale=.35]{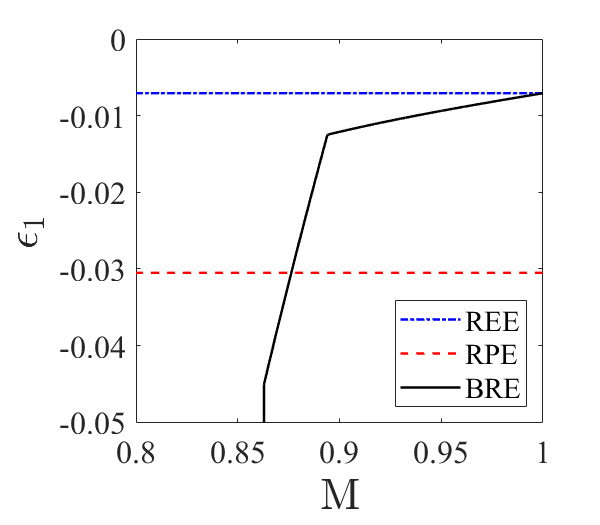}
		\caption*{(b) $p=0.965$}
         \label{fig: BR coherence0.9}
     \end{subfigure}
   \begin{subfigure}{2in}
        \includegraphics[scale=.35]{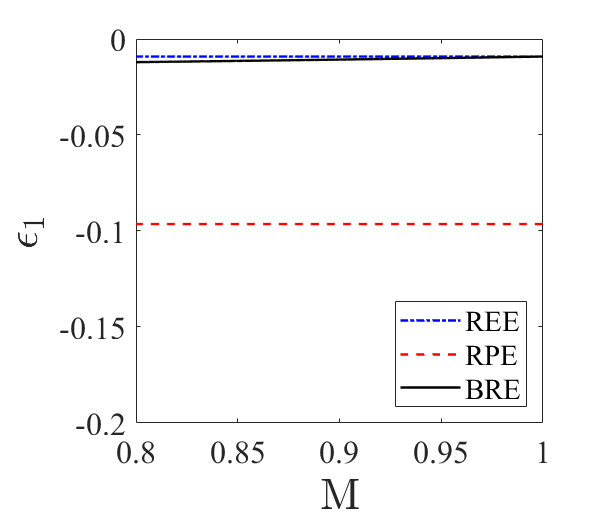}
		\caption*{(c) $\lambda=0.2$}
        \label{fig: BR coherencelambda8}
    \end{subfigure}

     \caption*{\footnotesize
        Note: The area above the blue (red) curve depicts values of $\epsilon_1$ for which at least one REE (RPE) exists. The area above the black curve depicts values of $\epsilon_1$ and $M=M_f$ for which at least one BRE exists. Other parameter values: $\beta=0.99$, $\sigma=1$, $\lambda=0.02$, $q=0.98$, $p=0.85$, $N=1$, $\epsilon_2=0.01$.}
\label{fig: BR cohe regions }
\end{figure}

To illustrate the importance of price rigidity, Figure \ref{fig: BR cohe regions } depicts different combinations of values for the negative shock, $\epsilon_1,$ and for the bounded rationality discount factor, $M$, that yield coherence in the REE, RPE and BRE cases. The blue and red lines depict $\bar{\epsilon}_{REE}$ and $\bar{\epsilon}_{RPE}$, respectively, and the black line depicts $\bar{\epsilon}_{BRE}$ for different values of $\epsilon_1$ and $M=M_f$. Panels (a), (b) and (c) shows that the difference between $\bar{\epsilon}_{REE}$, $\bar{\epsilon}_{RPE}$, and $\bar{\epsilon}_{BR}$ can be substantial. Panel (a) shows that larger values of $M$ can rule out existence of BRE in cases where a RPE exists. Panel (b) shows that the same result holds even if the expected duration of the low-demand state is calibrated to match the duration of the 2008-2015 U.S. ZLB episode (i.e. $p=0.965$ implies an expected duration of 28 quarters). However, if $M<0.86$ in the calibrated model then $(M-1)(1-M_f\beta)+\lambda\sigma N<0$ and $\bar{\epsilon}_{BRE}=-\infty$. Panel (c) reveals that in addition to small $M$, a high degree of price stickiness (small $\lambda$) is necessary for the BRE approach to provide a more complete solution to the incoherence problem than the RPE concept. For high values of $\lambda$ even heavy cognitive discounting in the Euler equation and Phillips curve will not resolve the problem of incoherence.\footnote{For any $M$, $M_f$, $N$, there is always a large enough value of the product $\lambda \sigma$ to ensure that $(M-1)(1-M_f\beta)+\lambda\sigma N>0$. Thus, price rigidity and the intertemporal elasticity of substitution play a key role in the existence of BRE.}  The so-called ``curse of flexibility'' is therefore a much more pronounced problem for both REE and BRE than for RPE.  
When considered alongside the theoretical literature on state-dependent models, and the empirical evidence on the time-variation of the frequency of price-setting, both of which indicate that the flexibility of prices might vary with economic conditions, one might expect that in deep recessions where the ZLB is binding persistently, prices should be more flexible and thus $\lambda$ should be high, making the solution provided by BRE less robust. 

BRE also may not exist if agents are assumed to choose their discount factors optimally. Thus far, in keeping with most of the literature on the bounded rationality approach by \cite{gabaix2020}, we have kept fixed the cognitive parameters $M,M_f,N.$ However, the degree of attention of agents should be endogenous, and agents might pay more attention when the economy is subject to large shocks, as in deep recessions where the ZLB is binding persistently. Appendix \ref{appe sec: EndoBR} employs the approach developed by \cite{Moberly} to endogenize the degree of attention in the \cite{gabaix2020} model. In \cite{Moberly}, firms and households face a cost of paying attention, as in \cite{gabaix2020}, and they choose discount factors, $M_{f,\epsilon_t},M_{\epsilon_t}$ in order to balance the loss of not paying attention with the cost of paying attention. Appendix \ref{appe sec: EndoBR} shows that in this case the shock must be bounded for a solution to exist. Intuitively, it is optimal to pay full attention ($M_{f,\epsilon_t}=M_{\epsilon_t}=1$) when the shock $\epsilon_1$ is sufficiently large in magnitude. However, a solution does not exist when the shock is large and discount factors are high (see Proposition \ref{prop:BR coherence}). Appendix \ref{appe sec: EndoBR} details this important caveat, showing that whether bounded rationality solves the problem of incoherence hinges on whether discount factors are predetermined or fixed.

Second, the results above cast doubt on whether the BRE concept can provide a robust solution to the coherence problem, motivating the consideration of alternative departures from RE,
that is, imperfect knowledge/adaptive learning. 
However, it is important to note that the two deviations are not mutually exclusive, and some recent papers have combined imperfect knowledge with myopia or versions of bounded rationality. For example, \cite{Hajdini2022} studies the expectations of myopic agents who have misspecified forecasting models; \cite{MegMilani2021} estimates a model that combines adaptive learning and myopia; and \cite{AudzeiSlobodyan} derives restricted perceptions equilibrium in an environment that combines adaptive learning and Gabaix's sparse rationality.
Similarly, it is possible to combine the two deviations from RE in our model.

\begin{definition} \label{def: BRIK}
Agents have \textbf{bounded rationality \textit{and} imperfect knowledge} if $\hat E\neq E$; $\max\{M,M_f,N\}<1$ in the NK model given by Equations (\ref{eq:IS})-(\ref{eq:MP}).
\end{definition}

The analysis in Appendix \ref{appe sec: BRRPE} shows that an environment with boundedly rational agents who have imperfect knowledge could admit a bounded rationality RPE.

\begin{definition} \label{def: BRRPE}
\textit{\textbf{Bounded rationality restricted perceptions equilibrium (BR-RPE)}.} $\mathbf{\hat{ Y}}=(\mathbf{\hat Y}_1',\mathbf{\hat Y}_2')'$ \textit{ is a restricted perceptions equilibrium if and only if (i)} $\mathbf{\hat Y}_j$ \textit{solves (\ref{eq:IS})-(\ref{eq:MP}) given} $M,M_f,N$, $E_tY_{t+1}=\bar{\mathbf{ Y}}:=\bar q \mathbf{\hat Y}_2+(1-\bar q) \mathbf{\hat Y}_1$ \textit{and } $\epsilon_t=\epsilon_j$ \textit{for $j=1,2$; and (ii)} $E(Y_t)=\bar{\mathbf{Y}}$.
\end{definition}

There are four possible BR-RPE of (\ref{eq:IS})-(\ref{eq:MP}) indexed by $i=PP,ZP,PZ,ZZ$, which are analogous to the BRE and RPE discussed earlier. Suitable restrictions on the model ensure existence of BR-RPE.

\begin{proposition}\label{prop: BRRPE existence}
Consider (\ref{eq:IS})-(\ref{eq:MP}) and suppose $min\{M, M_f, N\}<1$ and $\epsilon_2\ge0$. Then:
\begin{enumerate}
    \item[i.] A bounded-rationality restricted-perceptions equilibrium (BR-RPE) exists if and only if $\epsilon_1 \ge \bar{\epsilon}_{BR,RPE}$, for some constant $\bar{\epsilon}_{BR,RPE}$ that depends on the model's parameters, see Equation \eqref{eq: BRRPE cutoff} in Appendix \ref{appe sec: BRRPE}. 
    \item[ii.] If $(M-1)(1-M_f\beta)+\lambda\sigma N<0$, then $\bar{\epsilon}_{BR,RPE}=-\infty$.
    \item[iii.] If $(M-1)(1-M_f\beta)+\lambda\sigma N\ge 0$ and $p+q\ge1$ or if $(M-1)(1-M_f\beta)+\lambda\sigma N<0$, then $\bar{\epsilon}_{BR}\ge\bar{\epsilon}_{BR,RPE}$.
\end{enumerate}
\end{proposition}

The condition for BR-RPE existence in Proposition \ref{prop: BRRPE existence} is weaker than the condition for BRE existence when the shocks are persistent ($p+q>1$). Thus, the two deviations from RE are not redundant, and combining them leads to a less restricted resolution to the incoherence problem than either assumption alone 
given that standard calibrations in the literature assume persistent shocks.

Finally, it is well known that bounded rationality can attenuate the so-called ``forward guidance puzzle" which is the counter-intuitive prediction that the macroeconomic effects of a promise to cut the interest rate in some future period, $T$, are strictly increasing in $T$. Theorem \ref{Theo FG} in Appendix \ref{appe sec: FG} proves that the condition in Proposition \ref{prop:BR coherence}.ii that ensures coherence/completeness in the occasionally-binding constraint framework, also rules out the forward guidance puzzle. Moreover, Propositions \ref{prop:FGP1} and \ref{prop:FGP2} in Appendix \ref{appe sec: FG}  show that the forward guidance puzzle is also absent under imperfect knowledge with adaptive learning. Note that the forward guidance problem is a very different problem from the coherence problem highlighted in this section. First, forward-guidance is generated by a peg of the interest rate, while a peg would not be an issue for coherence, i.e., for the existence of an equilibrium. Second, forward guidance is often modelled as a fixed interest rate for a known duration (and a known duration of the negative deflationary shock) and then the policy would revert to a standard Taylor rule. Again, if the duration of the shock and of the peg is known, there is no issue of incoherence. Indeed, the model of forward guidance used in \cite{gabaix2020} and in Appendix \ref{appe sec: FG} are not susceptible to the problem of incoherence.\footnote{See also \cite{EGP2021}, \cite{ColeJMCB}, and \cite{GibbsMcClung} for more on forward guidance and adaptive learning considerations.}
Thus, both deviations from RE help resolve various puzzles and paradoxes of the New Keynesian ZLB, in addition to resolving the problem of incoherence.

\section{Learning to solve incompleteness: multiplicity of (MSV) solutions
}\label{sec: learning}
We just saw that a BRE can ensure coherence and completeness with sufficient discounting, without any restrictions on the support of the shock.
What about completeness in the REE and RPE cases? The coherence condition guarantees existence, but this generally implies a multiplicity of admissible MSV solutions in the case of RE \citep[e.g.,][]{AscariMavroeidis}. Incompleteness is a problem that can only be solved using some criterion for selecting an equilibrium. Here we investigate whether learning can provide any guidance, that is, whether the ``E-stability" criterion can select an equilibrium of the model as the outcome of an adaptive learning process.

\subsection{Learning the REE}
In order to derive the conditions under which a REE is E-stable, we first need to be precise about what it means for agents to be learning a REE. As in Section \ref{sec: model}, adaptive learning agents have imperfect knowledge and cannot compute an equilibrium analytically. However, these agents make use of a subjective forecasting model or ``perceived law of motion" (PLM) when making consumption, labor, savings and pricing decisions consistent with (\ref{eq:IS})-(\ref{eq:PC}). If the learning agents choose a PLM that is also consistent with how expectations are formed in a REE, then it is possible for learning agents to ``learn" a REE if their beliefs about the PLM converge to RE, as beliefs are updated recursively using some statistical scheme for estimating the coefficients of the PLM and observable macro data.

Recall from Section \ref{sec: ratnocoh} that our model admits four possible REE in which output and inflation follow a two-state process, which are indexed by superscript $i$ to $\textbf{Y},$ i.e. $\textbf{Y}^i$ where $i=PP,ZP,PZ,ZZ$. Agents could conceivably learn one of these REE if their PLM for output and inflation is a two-state process which is estimated recursively using least squares. Consider the following model of learning, in which agents' PLM is a two-state process for inflation and output, like the REE, and beliefs about the state-contingent means are updated recursively using least squares:
\begin{eqnarray}
Y^e_{j,t}&=&Y^e_{j,t-1}+t^{-1}\mathcal{I}_{j,t-1}\nu_{j,t-1}^{-1}\left(Y_{t-1}-Y^e_{j,t-1}\right), \label{REEalg1} \\
\nu_{j,t}&=&\nu_{j,t-1}+t^{-1}\left(\mathcal{I}_{j,t-1}-\nu_{j,t-1}\right),  \label{REEalg2}\\
\hat E_t Y_{t+1}&=&Pr(\epsilon_{t+1}=\epsilon_1|\epsilon_t)Y^e_{1,t}+(1-Pr(\epsilon_{t+1}=\epsilon_1|\epsilon_t))Y^e_{2,t},\label{REEPLM}
\end{eqnarray}
where $j=1,2$,  $k\nu_{j,k}$ is the number of periods for which $\epsilon_t=\epsilon_j$ up until time $k$, and $\mathcal{I}_{j,t}=1$ if $\epsilon_t=\epsilon_j$ and $\mathcal{I}_{j,t}=0$ otherwise (i.e. $\mathcal{I}_{j,t}=1$ is the indicator function for state $j$). $Y^e_{j,t}$ is the agents' most recent estimate of the state-contingent average of $Y_t$ when $\epsilon_t=\epsilon_j$.
According to equation (\ref{REEalg1}), agents revise their beliefs about the state-contingent average of $Y$ in state $j$ (i.e. $Y^e_{j,t}$) in the direction of their time-$t-1$ forecast error only if $\epsilon_{t-1}=\epsilon_j$ (otherwise, $Y^e_{j,t}=Y^e_{j,t-1}$). Equation (\ref{REEPLM}) then gives agents' time-$t$ forecast of period-ahead inflation and forecast. It is assumed that agents observe $\epsilon_t$ when forecasting at time-$t$ and also that $Pr(\epsilon_{t+1}|\epsilon_t)$ coincides with the actual transition probabilities---e.g. agents know $Pr(\epsilon_{t+1}=\epsilon_1|\epsilon_t=\epsilon_1)=p$ and $Pr(\epsilon_{t+1}=\epsilon_2|\epsilon_t=\epsilon_2)=q$. After agents form time-$t$ expectations, we obtain the time-$t$ market-clearing equilibrium, $Y_t$, by substituting equation (\ref{REEPLM}) into the model (\ref{eq:IS})-(\ref{eq:MP}). The process repeats itself at time $t+1$ and so on.\footnote{Closely related learning algorithms are used by \cite{Woodford1990}, \cite{EvansHonkapohjaJET94} and \cite[p.305-308]{evans2001learning} to study the E-stability of sunspot equilibria involving discrete-valued shocks, and by \cite{EvansHonkapohjaJournalMath98} to study learnability of fundamental equilibria with exogenous shocks following a finite state Markov chain. We arrive at identical E-stability results if we alternatively assume least squares estimation of a PLM of the form: $Y^e_t=\hat a+\hat b \mathcal{I}_t$ where $\mathcal{I}_t=1$ if $\epsilon_t=\epsilon_2$ and 0 otherwise.}

We are interested in knowing if $(Y^e_{1,t},Y^e_{2,t})\rightarrow (\textbf{Y}^i_1,\textbf{Y}^i_2)$ for some REE $i$ as time goes on ($t\rightarrow \infty$) and agents' expectations evolve according to (\ref{REEalg1})-(\ref{REEPLM}). We say that REE $i$ is ``stable under learning" if $(Y^e_{1,t},Y^e_{2,t})\rightarrow (\textbf{Y}^i_1,\textbf{Y}^i_2)$ almost surely. When might this convergence of subjective beliefs to RE occur? To make this question tractable, assume that $Y^e_{t}=(Y^{e'}_{1,t}$, $Y^{e'}_{2,t})'$ is sufficiently near REE $i$, such that the ZLB binds under adaptive learning if and only if the ZLB would bind in REE $i$. This implies the following actual law of motion for $Y$:
\begin{eqnarray}
Y_t&=&A^i_{t}\left(Pr(\epsilon_{t+1}=\epsilon_1|\epsilon_t)Y^e_{1,t}+(1-Pr(\epsilon_{t+1}=\epsilon_1|\epsilon_t))Y^e_{2,t}\right)+B^i_t, \label{REEALM}
\end{eqnarray}
for $i\in \{PP,PZ,ZP,ZZ\}$, where $A^{PP}_t=A_P$ and $B^{PP}_t=B_{P,t}$ for all $t$; $A^{ZZ}_t=A_Z$ and $B^{ZZ}_t=B_{Z,t}$ for all $t$; $A^{ZP}_t=A_P$ and $B^{ZP}_t=B_{P,t}$ if $\epsilon_t=\epsilon_2$ and $A^{ZP}_t=A_Z$ and $B^{ZP}_t=B_{Z,t}$ otherwise; $A^{PZ}_t=A_P$ and $B^{PZ}_t=B_{P,t}$ if $\epsilon_t=\epsilon_1$ and $A^{PZ}_t=A_Z$ and $B^{PZ}_t=B_{Z,t}$ otherwise, and 
\begin{align*}
A_P &:= 
\begin{pmatrix}
\frac{1}{\lambda \sigma \psi + 1} & \frac{\sigma - \beta \sigma \psi}{\lambda \sigma \psi + 1} \\
\frac{\lambda}{\lambda \sigma \psi + 1} & \frac{\beta + \lambda\sigma}{\lambda \sigma \psi + 1}
\end{pmatrix} &\quad
A_Z &:= 
\begin{pmatrix}
1 & \sigma \\
\lambda & \beta + \lambda\sigma
\end{pmatrix}
\\
B_{P,t} &:= 
\begin{pmatrix}
\frac{\epsilon_t}{1 + \lambda\psi\sigma} \\
\frac{\lambda\epsilon_t}{1 + \lambda\psi\sigma}
\end{pmatrix} &\quad
B_{Z,t} &:= 
\begin{pmatrix}
\epsilon_t + \sigma\mu \\
\lambda\epsilon_t + \lambda\sigma\mu
\end{pmatrix}
\end{align*}

Given beliefs that are local to RE beliefs, we assess the learnability of equilibrium using the E-stability principle. A REE $i$ is said to be E-stable if it is a locally stable fixed point of the ordinary differential equation (ODE):
\begin{eqnarray} \label{ODE2}
\frac{\partial \tilde{Y}^e}{\partial \tau}=H^i(\tilde{Y}^e),
\hspace{2em} \text{where} \hspace{2em} 
H^i(\tilde{Y}^e):=
\begin{pmatrix}
Y^i_1(Y^e_1,Y^e_2)\\
Y^i_2(Y^e_1,Y^e_2)
\end{pmatrix}
-
\begin{pmatrix}
Y^e_1\\
Y^e_2
\end{pmatrix},
\end{eqnarray}
where $\tau$ is ``notional'' time, $Y^i_j(Y^e_1,Y^e_2)$ is the value of $Y$ when $\epsilon_t=\epsilon_j$ as a function of expectations, $\tilde{Y}^e:=(Y^{e'}_1,Y^{e'}_2)'$. The relevant Jacobian for assessing the E-stability of REE $i$ is: $DT_{Y^i}:=\frac{\partial H^i(\tilde{Y}^e)}{\partial \tilde{Y}^e}|_{\tilde{Y}^e=\textbf{Y}^i}$. A REE $i$ is E-stable if the eigenvalues of $DT_{Y^i}$ have negative real parts, see \cite{evans2001learning}.

There is an intuition for the link between the E-stability condition and stability of beliefs. The ODE (\ref{ODE2}) is an approximation of the dynamics of $Y^e_{t}$ near the REE for large $t$, and it tells us that agents' expectations are revised in the direction of the forecast error, $\bar{Y}^i(Y^e)-Y^e$. If the roots of $DT_{\bar Y^i}$ have negative real parts, then agents' expectations about the unconditional means of inflation and output are also revised in the direction of their REE values.

We note the E-stability conditions applied to the REE of the occasionally binding constraint model are identical to the E-stability conditions applied to a model that features exogenous Markov-switching in the monetary policy stance driven entirely by $\epsilon_t$ \citep[e.g., see][]{BDM2013,McClung2020}.\footnote{\cite{MertensRavn} also derive E-stability conditions for an equilibrium of a simple New Keynesian model with ZLB constraint, assuming a two-state discrete sunspot shock with an absorbing regime.} For example, the E-stability condition associated with the ZP equilibrium of (\ref{eq:IS})-(\ref{eq:MP}) is the same condition associated with the MSV solution of a model that assumes $i_t=\psi \pi_t$ if $\epsilon_t=\epsilon_2$ and $i_t=-\mu$ if $\epsilon_t=\epsilon_1$ regardless of whether the ZLB binds.

Applying the E-stability conditions to the model at hand leads us to the conclusion that only one REE has the property of being E-stable (see Appendix \ref{appe sec: prop7} for the proof).

        \begin{proposition}\label{prop: REE E-stable}
Consider (\ref{eq:IS})-(\ref{eq:MP}) and suppose $M=M_f=N=1$, $\epsilon_2\ge0$. Then:
\begin{enumerate}
    \item[i.] If $\epsilon_1>\bar{\epsilon}_{REE}$, at most one E-stable rational expectations equilibrium (REE) exists.
    \item[ii.] The E-stable REE is either the PP REE or the ZP REE.
\end{enumerate}
        \end{proposition}
        
Proposition \ref{prop: REE E-stable} somewhat extends insights from \cite{ChristianoEichenbaumJohannsen} to models with recurring low demand states (i.e. $q<1$). Thus Proposition \ref{prop: REE E-stable} can be applied to study an economy such as the U.S.~economy, which has visited the ZLB twice since 2007, following two distinct negative shocks to the economy. The result in Proposition \ref{prop: REE E-stable} makes it clear that while multiple solutions exist, only one of them can be understood as the outcome of an adaptive learning process. Hence, incompleteness is resolved by E-stability.

\subsection{Learning the RPE}
We now turn to the question of learnability of RPE. Proposition \ref{prop: RPE existence} shows that a RPE can exist even if a REE does not. It turns out multiple RPE may exist when the restrictions in Proposition \ref{prop: RPE existence} hold. Can one or more of these RPE emerge as the outcome an econometric learning process, similar to what we considered in the case of REE? The answer is yes. Here we show that the model may still admit one unique learnable, self-confirming RPE.

First, we must assume agents have a subjective PLM for output and inflation that is consistent with how expectations are formed in a RPE, which is given by equation (\ref{RPEPLM}). If we substitute (\ref{RPEPLM}) into the model and assume $Y^e_{t}$ is sufficiently near RPE $i$ then we have the following actual law of motion for $Y$:
\begin{eqnarray}
Y_t&=&A^i_{t}Y^e_{t}+B^i_t,\label{RPEALM}
\end{eqnarray}
where $A^{i}_t$ and $B^{i}_t$ are defined below equation (\ref{REEALM}).

We say that RPE $i$ is stable under learning if $Y^e_{t}\rightarrow \bar{\textbf{Y}}^i$ almost surely, where $\bar{\textbf{Y}}^i$ denotes the unconditional mean of $Y_t^i$. Analogous to the discussion of E-stability of REE above, we say that RPE $i$ is said to be E-stable if it is a locally stable fixed point of the ODE, $\partial Y^e/\partial \tau=h^i(Y^e)$, where $h^i(Y^e)=\bar{Y}^i(Y^e)-Y^e$
and $\bar{Y}^i(Y^e)$ is the unconditional mean of $Y$ as a function of expectations, $Y^e$. Formally, E-stability obtains if the eigenvalues of the Jacobian, $DT_{\bar Y^i}:=\frac{\partial h^i(Y_e) }{\partial Y^e}|_{Y^e=\bar{\textbf{Y}}^i}$ have negative real parts. An E-stable RPE is stable under learning if agents estimate $Y^e_{t}$ using least squares, as in (\ref{RPEPLM}), or related estimation routines.
\begin{proposition} \label{prop: RPE E-stable}
Consider (\ref{eq:IS})-(\ref{eq:MP}) and suppose $M=M_f=N=1$, $\epsilon_2\ge0$. If $\epsilon_1>\bar{\epsilon}_{RPE}$, then:
\begin{enumerate}
    \item[i.] There is a unique E-stable restricted perceptions equilibrium (RPE).
    \item[ii.] The E-stable RPE is either the PP RPE or the ZP RPE.
\end{enumerate}
        \end{proposition}
 Appendix \ref{appe sec: estab-BR-RPE} shows that a unique E-stable BR-RPE exists in the case where agents both are boundedly rational and have imperfect knowledge and BR-RPE exist.
 
Proposition \ref{prop: RPE E-stable} indicates that agents can learn a unique RPE, but an attentive econometric agent might also detect that RPE beliefs are misspecified. Is the RPE therefore unreasonable? In the case of coherence we might doubt the plausibility of RPE on the basis that a learnable REE may exist (Proposition \ref{prop: REE E-stable}). However, incoherence precludes REE, and as shown in Appendix \ref{appe sec: prop8}, agents fail to form self-confirming expectations using a variety of different forecasting models that condition on the demand shock or lags of the endogenous variables in the case of incoherence. Further, the economy easily derails into a deflationary spiral when agents attempt to learn the RE-consistent dynamics of inflation and output when no REE exists, while RPE remain learnable (Proposition \ref{prop: RPE E-stable}).  
Consequently, RPE provide coherent alternatives to REE in the case of rational incoherence by relaxing conditions for existence of a self-confirming equilibrium. In particular, learnable RPE exist when demand shocks are too persistent or large in magnitude, or prices are too flexible, to permit existence of REE. For standard model calibrations, this means that RPE can feature (recurring) ZLB episodes that are expected to last for over a decade, similar to the persistent ZLB events observed in Japan, or even Europe or the US. In contrast, RE ZLB events are implausibly short-lived and usually expected to last for less than 2 years under standard calibrations. Appendix \ref{appe sec: RPEreasonable} provides the details of these results, alongside brief treatments of RPE in a model with continuous shocks, and an alternative equilibrium concept for incoherent models (Appendices \ref{appe sec: RPEcontinuous} and \ref{appe sec: lagged}, respectively). A complete treatment of alternative learnable non-rational equilibria is beyond the scope of this paper, but the existence of such equilibria is not relevant for our main result: rationally incoherent models can be non-rationally coherent.

\section{Concluding remarks}\label{sec: conclusion}

Standard RE models with an occasionally binding zero lower bound (ZLB) constraint either admit no solutions (incoherence) or multiple solutions (incompleteness). This paper shows that the problem of incompleteness and incoherence hinges on the assumption of RE. 

Models with no rational equilibria may admit self-confirming equilibria involving the use of simple mis-specified forecasting models. The main message of the paper from the existence analysis is that when negative shocks are
sufficiently large in magnitude or sufficiently persistent, the baseline NK model is incoherent, but can admit RPE or BRE. Completeness and coherence can be restored if expectations are adaptive or if agents are less forward-looking due to some informational or behavioral friction. 

In the case of multiple solutions, the E-stability criterion selects an equilibrium.  A RPE can exist as a self-confirming equilibrium, even if the underlying model does not admit a REE. Thus, non-rationality of agents’ beliefs can save the economy from blowing up into infinite deflationary spirals, while it yields persistent liquidity traps. These results highlight how deviations from RE help us understand persistent liquidity traps in theoretical models and interpret the recent episodes of liquidity traps in Japan, the Euro Area, and the U.S.

We leave room for future work. In particular, we used the RPE and BRE concepts to make our point simple and clear, and consequently we abstracted from other self-confirming equilibria that could emerge under adaptive learning, such as consistent expectations equilibrium or stochastic consistent expectations equilibrium. Similarly, we excluded other popular forms of non-rationality from our analysis, such as level-$k$ reasoning, or social memory frictions as in \cite{angeletos2023}.

Finally, we put a premium on analytical results and therefore we focus on a simple theoretical model. Future work could examine related issues in larger, empirically-relevant DSGE models. In that regard, the findings of this paper complement the conclusions of \citetalias{AscariMavroeidis} about the potential implications of incoherency for estimating models with occasionally binding constraints. In particular, \citetalias{AscariMavroeidis} discuss the potential identification and misspecification issues arising from using estimation methods that neglect incoherent or incomplete regions of the parameter space under RE. Convergence issues due to incoherence may lead researchers to impose overly restrictive prior distributions, further exacerbating these concerns. Estimating models under deviations from RE may alleviate incoherence and incompleteness issues, thus providing an argument for their use in applied work. It is, therefore, worth studying this issue further in empirical applications including the ZLB, such as \citet{AruobaCubaBordaSchorfheide2018}. 

{\footnotesize
\bibliography{references} }

\newpage

\begin{subappendices}
\setcounter{equation}{0} \renewcommand{\theequation}{A\arabic{equation}}

\section*{Appendix A}\label{sec: Appendix}
\renewcommand{\thesubsection}{A.\arabic{subsection}}

We use the following definitions throughout the proofs: $a:=\lambda\sigma$, $\hat{\pi}^i:=(\pi^i_1,\pi^i_2)'$, $\rho:=p+q-1$, and $e_j$ is the $j$-th column of the $2\times2$ identity matrix, $I_2$. 

\subsection{Proof of Proposition \ref{prop1}} \label{appe sec: prop1}

Define $Q:=I_2-(1+\beta+\lambda\sigma)K+\beta K^2$.
\paragraph{Case $q<1$.} Because $det(Q+\lambda\sigma\psi I_2)=a  (\psi -1) (a  (\psi -\rho)+(1-\rho) (1-\beta \rho))> 0$, the PP solution is given by:
\begin{eqnarray*}
\hat{\pi}^{PP}=\left(Q+\lambda\sigma\psi I_2\right)^{-1}\begin{pmatrix} \lambda \epsilon_1\\\lambda \epsilon_2
\end{pmatrix}. \label{fpPP}
\end{eqnarray*}
The $PP$ solution exists if and only if $\psi\pi^{PP}_{j}>-\mu$ for $j=1,2$. We have:
\begin{eqnarray*}
\frac{\partial\pi^{PP}_1}{\partial \epsilon_1}&=&\frac{\lambda ((1 - q) (1 + a - \rho \beta) + a (\psi - 1)) }{a  (\psi -1) (a  (\psi -\rho)+(1-\rho) (1-\beta \rho))}>0,\\
\frac{\partial \pi^{PP}_2}{\partial \epsilon_1}&=&\frac{\lambda(1-q) (a-\beta   \rho+1)}{a(\psi -1) (a (\psi-\rho)+(1-\rho) (1-\beta   \rho))}>0.
\end{eqnarray*}
Thus, PP exists if and only if $\epsilon_1>\epsilon^{PP}=\max\{\epsilon^{PP}_{1},\epsilon^{PP}_{2}\}$, where $\epsilon^{PP}_{1}$ and $\epsilon^{PP}_{2}$ solve $\psi\pi^{PP}_{1}=-\mu$ and $\psi\pi^{PP}_{2}=-\mu$, respectively. We have
\begin{eqnarray*}
\epsilon^{PP}_{1}-\epsilon^{PP}_{2}&=&\frac{a (\psi -1) (a \mu  (\psi -1)+\lambda\epsilon_2 \psi ) (a (\psi-\rho)+(1-\rho) (1-\beta   \rho))}{\lambda(1-q) \psi  (a-\beta   \rho+1) (a (\psi -q)+(1-q) (1-\beta   \rho))}
\end{eqnarray*}
and hence $\epsilon^{PP}_{1}>\epsilon^{PP}_{2}$. Therefore, the PP solution exists if and only if $\epsilon_1>\epsilon^{PP}=\epsilon^{PP}_{1}$, where
\begin{align}\nonumber
&\epsilon^{PP}=\frac{a^2 \mu  (\psi -1) (\rho-\psi)}{\lambda\psi  (1-(a+1) q+a \psi +\beta (q-1)\rho)}\\ 
   &+\frac{a (\lambda\epsilon_2 (p-1) \psi +\mu  (\psi -1) (1-\rho) (\beta   \rho-1))-\lambda\epsilon_2
   (p-1) \psi  (\beta   \rho-1)}{\lambda\psi  (1-(a+1) q+a \psi +\beta (q-1)\rho)}.\label{eq: eps_PP}
\end{align}

From above, $\left(Q+\lambda\sigma\psi I_2\right)^{-1}((\lambda \epsilon_1,\lambda \epsilon_2)')$ is a ZP solution if $\epsilon_1=\epsilon^{PP}$. If $det(Q+\lambda\sigma\psi e_2e_2')\neq 0$, then the ZP solution is given by
\begin{eqnarray*}
\hat{\pi}^{ZP}=\left(Q+\lambda\sigma\psi e_2e_2'\right)^{-1}\begin{pmatrix} \lambda \epsilon_1+\lambda\sigma\mu\\\lambda \epsilon_2
\end{pmatrix}.\label{fpZP}
\end{eqnarray*}
The $ZP$ solution exists if and only if $\psi\pi^{ZP}_{2}>-\mu \ge \psi\pi^{ZP}_{1}$.
From $\hat \pi^{ZP}$ we see that $\pi^{ZP}_1$ and $\pi^{ZP}_2$ are linear in $\epsilon_1$ and 
\begin{eqnarray*}
\frac{\partial \pi^{ZP}_1}{\partial \epsilon_1}&=&\frac{-\lambda ((1-q) (a-\beta   \rho+1)+a (\psi -1))}{a (a (p \psi-\rho) -(\beta   \rho-1) ((p-1)\psi+1-\rho)},\\
\frac{\partial \pi^{ZP}_2}{\partial \epsilon_1}&=&\frac{\lambda(q-1)  (a-\beta   \rho+1)}{a (a (p \psi-\rho) -(\beta   \rho-1) ((p-1)\psi+1-\rho)}.
\end{eqnarray*}

Hence, $\frac{\partial\pi^{ZP}_1}{\partial \epsilon_1}>0$ and $\frac{\partial\pi^{ZP}_2}{\partial \epsilon_1}>0$ if and only if 
$den^{ZP}:=- (a (p \psi-\rho) -(\beta   \rho-1) ((p-1)\psi+1-\rho))=a^{-1}det(Q+\lambda\sigma\psi e_2e_2')>0$. Solving for $\epsilon^{ZP}_1$ and $\epsilon^{ZP}_2$ such that $\psi\pi^{ZP}_{1}=-\mu$ and $\psi\pi^{ZP}_{2}=-\mu$, respectively, we have
\begin{eqnarray*}
\epsilon^{ZP}_{1}-\epsilon^{ZP}_{2}&=&\frac{a(a\mu(\psi-1)+\lambda\epsilon_2\psi)den^{ZP}}{\epsilon_{\Delta ZP,den}},\\
\epsilon_{\Delta ZP,den}&:=&(1-q) \lambda \psi  (a-\beta  \rho+1) ((1-q) (a-\beta   \rho+1)+a (\psi -1))> 0.
\end{eqnarray*}
Therefore, if $den^{ZP}>0$ ($den^{ZP}<0$) then $\epsilon^{ZP}_{2}<\epsilon_1\le \epsilon^{ZP}_{1}$ ($\epsilon^{ZP}_{1}\le \epsilon_1<\epsilon^{ZP}_{2}$) is necessary and sufficient for ZP existence. Further, $\epsilon^{ZP}_{1}=\epsilon^{PP}$ and 
\begin{eqnarray}
\epsilon^{ZP}_{2}=\frac{a^2 \mu  (\psi -1)  \rho-\lambda\epsilon_2 (p-1) \psi  (\beta   \rho-1)+a (\lambda\epsilon_2 p \psi +\mu  (\psi -1) (1-\rho) (\beta   \rho-1))}{\lambda(q-1) \psi  (\beta   \rho-a-1)}.\label{eq: eps_2^ZP}
\end{eqnarray}
Finally, if $det(Q+\lambda\sigma \psi e_2e_2')=0$ ($den^{ZP}=0$) then $\epsilon^{PP}=\epsilon_2^{ZP}$, and a continuum of ZP solutions exist if $\epsilon_{1}=\epsilon^{PP}$ and a ZP solution does not exist if $det(Q+\lambda\sigma \psi e_2e_2')=0$ ($den^{ZP}=0$) and $\epsilon_{1}\neq \epsilon^{PP}$.

One can show that the PZ solution does not exist if $den^{PZ}:= det(Q+\lambda\sigma\psi e_1e_1')= 0$. If $det(Q+\lambda\sigma\psi e_1e_1')\neq 0$, the PZ solution is given by
\begin{eqnarray*}
\hat{\pi}^{PZ}=\left(Q+\lambda\sigma\psi e_1e_1'\right)^{-1}\begin{pmatrix} \lambda \epsilon_1\\\lambda \epsilon_2+\lambda\sigma\mu
\end{pmatrix}.
\end{eqnarray*}
The PZ solution exists if and only if $\psi\pi^{PZ}_{1}>-\mu \ge \psi\pi^{PZ}_{2}$. One can show 
\begin{eqnarray*}
\frac{\partial\pi^{PZ}_1}{\partial \epsilon_1}&=&\frac{\lambda  (1-(a+1) q+\beta  (q-1)  \rho)}{a (a (\rho-q \psi)-(\beta   \rho-1) (\rho-1-q \psi +\psi))}=\frac{\lambda  num^{PZ}_{1}}{den^{PZ}},\\
\frac{\partial \pi^{PZ}_2}{\partial \epsilon_1}&=&\frac{\lambda(1-q) (a-\beta   \rho+1)}{a (a (\rho-q \psi)-(\beta   \rho-1) (\rho-1-q \psi +\psi))}=\frac{\lambda num^{PZ}_{2}}{den^{PZ}}.
\end{eqnarray*}
Clearly $num^{PZ}_{2}>0$. Furthermore, if $num^{PZ}_1 = 0$ then the PZ solution does not exist. Suppose $num^{PZ}_1\neq 0$ and $den^{PZ}\neq 0$. Solving for $\epsilon^{PZ}_1$ and $\epsilon^{PZ}_2$ such that $\psi\pi^{PZ}_{1}=-\mu$ and $\psi\pi^{PZ}_{2}=-\mu$, respectively, we have
\begin{eqnarray*}
\epsilon^{PZ}_{1}-\epsilon^{PZ}_{2}&=&\frac{(a \mu  (\psi -1)+\lambda\epsilon_2 \psi)den^{PZ}}{\lambda((1-q) (a-\beta   \rho+1))\psi num^{PZ}_{1} }.
\end{eqnarray*}
There are three cases to consider. First, if $den^{PZ}>0$ (which implies $num^{PZ}_1>0$ since $num^{PZ}_1=(a(\psi-1))^{-1}\left(den^{PZ}+a(1-p)(1-\beta \rho+a)\right)>0$), then  $\epsilon_1>\epsilon^{PZ}_{1}>\epsilon^{PZ}_{2}\ge \epsilon_1$ is necessary for PZ existence, but not possible. Second, if $den^{PZ}<0$ and $num^{PZ}_{1}>0$, then $\epsilon_1<\epsilon^{PZ}_{1}<\epsilon^{PZ}_{2}\le \epsilon_1$ is necessary for PZ existence, but not possible. In the third case, $den^{PZ}<0$ and $num^{PZ}_{1}<0$, which implies $\epsilon^{PZ}_{2}<\epsilon^{PZ}_{1}< \epsilon_1$ is necessary and sufficient for PZ existence. One can show:
\begin{eqnarray*}
\epsilon^{PZ}_{1}-\epsilon^{PP}=\frac{a (p-1) (a \mu  (\psi -1)+\lambda \epsilon_2 \psi ) (a-\beta   \rho+1)}{(num^{PZ}_{1})\lambda((1 - q) (1 + a - \rho \beta) + a ( \psi - 1))}\ge0,
\end{eqnarray*}
if PZ exists (since this requires $num^{PZ}_{1}<0$). Therefore, if PZ exists then $\epsilon_1\ge \epsilon^{PP}$ and hence the PP or ZP solution also exists. 
 
From above, $\left(Q+\lambda\sigma\psi e_2 e_2'\right)^{-1}((\lambda \epsilon_1+\lambda\sigma \mu,\lambda \epsilon_2)')$ is a ZZ solution if $\epsilon_1=\epsilon^{ZP}_2$ and $det(Q+\lambda\sigma\psi e_2 e_2')\neq 0$. If $det(Q)\neq 0$, the ZZ solution is given by 
 \begin{eqnarray*}
\hat{\pi}^{ZZ}=\left(Q\right)^{-1}\begin{pmatrix} \lambda \epsilon_1+\lambda\sigma\mu\\\lambda \epsilon_2+\lambda\sigma\mu
\end{pmatrix}.
\end{eqnarray*}
The $ZZ$ solution exists if and only if $\psi\pi^{ZZ}_{j}\le -\mu$ for $j=1,2$. One can show 
\begin{eqnarray*}
\frac{\partial\pi^{ZZ}_1}{\partial \epsilon_1}&=&\frac{\lambda  (1-(a+1) q+\beta  (q-1)  \rho)}{a (a  \rho-(\rho-1) (\beta   \rho-1))}=\frac{\lambda num^{ZZ}_{1}}{a den^{ZZ}},\\
\frac{\partial \pi^{ZZ}_2}{\partial \epsilon_1}&=&\frac{\lambda(1-q)  (a-\beta   \rho+1)}{a (a  \rho-(\rho-1) (\beta   \rho-1))}=\frac{\lambda num^{ZZ}_{2}}{a den^{ZZ}}.
\end{eqnarray*}
where $det(Q)=a den^{ZZ}.$ Clearly, $num^{ZZ}_{2}>0$. We can further show that $-num^{ZZ}_{1}=den^{ZZ}+(1-p)(1+a- \rho\beta)\ge den^{ZZ}$. Hence $den^{ZZ}>0$ implies $num^{ZZ}_{1}<0$. Solving for $\epsilon^{ZZ}_{1}$ and $\epsilon^{ZZ}_{2}$ such that $\psi\pi^{ZZ}_{1}=-\mu$ and $\psi\pi^{ZZ}_{2}=-\mu$, respectively, we have
\begin{eqnarray*}
\epsilon^{ZZ}_{1}-\epsilon^{ZZ}_{2}&=&\frac{a den^{ZZ} (a \mu  (\psi -1)+\lambda \epsilon_2 \psi )}{\lambda num^{ZZ}_{1} \psi (1-q) (a-\beta   \rho+1)},
\end{eqnarray*}
if $num^{ZZ}_1\neq 0$. There are the following cases to consider. First, if $den^{ZZ}>0$ (which implies $num^{ZZ}_{1}<0$) then ZZ existence requires $\epsilon^{ZZ}_{2}\ge\epsilon_1\ge\epsilon^{ZZ}_{1}$. Second, if $den^{ZZ}<0$ and $num^{ZZ}_{1}>0$ then ZZ existence requires $\epsilon_1\ge\epsilon^{ZZ}_{2}>\epsilon^{ZZ}_{1}$. In the third case,  $den^{ZZ}<0$ and $num^{ZZ}_{1}<0$ then ZZ existence requires $\epsilon^{ZZ}_{1}\ge\epsilon_1\ge \epsilon^{ZZ}_{2}$.  If $num^{ZZ}_1=0$ and $det(Q)\neq 0$ then a ZZ exists if and only if $\epsilon_1 \ge \epsilon^{ZZ}_2$. Finally, if $det(Q)=0$ ($den^{ZZ}=0$) and $\epsilon_1=\epsilon^{ZP}_{2}$ then a continuum of ZZ solutions exist, and if $det(Q)=0$ ($den^{ZZ}=0$) and $\epsilon_1\neq \epsilon^{ZP}_{2}$ then a ZZ solution does not exist. Now it can be shown that $\epsilon^{ZZ}_{2}=\epsilon^{ZP}_{2}$ and 
\begin{eqnarray*}
\epsilon^{ZZ}_{1}-\epsilon^{PP}=\frac{a (p-1) (a \mu  (\psi -1)+\lambda\epsilon_2 \psi ) (a-\beta   \rho+1)}{\lambda num^{ZZ}_{1} ((1 - q) (1 + a - \rho \beta) + a (\psi - 1))}\ge 0,
\end{eqnarray*}
if $num^{ZZ}_{1}<0$. Hence ZZ existence and $\epsilon_1>\min\{\epsilon^{PP},\epsilon^{ZP}_{2}\}$ implies ZP or PP existence.

From the analysis above, a REE exists only if   $\epsilon_1\ge \min\{\epsilon^{PP},\epsilon^{ZP}_{2}\}$. Further, if $\epsilon_1\ge \epsilon^{PP}$ then a PP or ZP exists because $det(Q+\lambda\sigma \psi I_2)>0$. If $\epsilon^{PP}>\epsilon^{ZP}_{2}$, then $det(Q+a \psi e_2e_2')=a den^{ZP}\neq 0$ and therefore a PP, ZP or ZZ solution exists if, in addition, $\epsilon_1\ge \epsilon^{ZP}_2$. We conclude that a REE exists if and only if
\begin{equation}\label{eq: cutoff REE}
    \epsilon_1\ge\bar{\epsilon}_{REE}:=\min\{\epsilon^{PP},\epsilon^{ZP}_{2}\},
\end{equation}
where $\epsilon^{PP}$ and $\epsilon^{ZP}_{2}$ are defined in \eqref{eq: eps_PP} and \eqref{eq: eps_2^ZP}, respectively.

\paragraph{Case $q=1$.}
 Here we show that Proposition 1 nests Proposition 5 of \citetalias{AscariMavroeidis} as a special case. Specifically, we compute the condition from $\lim_{q\rightarrow 1}\bar{\epsilon}_{REE}$ and show that this recovers the result in Proposition 5 of \citetalias{AscariMavroeidis}.\footnote{ Alternatively, we could repeat the preceding analysis in the model with $q=1$, but this gives the same result.. Mathematica routine available on request.} Define $\theta:=\frac{(1-p)(1-p\beta)}{\lambda \sigma p}=\frac{(1-p)(1-p\beta)}{a p}$. From the preceding analysis, a REE exists if and only if $\epsilon_1\ge\bar{\epsilon}_{REE}=\min\{\epsilon^{PP},\epsilon^{ZP}_{2}\}$ where $\epsilon^{ZP}_{2}$ can be expressed as $\epsilon^{ZP}_{2}=\chi(1-q)^{-1}$. In the limit $q\rightarrow 1$ we have:
\begin{eqnarray*}
\epsilon^{PP}&=&\mu  \left(\frac{a (p-\psi )}{\lambda\psi }-\frac{p a \theta}{\lambda\psi }\right)+\frac{\lambda\epsilon_2 (p-1) (a-\beta  p+1)}{a \lambda(\psi -1)},\\
\chi&:=&\frac{(p(1+a+\beta)-p^2\beta-1)(a\mu(\psi-1)+\lambda\epsilon_2\psi)}{(1+a-p\beta)\psi\lambda}.
\end{eqnarray*}
Now, $p(1+a+\beta)-1-p^2\beta<0$ if and only if $\theta>1$. Therefore, $\bar{\epsilon}_{REE}=\epsilon^{ZP}_2\rightarrow -\infty$ as $q\rightarrow 1$ if $\theta>1$. We conclude that any value of $\epsilon_1$ ensures existence of a solution when $\theta>1$ and $q=1$. If $\theta<1$, then $\chi\rightarrow +\infty$ and $\bar{\epsilon}_{REE}=\epsilon^{PP}$, and $\epsilon^{ZP}_2 \ge \epsilon^{PP}=\bar{\epsilon}_{REE}$ if $\theta=1$.\footnote{The $\theta=1$ case arises if $a=\frac{(1-p)(1-\beta\rho)}{p}$ and $q=1$. To compute $\epsilon^{ZP}_2$, set $a=\frac{(1-p)(1-\beta\rho)}{p}$ and compute $\lim_{q\rightarrow 1}\epsilon^{ZP}_2$.}

Now we show that our conditions recover Proposition 5 in \citetalias{AscariMavroeidis}. First, we have $\mu=log(r\pi_{*})>0$ which implies $r^{-1}\le \pi_{*}$ where $r$ and $\pi_{*}$ are the steady state gross real interest rate and inflation rate, respectively. Further, we set $\epsilon_2=0$ and $\epsilon_1=-\sigma \hat M_{t+1|t}=\sigma p r_L$. The critical threshold, $\epsilon^{PP}$ becomes: $-r_L\le \mu  \left(\frac{\theta}{\psi }  +\frac{(\psi-p)}{p\psi }\right)$. Thus, a solution exists if and only if $\theta>1$ or $\theta\le1$ and $-r_L\le \mu  \left(\frac{\theta}{\psi }  +\frac{(\psi-p)}{p\psi }\right)$ as in \citetalias{AscariMavroeidis}.

\subsection{Proof of Proposition \ref{prop: CC temp RPE}} \label{appe sec: prop2}

Define $z_{t}:=\pi_{t}+\mu/\psi$ and assume $\psi>0$, so that the positive
interest rate regime arises when $z_{t}>0$ (equivalent to $\psi\pi
_{t}>-\mu$), and the zero interest rate regime when $z_{t}\leq0$.
Substituting out $i_{t}$, and $\pi_{t}=z_{t}-\mu/\psi,$ equations
(\ref{eq:IS})-(\ref{eq:MP}) can be written as%
\begin{align*}
x_{t}  & =x_{t}^{e}-\sigma\left(  \psi z_{t}\mathbf{1}\left\{  z_{t}>0\right\}-\mu-\pi_{t}^{e}\right)
+\epsilon_{t},\\
z_{t}  & =\mu/\psi+\lambda x_{t}+\beta\pi_{t}^{e},
\end{align*}
or, compactly, as%
\begin{equation}%
\begin{pmatrix}
1 & \sigma\psi\mathbf{1}\left\{  z_{t}>0\right\}  \\
-\lambda & 1
\end{pmatrix}%
\begin{pmatrix}
x_{t}\\
z_{t}%
\end{pmatrix}
=%
\begin{pmatrix}
1 & \sigma\\
0 & \beta
\end{pmatrix}
Y_{t}^{e}+%
\begin{pmatrix}
\sigma\mu+\epsilon_{t}\\
\mu/\psi
\end{pmatrix}
,\label{eq: SEM}%
\end{equation}
where $\mathbf{1}\left\{  \cdot\right\}  $ is the indicator function that
takes the value 1 when its argument is true and zero otherwise. With $k=1$ in
(\ref{RPEPLM}), the variable $Y_{t}^{e}$ is predetermined. Coherence and completeness of
(\ref{eq: SEM}) means that the model can be solved uniquely for $x_{t},z_{t}$
(equivalently $x_{t},\pi_{t}$). Equation (\ref{eq: SEM}) is a piecewise-linear
continuous simultaneous equations model for $\left(  x_{t},z_{t}\right)
^{\prime}$ whose coherence conditions (existence and uniqueness of
equilibrium) are given by \cite[Theorem 1]{GourierouxLaffontMonfort1980}.
Specifically,
\[%
\det%
\begin{pmatrix}
1 & \sigma\psi\\
-\lambda & 1
\end{pmatrix}
\det%
\begin{pmatrix}
1 & 0\\
-\lambda & 1
\end{pmatrix}
=1+\sigma\lambda\psi>0,
\]
which always holds when $\sigma,\lambda,\psi>0$. 

\subsection{Proof of Proposition \ref{prop: RPE existence}} \label{appe sec: prop3}

The proof of Proposition \ref{prop: RPE existence} is a straightforward extension of the proof of Proposition \ref{prop1}. Define $\bar q:=Pr(\epsilon_t=2)=(1-p)/(2-p-q)$. The regime-specific levels of inflation in RPE $i$, $\hat{\pi}^i=(\pi^i_1,\pi^i_2)'$, are given by fixed point restrictions that have the same basic form as the REE fixed point restrictions except we replace $q$ with $\bar q$ and $p$ with $1-\bar q$. Therefore, RPE will exist if and only if
\begin{equation}\label{eq: RPE cutoff}
    \epsilon_1\ge\bar{\epsilon}_{RPE}=\min\{ \epsilon^{PP,RPE},\epsilon^{ZP,RPE}_2\},
\end{equation} 
where  $\epsilon^{PP,RPE},\epsilon^{ZP,RPE}_2$ have the same form as $ \epsilon^{PP},\epsilon^{ZP}_{2}$ given in \eqref{eq: eps_PP},\eqref{eq: eps_2^ZP} except we replace $q$ and $p$ with $\bar q$ and $1-\bar q$, respectively. In the special case $q=1$ (which implies $\bar q=1$), we have $\bar{\epsilon}_{RPE}=-\infty$, as the PP solution exists if and only if $\epsilon_1> -\mu(1+\lambda\sigma\psi)(\lambda\psi)^{-1}+(1+\lambda\sigma)(\lambda\sigma(1-\psi))^{-1}\epsilon_2=\epsilon^{PP,RPE}$ and the ZP exists if and only if $\epsilon_1\le \epsilon^{PP,RPE}$. For $q<1$, one can show: $\epsilon^{PP}-\epsilon^{PP,RPE}=-\Xi_{PP}\rho$ and $\epsilon^{ZP}-\epsilon^{ZP,RPE}=-\Xi_{ZP}\rho$ where $\Xi_{PP}:=\frac{a(1+a-\beta(\rho-1))(1-p)(\psi-1) (a \mu  (\psi -1)+\lambda\epsilon_2 \psi )}{\lambda \psi  (a (1-\psi )(1-\rho)+(a+1)(q-1))((1-q)(1+a-\beta \rho)+a(\psi-1))}\le 0$ and $\Xi_{ZP}:=\frac{a (a \mu  (\psi -1)+\lambda\epsilon_2 \psi ) (1+a-\beta  (\rho-1))}{\lambda(a+1) (q-1) \psi  (1+a-\beta   \rho)}<0$. Hence, $\bar{\epsilon}_{REE}\ge\bar{\epsilon}_{RPE}$ if and only if $p+q\ge1$. 

\subsection{Proof of Proposition \ref{prop:BR coherence}} \label{appe sec: prop4}

Define $\delta:=(M-1)(1-M_f\beta)+\lambda\sigma N$ and $Q:=I_2-(M+M_f\beta+\lambda\sigma N)K+\beta M M_f K^2$.
\paragraph{Case $q<1$.}
Since $den^{PP,BR}:=det\left(Q+\lambda\sigma\psi I_2\right)=((1-M \rho)(1-M_f \beta  \rho)+a(\psi-N\rho))((1-M)(1-M_f\beta)+a(\psi-N))>0$, the PP solution is given by:
\begin{eqnarray*}
\hat{\pi}^{PP,BR}=\left(Q+\lambda\sigma\psi I_2\right)^{-1}\begin{pmatrix} \lambda \epsilon_1\\\lambda \epsilon_2
\end{pmatrix}. 
\end{eqnarray*}
The $PP$ solution exists if and only if $\psi\pi^{PP,BR}_{j}>-\mu$ for $j=1,2$. We have:
\begin{align*}
\frac{\partial\pi^{PP,BR}_1}{\partial \epsilon_1}=\frac{num^{PP,BR}_1}{den^{PP,BR}}>0, && 
\frac{\partial \pi^{PP,BR}_2}{\partial \epsilon_1}=\frac{num^{PP,BR}_2}{den^{PP,BR}}>0,
\end{align*}
where
\begin{eqnarray*}
num^{PP,BR}_1&:=&\lambda   (a  \psi +\beta  M M_f (q(p+q)-\rho)-M q-q (\beta  M_f+a N )+1)>0,\\
num^{PP,BR}_2&:=&\lambda  (q-1)   (\beta  M_f (M (p+q)-1)-M-a  N  )> 0
\end{eqnarray*}
Thus, PP exists if and only if $\epsilon_1>\epsilon^{PP,BR}=\max\{\epsilon^{PP,BR}_{1},\epsilon^{PP,BR}_{2}\}$ where $\epsilon^{PP,BR}_{1}$ and $\epsilon^{PP,BR}_{2}$ solve $\psi\pi^{PP,BR}_{1}=-\mu$ and $\psi\pi^{PP,BR}_{2}=-\mu$, respectively. We have
\begin{eqnarray}\label{eq: eps^PP,BR}
\epsilon^{PP,BR}&=&\frac{\eta_1 \eta_2 \eta_3}{\psi num^{PP,BR}_1},\\\nonumber
    \eta_1&:=&a(\psi-N)+(1-M)(1-M_f\beta)>0,\\\nonumber
    \eta_2&:=&(a(N+\psi)-(p+q) (a N+\beta  M_f) +M  \rho (\beta  M_f  \rho-1)+\beta  M_f+1),\\\nonumber
\eta_3&:=&\frac{\lambda \epsilon_2 (1-p) \psi  (\beta  M_f (M (p+q)-1)-a N-M)}{den^{PP,BR}}-\mu.
\end{eqnarray}

From above, $(Q+\lambda\sigma\psi I_2)^{-1}((\lambda\epsilon_1,\lambda\epsilon_2)')$ is a ZP solution if $\epsilon_1=\epsilon^{PP,BR}$. If $det(Q+\lambda\sigma \psi e_2e_2')\neq 0$, then the ZP solution is given by
\begin{eqnarray*}
\hat{\pi}^{ZP,BR}=\left(Q+\lambda\sigma\psi e_2e_2'\right)^{-1}\begin{pmatrix} \lambda \epsilon_1+\lambda\sigma\mu\\\lambda \epsilon_2
\end{pmatrix}.
\end{eqnarray*}
The $ZP$ solution exists if and only if $\psi\pi^{ZP,BR}_{2}>-\mu\ge\psi\pi^{ZP,BR}_{1}$.
We have:
\begin{eqnarray*}
\frac{\partial \pi^{ZP,BR}_1}{\partial \epsilon_1}&=&\frac{\lambda (a (\psi -N q)+\beta  M_f (M (q(p+q)-\rho)-q)-M q+1)}{den^{ZP,BR}},\\
\frac{\partial \pi^{ZP,BR}_2}{\partial \epsilon_1}&=&\frac{\lambda(1-q)   (a N+\beta  M_f (1-M (p+q))+M)}{den^{ZP,BR}}.
\end{eqnarray*}

From the last equations, $\frac{\partial\pi^{ZP}_1}{\partial \epsilon_1}>0$ and $\frac{\partial\pi^{ZP}_2}{\partial \epsilon_1}>0$ if and only if 
$den^{ZP,BR}:=\delta^2-\delta(a \psi+(1-\rho)(M+a N+M_f\beta(1-M(p+q))))+(1-p)a \psi(M+a N+M_f\beta(1-M(p+q)))>0$. Solving for $\epsilon^{ZP,BR}_1$ and $\epsilon^{ZP,BR}_2$ such that $\psi\pi^{ZP,BR}_{1}=-\mu$ and $\psi\pi^{ZP,BR}_{2}=-\mu$, respectively, we have
\begin{eqnarray*}
\epsilon^{ZP,BR}_{1}-\epsilon^{ZP,BR}_{2}&=&\frac{(\mu((1-M)(1-M_f\beta)+a(\psi-N))+\lambda\epsilon_2\psi)den^{ZP,BR}}{\lambda\epsilon_{\Delta ZP,BR}},\\
\epsilon_{\Delta ZP,BR}&:=&(1-q)num^{ZP,BR}_1(M+a N+M_f\beta(1-M(p+q)))> 0,\\
num^{ZP,BR}_1&:=&\psi (a (\psi -N q)+\beta  M_f (M (q(p+q)-\rho)-q)-M q+1)>0.
\end{eqnarray*}
Therefore, if $den^{ZP,BR}>0$ ($den^{ZP,BR}<0$) then $\epsilon^{ZP,BR}_{2}<\epsilon_1\le \epsilon^{ZP,BR}_{1}$ ($\epsilon^{ZP,BR}_{1}\le \epsilon_1<\epsilon^{ZP,BR}_{2}$) is necessary and sufficient for existence of ZP. Further, we can show: $\epsilon^{ZP,BR}_{1}=\epsilon^{PP,BR}$
and 
\begin{eqnarray}
\epsilon^{ZP,BR}_{2}&=&\frac{ \mu  \eta_1  (a N-(p+q) (a N+\beta  M_f)+M  \rho (\beta  M_f  \rho-1)+\beta  M_f+1)}{\lambda (q-1) \psi  (a N-\beta  M M_f (p+q)+M+\beta  M_f)}\nonumber\\\label{eq: eps^ZP,BR}
&-&\frac{\epsilon_2 \lambda  (a N p+\beta  M_f (M (q-p  \rho-1)+p)+M p-1)}{\lambda (q-1) (a N-\beta  M M_f (p+q)+M+\beta  M_f)}.
\end{eqnarray}

Finally, if $det(Q+\lambda\sigma \psi e_2e_2')=0$ ($den^{ZP,BR}=0$) then $\epsilon^{PP,BR}=\epsilon_2^{ZP,BR}$, and a continuum of ZP solutions exist if $\epsilon_{1}=\epsilon^{PP,BR}=\epsilon_2^{ZP,BR}$ and no ZP solution exists if $det(Q+\lambda\sigma \psi e_2e_2')=0$ ($den^{ZP,BR}=0$) and $\epsilon_{1}\neq \epsilon^{PP,BR}$.

It is straightforward to show that the PZ solution does not exist if $det(Q+\lambda\sigma\psi e_1e_1')=0$. If $det(Q+\lambda\sigma\psi e_1e_1')\neq0$, the PZ solution is given by
\begin{eqnarray*}
\hat{\pi}^{PZ,BR}=\left(Q+\lambda\sigma\psi e_1e_1'\right)^{-1}\begin{pmatrix} \lambda \epsilon_1\\\lambda \epsilon_2+\lambda\sigma\mu
\end{pmatrix}.\label{fpPZ}
\end{eqnarray*}
The $PZ$ solution exists if and only if $\psi\pi^{PZ,BR}_{1}>-\mu\ge \psi\pi^{PZ,BR}_{2}$. One can show 
\begin{eqnarray*}
\frac{\partial\pi^{PZ,BR}_1}{\partial \epsilon_1}&=&\frac{\lambda  (1 - (M + a N) q + 
    M_f (M + M p (q-1) - q + M (q-1) q) \beta) )}{den^{PZ,BR}}\\
&=&\frac{\lambda  num^{PZ,BR}_{1}}{den^{PZ,BR}},\\
\frac{\partial \pi^{PZ,BR}_2}{\partial \epsilon_1}&=&\frac{\lambda(1-q) (M + a N + M_f (1 - M (p + q)) \beta) }{den^{PZ,BR}}=\frac{\lambda num^{PZ,BR}_{2}}{den^{PZ,BR}}.
\end{eqnarray*}
where $den^{PZ,BR}:=det(Q+\lambda\sigma\psi e_1e_1')=-M (a N  \rho (\beta  M_f (p+q)-2)+a \psi  (\beta  M_f (p-1)-\beta M_f q  \rho+q)+(\beta  M_f-1) (p+q) (\beta  M_f  \rho-1))+(a N+\beta  M_f-1) (a N
    \rho+\beta  M_f  \rho-1)-a \psi  (a N q+\beta  M_f q-1)+M^2 (\beta  M_f-1)  \rho (\beta  M_f  \rho-1)$. Clearly $num^{PZ,BR}_{2}>0$.  Furthermore, it is straightforward to show that $num^{PZ,BR}_1 \neq 0$ is necessary for existence of PZ solution. Solving for $\epsilon^{PZ,BR}_1$ and $\epsilon^{PZ,BR}_2$ such that $\psi\pi^{PZ,BR}_{1}=-\mu$ and $\psi\pi^{PZ,BR}_{2}=-\mu$, respectively, we have
\begin{eqnarray*}
\epsilon^{PZ,BR}_{1}-\epsilon^{PZ,BR}_{2}&=&\frac{ (\eta_1\mu+\psi\lambda\epsilon_2)den^{PZ,BR}}{\lambda (1-q)\psi(M+a N+M_f\beta(1-M(p+q)))num^{PZ,BR}_1},
\end{eqnarray*}
if $num^{PZ,BR}\neq 0$. There are three cases to consider. First, if $den^{PZ,BR}>0$ and $num^{PZ,BR}_{1}>0$ then $\epsilon_1>\epsilon^{PZ,BR}_{1}>\epsilon^{PZ,BR}_{2}\ge \epsilon_1$ is necessary for PZ existence, but not possible. Second, if $den^{PZ,BR}<0$ and $num^{PZ,BR}_{1}>0$ then $\epsilon_1< \epsilon^{PZ,BR}_{1}<\epsilon^{PZ,BR}_{2}\le \epsilon_1$ is necessary for PZ existence, but not possible. In the third case, $den^{PZ,BR}<0$ and $num^{PZ,BR}_{1}<0$, which implies $\epsilon^{PZ,BR}_{2}<\epsilon^{PZ,BR}_{1}< \epsilon_1$ is necessary and sufficient for PZ existence. Note that $den^{PZ,BR}>0$ and $num^{PZ,BR}_{1}<0$ cannot hold simultaneously because
\begin{eqnarray*}
den^{PZ,BR}&=&\delta(p-1)(M+a N+M_f\beta(1-M(p+q)))+num^{PZ,BR}_1\eta_1>0,
\end{eqnarray*}
requires $\delta<0$ if $num^{PZ,BR}_{1}<0$, but
\begin{eqnarray*}
num^{PZ,BR}_1=-\delta+(1-q)(M+aN+M_f\beta(1-(p+q)M))<0,
\end{eqnarray*}
requires $\delta> 0$. Hence, a PZ solution can only exist if $den^{PZ,BR}<0$ and $num^{PZ,BR}_{1}<0$ and $\epsilon_1> \epsilon^{PZ,BR}_1$. One can show:
\begin{eqnarray*}
\epsilon^{PZ,BR}_{1}-\epsilon^{PP,BR}=\frac{\psi a(1-p)(a N+M+M_f\beta(1-M(p+q)))(\lambda \psi \epsilon_2+\mu \eta_1)}{-\lambda num^{ZP,BR}_1 num^{PZ,BR}_1}\ge 0,
\end{eqnarray*}
if PZ exists (since this requires $num^{PZ,BR}_{1}<0$). Therefore, if the PZ exists then $\epsilon_1\ge \epsilon^{PP,BR}$ and hence the PP or ZP solution also exists.

From above, $(Q+\lambda\sigma\psi e_2e_2')^{-1}((\lambda\epsilon_1+\lambda\sigma \mu,\lambda\epsilon_2)')$ is a ZZ solution if $\epsilon_1=\epsilon^{ZP,BR}_2$ and $det(Q+\lambda\sigma\psi e_2e_2')\neq 0$ ($den^{ZP,BR}\neq0$). If $det(Q)\neq 0$ then the ZZ solution is given by 
 \begin{eqnarray*}
\hat{\pi}^{ZZ,BR}=\left(Q\right)^{-1}\begin{pmatrix} \lambda \epsilon_1+\lambda\sigma\mu\\\lambda \epsilon_2+\lambda\sigma\mu
\end{pmatrix}.\label{fpZZ}
\end{eqnarray*}
The $ZZ$ solution exists if and only if $\psi\pi^{ZZ,BR}_{j}\le-\mu$ for $j=1,2$. One can show that
\begin{eqnarray*}
\frac{\partial\pi^{ZZ,BR}_1}{\partial \epsilon_1}&=&\frac{\lambda  ((1 - (M + a N) q + 
   M_f (M + M p (q-1) - q + M (q-1) q) \beta) )}{den^{ZZ,BR}}\\
&=&\frac{\lambda num^{ZZ,BR}_{1}}{den^{ZZ,BR}},\\
\frac{\partial \pi^{ZZ,BR}_2}{\partial \epsilon_1}&=&\frac{\lambda(1 - q) (M + a N + M_f\beta (1 - M (p + q))) }{{den^{ZZ,BR}}}=\frac{\lambda num^{ZZ,BR}_{2}}{den^{ZZ,BR}}.
\end{eqnarray*}
where $den^{ZZ,BR}:=-\delta(-\delta+(1-\rho)(M+a N+M_f\beta(1-(p+q)M)))=det(Q)$ and clearly $num^{ZZ,BR}_2>0$. Solving for $\epsilon^{ZZ,BR}_{1}$ and $\epsilon^{ZZ,BR}_{2}$ such that $\psi\pi^{ZZ,BR}_{1}=-\mu$ and $\psi\pi^{ZZ,BR}_{2}=-\mu$, respectively, we have
\begin{eqnarray*}
\epsilon^{ZZ,BR}_{1}-\epsilon^{ZZ,BR}_{2}&=&\frac{ den^{ZZ,BR} (\eta_1\mu+\lambda\psi\epsilon_2)}{\lambda(1-q)\psi(M+a N+M_f\beta(1-M(p+q)))num^{ZZ,BR}_1},
\end{eqnarray*}
if $num^{ZZ,BR}_1\neq 0$. There are the following cases to consider. First, if $den^{ZZ,BR}>0$ and $num^{ZZ,BR}_{1}<0$ then ZZ existence requires $\epsilon^{ZZ,BR}_{2}\ge\epsilon_1\ge\epsilon^{ZZ,BR}_{1}$. Second, if $den^{ZZ,BR}<0$ and $num^{ZZ,BR}_{1}>0$ then ZZ existence requires $\epsilon_1\ge\epsilon^{ZZ,BR}_{2}>\epsilon^{ZZ,BR}_{1}$. In the third case,  $den^{ZZ,BR}<0$ and $num^{ZZ,BR}_{1}<0$ then ZZ existence requires $\epsilon^{ZZ,BR}_{1}\ge\epsilon_1\ge\epsilon^{ZZ,BR}_{2}$. Now it can be shown that $\epsilon^{ZZ,BR}_{2}=\epsilon^{ZP,BR}_{2}$ and 
\begin{eqnarray*}
\epsilon^{ZZ,BR}_{1}-\epsilon^{PP,BR}&=&\frac{- a(1-p)(M+a N+M_f\beta(1-M(p+q)))(\psi\lambda\epsilon_2+\eta_1\mu)}{\lambda num^{ZZ,BR}_1\eta_4}\ge 0,
\end{eqnarray*}
if $num^{ZZ,BR}_{1}<0$, where $\eta_4:=(1-q) (a N+\beta M_f (1-M (p+q))+M)+a (\psi -N)+(1-M) (1-\beta  M_f)>0$. Since $\epsilon^{ZZ,BR}_{2}=\epsilon^{ZP,BR}_{2}$ and existence of ZZ in the first three cases only hinges on $\epsilon_1\ge\epsilon^{ZZ,BR}_{1}$ if $num^{ZZ,BR}_{1}<0$ it follows that the ZP or PP solution will exist if the ZZ solution exists in the first three cases and $\epsilon_1>\max\{\epsilon^{PP,BR},\epsilon^{ZP,BR}_2\}$.

In the fourth case, $den^{ZZ,BR}>0$ and $num^{ZZ,BR}_{1}>0$. One can show that:
\begin{eqnarray*}
den^{ZZ,BR}&=&-\delta(-\delta+(1-\rho)(M+aN+M_f\beta(1-(p+q)M))),\\
num^{ZZ,BR}_1&=&-\delta^{-1}den^{ZZ,BR}+\eta_5\\
&=&-\delta+(1-q)(M+aN+M_f\beta(1-(p+q)M)),\\
\eta_5&:=&(p-1)(M+aN+M_f\beta(1-(p+q)M))\le0.
\end{eqnarray*}
Therefore, $\delta<0$ if and only if the fourth case ($num^{ZZ,BR}_1>0$ and $den^{ZZ,BR}>0$) applies. In the fourth case, ZZ existence requires $\epsilon^{ZP,BR}_{2}\ge \epsilon_1.$ It is furthermore straightforward to show that if $num^{ZZ,BR}_1=0$ and $det(Q)\neq 0$ then a ZZ exists if and only if $\epsilon_1\ge \epsilon^{ZZ,BR}_2=\epsilon^{ZP,BR}_2$. Finally, if $det(Q)=0$ ($den^{ZZ,BR}=0$) and $\epsilon_1=\epsilon^{ZP,BR}_2$ then a continuum of ZZ solutions exist and if $det(Q)=0$ ($den^{ZZ,BR}=0$) and $\epsilon_1\neq \epsilon^{ZP,BR}_2$ then a ZZ solution does not exist.

From the analysis above, if a BRE exists then $\delta\ge0$ and $\epsilon_1\ge \min\{\epsilon^{PP,BR},\epsilon^{ZP,BR}_{2}\}$ or $\delta<0$. Further, if $\epsilon_1\ge \epsilon^{PP,BR}$ then a PP or ZP exists because $det(Q+\lambda\sigma\psi I_2)>0$. If $\epsilon^{PP,BR}>\epsilon^{ZP,BR}_{2}$, then $den^{ZP,BR}=det(Q+\lambda\sigma e_2e_2')\neq 0$ and therefore a PP, ZP or ZZ solution exists if, in addition, $\epsilon_1\ge \epsilon^{ZP,BR}_2$. If $\delta<0$, then a ZZ exists for $\epsilon_1\le \epsilon^{ZP,BR}_2$. We conclude that a BRE exists if and only if
\begin{equation}\label{eq: BRE cutoff}
    \epsilon_1\ge\bar{\epsilon}_{BR} := \left\{
\begin{array}
[c]{ll}%
\min\left\{\epsilon^{PP,BR},\epsilon^{ZP,BR}_2\right\}, & \text{if }\delta\ge0\\
-\infty, & \text{if }\delta<0,
\end{array}
\right.
\end{equation} where $\epsilon^{PP,BR}$ and $\epsilon^{ZP,BR}_2$ are defined in \eqref{eq: eps^PP,BR} and \eqref{eq: eps^ZP,BR}, respectively.

\paragraph{Case $q=1$.} 
Note that $\epsilon^{ZP,BR}_2$ from (\ref{eq: eps^ZP,BR}) can be expressed as $\epsilon^{ZP,BR}_2=(q-1)^{-1} \chi_{BR}$ where, if $q=1$, and $\chi^1:=-\delta+(1-p)(a N+M(1-M_f\beta p)+M_f\beta(1-M))\neq 0$:
 \begin{eqnarray*}
 \chi_{BR}&:=&\frac{\chi^1(\psi\lambda\epsilon_2+\mu((1-M)(1-M_f\beta)+a(\psi-N)))}{\lambda\psi(a N+M_f\beta(1-M)+M(1-M_fp\beta))} .\end{eqnarray*}
 
 For the PP solution, we have $\pi^{PP,BR}_2=\frac{\lambda \epsilon_2}{(1-M)(1-M_f\beta)+a(\psi-N)}\ge 0$ and therefore $\psi \pi^{PP,BR}_2>-\mu$.\footnote{It can be shown that $(Q+\lambda\sigma\psi I_2)^{-1}$ exists if $q=1$.} Further, $\partial \pi^{PP,BR}_1/\partial \epsilon_1=\lambda/((1-M p)(1-M_f\beta p)+a(\psi-N p))>0$
 and $\psi\pi^{PP,BR}_1=-\mu$ if and only if $\epsilon_1=\epsilon^{PP,BR}$ where $\epsilon^{PP,BR}$ is defined in (\ref{eq: eps^PP,BR}) with $q=1$. Therefore, PP exists if and only if $\epsilon_1>\epsilon^{PP,BR}$,  and a ZP solution always exists if $\epsilon_1=\epsilon^{PP}$.
 For the ZP solution, we have $\pi^{ZP,BR}_2=\pi^{PP,BR}_2$ and therefore $\psi \pi^{ZP,BR}_2>-\mu$. If $\chi^1\neq 0$, then: $\partial \pi^{ZP,BR}_1/\partial \epsilon_1=\lambda/\chi^1$ and $\psi\pi^{ZP,BR}_1=-\mu$ if and only if $\epsilon_1=\epsilon^{PP,BR}$ where $\epsilon^{PP,BR}$ is defined in (\ref{eq: eps^PP,BR}) with $q=1$. Therefore if $\chi^1>0$ then $\epsilon^{PP,BR}\ge \epsilon_1>\epsilon^{ZP,BR}_2=-\infty$ is necessary and sufficient for existence of the ZP solution. Otherwise, if $\chi^1<0$ then $\epsilon^{ZP,BR}_2=+\infty$ and $\epsilon_1\ge \epsilon^{PP,BR}$ is necessary and sufficient for existence of the ZP solution. Note that $\delta<0$ implies $\chi^1>0$. Finally, $\chi^1=0$ implies:\footnote{The $\chi^1=0$ case arises if $a=\frac{(1-M p)(1-M_f \beta\rho)+M_f(q-1)(1-M)\beta}{N p}$ and $q=1$. To compute $\epsilon^{ZP,BR}_2$, set $a=\frac{(1-M p)(1-M_f \beta\rho)+M_f(q-1)(1-M)\beta}{N p}$ and compute $\lim_{q\rightarrow 1}\epsilon^{ZP,BR}_2$.}
\begin{eqnarray*}&&\epsilon^{ZP,BR}_2-\epsilon^{PP,BR}=\frac{(1-p)(1-M M_f \beta p) }{\lambda p \psi}\mu +\\&&\frac{(1-p)(1-M M_f \beta p) N \epsilon_2}{(1-M)(1-M_f\beta)p+(1-N)(1-p)(1-M M_f\beta p)+(1-M p)(1-M_f p\beta)(\psi-1)}\ge 0.
\end{eqnarray*}
and that a continuum of ZP solutions exist if $\epsilon_1=\epsilon^{PP,BR}$, and no ZP solution exists if $\epsilon_1\neq\epsilon^{PP,BR}$. 

For the PZ solution, we have $\pi^{PZ,BR}_2=-\frac{\lambda\epsilon_2+a \mu}{\delta}$. If $\delta<0$, then $\pi^{PZ,BR}_2\ge0$, and if $\delta>0$, then $\psi\pi^{PZ,BR}_2=-\psi\frac{\lambda\epsilon_2+a \mu}{\delta} \le -\psi \mu<-\mu$, since $\delta\le a$ and $\epsilon_2\ge0$. If $\delta=0$ then a PZ solution does not exist. Therefore, $\psi\pi^{PZ,BR}_2+\mu<0$ if and only if $\delta>0$. Further, $\partial \pi^{PZ,BR}_1/\partial \epsilon_1=\lambda/((1-M p)(1-M_f\beta p)+a(\psi-N p)) > 0$ and $\psi\pi^{PZ,BR}_1=-\mu$ if and only if $\epsilon_1=\epsilon^{PZ,BR}_1$ where 
\begin{eqnarray*}
    \epsilon^{PZ,BR}_1&=& \epsilon^{PP,BR}+\frac{a(1-p)(M(1-M_f\beta p)+a N+M_f\beta(1-M))(\lambda\psi \epsilon_2+\mu \eta_1)}{\lambda \delta(a\psi-\delta)}\ge \epsilon^{PP,BR},
\end{eqnarray*}
and $\epsilon^{PP,BR}$ is defined in (\ref{eq: eps^PP,BR}) with $q=1$. It follows that PZ exists if and only if $\delta>0$ and $\epsilon_{1}>\epsilon^{PZ,BR}_1\ge \epsilon^{PP,BR}$.

 For the ZZ solution, we have $\pi^{ZZ,BR}_2=\pi^{PZ,BR}_2$, and therefore $\psi\pi^{ZZ,BR}_2+\mu\le 0$ if and only if $\delta>0$. Furthermore, if $\chi^1\neq 0$ then $\partial \pi^{ZZ,BR}_1/\partial \epsilon_1=\lambda/\chi^1$ and $\psi\pi^{ZZ,BR}_1=-\mu$ if and only if $\epsilon_1=\epsilon^{ZZ,BR}_1=\epsilon^{PZ,BR}_1\ge \epsilon^{PP,BR}$ where $\epsilon^{PP,BR}$ is defined in (\ref{eq: eps^PP,BR}) with $q=1$. Therefore if $\chi^1>0$ and $\delta>0$ then $\epsilon^{ZZ,BR}_1\ge \epsilon_1>\epsilon^{ZP,BR}_2=-\infty$ is necessary and sufficient for existence of the ZZ solution. Otherwise, if $\chi^1<0$ then $\epsilon^{ZP,BR}_2=+\infty$ and $\epsilon_1\ge\epsilon^{ZZ,BR}_1\ge\epsilon^{PP,BR}$ is necessary and sufficient for existence of the ZZ solution. If $\chi^1=0$ and $\delta>0$ then $\epsilon^{ZP,BR}_2-\epsilon^{PP,BR}\ge0$ as shown above and a continuum of ZZ solutions exist if and only if
\begin{eqnarray*}\epsilon_1&=&\epsilon^{PP,BR}+\epsilon_2+\frac{(1-M p)(1-M_fp \beta)\mu}{\lambda N p}+\\
&& \frac{\epsilon_2(1-p)N(1-M_f M p \beta)}{(1-p)(1-M M_f p \beta)(\psi-N)+(1-M)p(1-M_f\beta)\psi} \ge \epsilon^{PP,BR}.
\end{eqnarray*}

We conclude that a BRE exists if and only if
\begin{equation}\label{eq: BRE cutoff2}
    \epsilon_1\ge\bar{\epsilon}_{BR} := \left\{
\begin{array}
[c]{ll}%
\min\left\{\epsilon^{PP,BR},\epsilon^{ZP,BR}_2\right\}, & \text{if }\delta\ge 0\\
-\infty, & \text{if }\delta<0,
\end{array}
\right.
\end{equation} where $\epsilon^{PP,BR}$ and $\epsilon^{ZP,BR}_2$ are defined in \eqref{eq: eps^PP,BR} and \eqref{eq: eps^ZP,BR}, respectively, with $q=1$. 
\subsection{Proof of Proposition \ref{prop: BR completeness}} \label{appe sec: prop5}

Suppose $\delta=(M-1)(1-M_f\beta)+a N<0$, which implies $det(Q)=den^{ZZ,BR}=-\delta(-\delta+(1-\rho)(M+a N+M_f\beta(1-(p+q)M)))>0$, $det(Q+\lambda\sigma\psi e_2e_2')=den^{ZP,BR}=\delta^2-\delta(a \psi+(1-\rho)(M+a N+M_f\beta(1-M(p+q))))+(1-p)a \psi(M+a N+M_f\beta(1-M(p+q)))>0$, and $num^{PZ,BR}_1=((1-q)(M+a N+M_f\beta(1-M(p+q)))-\delta)>0$, from Proposition \ref{prop:BR coherence}. Also by Proposition \ref{prop:BR coherence}: $num^{PZ,BR}_1>0$ implies no PZ; ZZ exists under $\delta<0$ if and only if $q<1$ and $\epsilon_1\le \epsilon^{ZP,BR}_2$; $den^{ZP,BR}>0$ implies $\epsilon^{PP,BR}>\epsilon^{ZP,BR}_{2}$, $\epsilon^{ZP,BR}_2=-\infty$ if $q=1$, and ZP exists if and only if $\epsilon^{PP,BR}\ge \epsilon_1>\epsilon^{ZP,BR}_{2}$. Define $\epsilon^{ZP,BR}:=\epsilon^{ZP,BR}_2$. We conclude that the PP solution is the unique BRE when $\epsilon_1>\epsilon^{PP,BR}$, the ZP solution is the unique BRE when $\epsilon^{PP,BR}\ge\epsilon_1>\epsilon^{ZP,BR}$. Otherwise, the ZZ solution is the unique solution if $q<1$ and $\epsilon_1\le\epsilon^{ZP,BR}$. If $\delta\ge 0$ then by Proposition 4 there exist $p,q,\epsilon_1$ and $\epsilon_2\ge 0$ for which there are no solutions or multiple solutions.\footnote{Alternatively, one can show that $(M-1)(1-M_f\beta)+\lambda\sigma N<0$ ensures completeness and coherence using techniques developed by \citetalias{AscariMavroeidis}. Results available on request.}

\subsection{Proof of Proposition \ref{prop: BRRPE existence}} \label{appe sec: BRRPE}

The proof of Proposition \ref{prop: BRRPE existence} is a straightforward extension of the proof of Proposition \ref{prop:BR coherence}. Define $\bar q:=Pr(\epsilon_t=2)=(1-p)/(2-p-q)$. The regime-specific levels of inflation in BR-RPE $i$, $\hat{\pi}^i=(\pi^i_1,\pi^i_2)'$, are given by fixed point restrictions that have the same basic form as the BRE fixed point restrictions except we replace $q$ with $\bar q$ and $p$ with $1-\bar q$. Therefore, BR-RPE will exist if and only if
\begin{equation}\label{eq: BRRPE cutoff}
    \epsilon_1\ge\bar{\epsilon}_{BR,RPE} := \left\{
\begin{array}
[c]{ll}%
\min\left\{\epsilon^{PP,BR,RPE},\epsilon^{ZP,BR,RPE}_2\right\}, & \text{if }\delta\ge0\\
-\infty, & \text{if }\delta<0,
\end{array}
\right.
\end{equation} where $\delta=(M-1)(1-M_f\beta)+\lambda \sigma N$, and $\epsilon^{PP,BR,RPE}$ and $\epsilon^{ZP,BR,RPE}_2$ are defined in \eqref{eq: eps^PP,BR} and \eqref{eq: eps^ZP,BR}, respectively, assuming $p=1-\bar q$, and $q=\bar q.$ In the special case $q=1$ (which implies $\bar q=1$), we have $\bar{\epsilon}_{BR,RPE}=-\infty$ for any $\delta$, as the PP solution exists if and only if $\epsilon_1> -\mu(1+\lambda\sigma\psi)(\lambda\psi)^{-1}+(M(1-M_f\beta)+M_f\beta+\lambda\sigma N)((M-1)(1-M_f\beta)+\lambda\sigma(N-\psi))^{-1}\epsilon_2=\epsilon^{PP,BR,RPE}$ and the ZP exists if and only if $\epsilon_1\le \epsilon^{PP,BR,RPE}$. For $q<1$, one can show: $\epsilon^{PP,BR}-\epsilon^{PP,BR,RPE}=-\Xi^B_{PP}\rho$ and $\epsilon^{ZP,BR}-\epsilon^{ZP,BR,RPE}=-\Xi^B_{ZP}\rho$ where
\begin{eqnarray*}
\Xi^B_{PP}&:=& \frac{(p-1)(\eta_6+ M M_f \beta)\eta_7 (\lambda \epsilon_2 \psi+\mu \eta_7)}{\lambda\psi (\eta_7+(1-q)\eta_6)((1-p)(\lambda\sigma-\delta)+((1+\lambda\sigma)(1-q)+\lambda\sigma(1-\rho)(\psi-1)))},\\
\Xi^B_{ZP}&:=&\frac{\delta(\eta_6+M M_f\beta)(\lambda \epsilon_2 \psi+\mu \eta_7)}{\lambda(q - 1) (M + \lambda\sigma N + M_f \beta (1 - M )) \eta_6 \psi},
\end{eqnarray*}

\noindent and $\eta_6:=M(1-M_f \beta p)+\lambda\sigma N+M_f\beta(1-q M)>0$, and $\eta_7:=(a(\psi-N)+(1-M)(1-M_f\beta))>0$. Since $\delta\le \lambda\sigma$, it is straightforward to show that $\Xi^B_{PP}\le0$. Further, if $\delta \ge0$ then $\Xi^B_{ZP}\le0$. It follows that  $\bar \epsilon_{BR} \ge \bar{\epsilon}_{BR,RPE}$ if $\delta\ge 0$ and $p+q-1\ge0$ or $\delta<0$. 

\subsection{Proof of Proposition \ref{prop: REE E-stable}} \label{appe sec: prop7}

Consider Proposition \ref{prop: REE E-stable}. To assess E-stability of a REE, we express $Y^i=(Y^{i'}_1,Y^{i'}_2)'$ as a function of agents' expectations, $\tilde{Y}^e=(Y^{e'}_1,Y^{e'}_2)'$:
\begin{eqnarray*}
Y^{PP}(\tilde{Y}^e)&:=&\begin{pmatrix}
p A_P & (1-p) A_P\\
(1-q) A_P & q A_P
\end{pmatrix} \tilde{Y}^e+\Gamma^{PP},\\
Y^{ZP}(\tilde{Y}^e)&:=&\begin{pmatrix}
p A_Z & (1-p) A_Z\\
(1-q) A_P & q A_P
\end{pmatrix} \tilde{Y}^e+\Gamma^{ZP},\\
Y^{PZ}(\tilde{Y}^e)&:=&\begin{pmatrix}
p A_P & (1-p) A_P\\
(1-q) A_Z & q A_Z
\end{pmatrix} \tilde{Y}^e+\Gamma^{PZ},\\
Y^{PP}(\tilde{Y}^e)&:=&\begin{pmatrix}
p A_Z & (1-p) A_Z\\
(1-q) A_Z & q A_Z
\end{pmatrix} \tilde{Y}^e+\Gamma^{ZZ},
\end{eqnarray*}
where $\Gamma^i$ collect terms that do not depend on beliefs, $\tilde{Y}^e$. It immediately follows that 
\begin{align*}
DT_{ Y^{PP}}=K \otimes A_P-I, &&
DT_{ Y^{ZP}}=\begin{pmatrix}
p A_Z & (1-p) A_Z\\
(1-q) A_P & q A_P
\end{pmatrix}-I, \\
DT_{ Y^{ZZ}}= K\otimes A_Z-I, &&
DT_{Y^{PZ}}=\begin{pmatrix}
p A_P & (1-p) A_P\\
(1-q) A_Z & q A_Z
\end{pmatrix}-I.
\end{align*}

REE $i$ is E-stable if the real parts of the eigenvalues of $DT_{ Y^i}$ are negative. Since the real parts of the eigenvalues of $DT_{Y^{PP}}$ are negative and the real part of an eigenvalue of $DT_{ Y^{ZZ}}$ is positive, the PP (ZZ) solution is always (never) E-stable. The following condition is necessary for E-stability of the ZP solution: $Det(DT_{ Y^{ZP}})=\frac{a}{1+a\psi}den^{ZP}>0$, where $den^{ZP}$ is defined in the proof of Proposition \ref{prop1}. By  Proposition \ref{prop1}, $den^{ZP}>0$ implies $\epsilon^{PP}>\epsilon^{ZP}_{2}$,
where $\epsilon^{PP},\epsilon^{ZP}_{2}$ are defined in the proof of Proposition \ref{prop1}, and hence $\epsilon_1>\epsilon^{PP}$ is necessary for existence of PP and $\epsilon_1\le \epsilon^{PP}$ is necessary for existence of ZP. It follows that the E-stability and existence of the ZP solution precludes existence of the PP solution. The following condition is necessary for E-stability of the PZ solution:
$Det(DT_{ Y^{PZ}})=\frac{1}{1+a\psi}den^{PZ}>0$, 
where $den^{PZ}$ is defined in the proof of Proposition \ref{prop1}. By Proposition \ref{prop1}, $den^{PZ}<0$ is necessary for PZ existence. We conclude that the PZ solution can never be E-stable.\footnote{If $q=1$, the PP exists and is E-stable if and only if $\epsilon_1>\epsilon^{PP}$ and if $Det(DT_{ Y^{ZP}})>0$ then $\theta>1$, such that ZP exists if and only if $\epsilon_1\le\epsilon^{PP}$ by Proposition \ref{prop1}. The ZZ and PZ solutions cannot be E-stable. } 

In sum, if the PP solution exists it is E-stable. If the ZP solution exists and is E-stable then the PP solution does not exist. The ZZ and PZ solutions are never E-stable.  

\subsection{Proof of Proposition \ref{prop: RPE E-stable}} \label{appe sec: prop9}

To assess E-stability of each RPE, we express the RPE unconditional mean of inflation and output as a function of agents' expectations, $Y^e$:
\begin{align*}
\bar Y^{PP}(Y^e):=A_PY^e+\bar \Gamma^{PP}, &&
\bar Y^{ZP}(Y^e):=\left(\bar q A_P+(1-\bar q)A_Z\right)Y^e+\bar \Gamma^{ZP},\\
\bar Y^{ZZ}(Y^e):=A_ZY^e+\bar \Gamma^{ZZ} &&
\bar Y^{PZ}(Y^e):=\left((1-\bar q)A_P+\bar q A_Z\right)Y^e+\bar \Gamma^{PZ}.
\end{align*}
where $\bar \Gamma^i$ collect terms that do not depend on beliefs, $Y^e$. It immediately follows that 
\begin{align*}
DT_{\bar Y^{PP}}=A_P-I,  &&
DT_{\bar Y^{ZP}}=\bar q A_P+(1-\bar q)A_Z-I,\\
DT_{\bar Y^{ZZ}}=A_Z-I, &&
DT_{\bar Y^{PZ}}=(1-\bar q)A_P+\bar q A_Z-I.
\end{align*}

It is straightforward to show that the real parts of the eigenvalues of $DT_{\bar Y^{PP}}$ are negative and the real part of an eigenvalue of $DT_{\bar Y^{ZZ}}$ is positive. Therefore, the PP (ZZ) RPE is always (never) E-stable. The ZP RPE is E-stable if and only if
\begin{eqnarray*}
tr(DT_{\bar Y^{ZP}})&=& \beta +a -\frac{a \bar q  \psi  (\beta +a +1)}{a  \psi +1}-1<0,\\
Det(DT_{\bar Y^{ZP}})&=&\frac{\bar q a (a \psi +\psi )}{a  \psi +1}-a>0,
\end{eqnarray*}
where $tr(B)$ denotes the trace of matrix $B$. We have $tr(DT_{\bar Y^{ZP}})<0<Det(DT_{\bar Y^{ZP}})$ if and only if $\bar q (1+a)\psi-1-a\psi>0$. From the proofs of Propositions \ref{prop1} and \ref{prop: RPE existence}:
\begin{eqnarray*}
\epsilon^{PP,RPE}-\epsilon^{ZP,RPE}_2&=&v(\bar q (1+a)\psi-1-a\psi)),\\
v&:=&\frac{a (\lambda\epsilon_2 \psi +a\mu  (\psi -1))}{\lambda(1-\bar q) \psi  (a +1)
   (a (\psi -\bar q)+ 1-\bar q)}>0.
\end{eqnarray*}
Therefore, if the ZP RPE is E-stable then $\epsilon^{PP,RPE}>\epsilon^{ZP,RPE}_2$ and the condition for PP existence becomes $\epsilon_1>\epsilon^{PP,RPE}$ and the condition for ZP existence becomes $\epsilon^{PP,RPE}\ge\epsilon_1>\epsilon^{ZP,RPE}_2$ as demonstrated in the proofs of Propositions \ref{prop1} and \ref{prop: RPE existence}.\footnote{If $\bar q=1$, the PP exists and is E-stable if and only if $\epsilon_1>\epsilon^{PP,RPE}$ and the ZP exists and is E-stable if and only if $\epsilon_1\le\epsilon^{PP,RPE}$. The ZZ and PZ solutions cannot be E-stable. } Hence, if the ZP RPE exists and is E-stable then the PP solution does not exist. The PZ solution is E-stable if and only if 
\begin{eqnarray*}
tr(DT_{\bar Y^{PZ}})&=& \frac{\beta -2 a \psi +a -1}{a  \psi +1}+\frac{\bar q \left(\beta 
   a  \psi +a^2 \psi +a  \psi \right)}{a  \psi +1}<0,\\
Det(DT_{\bar Y^{PZ}})&=&-\frac{a (1-\psi )}{a  \psi +1}-\frac{a \bar q  (a  \psi
   +\psi )}{a  \psi +1}>0,
\end{eqnarray*}
which holds if and only if $0<\psi-1-\bar q \psi(1+a)=den^{PZ,RPE}a^{-1}$ where $den^{PZ,RPE}$ is equal to $den^{PZ}$ defined in the Proposition \ref{prop1} proof when $q=\bar q$ and $p=1-\bar q$. From the proof of Proposition \ref{prop: RPE existence}, the PZ RPE only exists if $den^{PZ,RPE}<0$. Hence the PZ RPE is never E-stable.

Therefore, the PP RPE is the only E-stable RPE solution when $\epsilon_{1}>\epsilon^{PP,RPE}$, and the ZP RPE is the only E-stable RPE solution when $\epsilon^{PP,RPE}\ge\epsilon_1>\epsilon^{ZP,RPE}_2$. It follows that a unique E-stable RPE solution exists when $\epsilon_{1}>\bar{\epsilon}_{RPE}$.

\end{subappendices}

\newpage

\begin{subappendices}
\setcounter{equation}{0} \renewcommand{\theequation}{B\arabic{equation}}

\section*{Appendix B}\label{sec: OAppendix}
\renewcommand{\thesubsection}{B.\arabic{subsection}}

\setcounter{subsection}{0}
\setcounter{page}{1}

\subsection{RPE under Infinite Horizon Learning}\label{oappesec: RPE IH}

Consider the following infinite horizon New Keynesian model:\footnote{See \cite{EGP2021} for a recent derivation of the model (\ref{eq:IHAD})-(\ref{eq:MP2}). Note that this model collapses to the standard 3-equation model in our paper if we impose RE. Consequently, a stochastic process for inflation, output and the interest rate is a REE of (\ref{eq:IHAD})-(\ref{eq:MP2}) if and only if said stochastic process is a REE of (\ref{eq:IS})-(\ref{eq:MP}).}
\begin{eqnarray}
x_t&=&-\sigma i_t+ \hat E_t \sum_{T\ge t}\beta^{T-t}\left((1-\beta)x_{T+1}+\sigma \pi_{T+1}-\sigma\beta i_{T+1}+\epsilon_T\right) \label{eq:IHAD}, \\
\pi_t&=&\lambda x_t+ \hat E_t \sum_{T\ge t}(\xi\beta)^{T-t}\left(\xi \beta \lambda x_{T+1}+(1-\xi)\beta \pi_{T+1}\right) \label{eq:IHAS}, \\
i_t&=&\max\{\psi \pi_t,-\mu\} \label{eq:MP2},
\end{eqnarray}
where $\lambda:=(1-\xi\beta)(1-\xi)/\xi$. Under infinite horizon learning, agents need to forecast the paths of the nominal interest rate and the shock, in addition to the paths of inflation and output. Consistent with the RPE studied in section \ref{sec: cohnorat}, we assume that agents set endogenous and exogenous variable forecasts at all horizons equal to the unconditional means of each variable (i.e. $\hat E_t z_T=E(z_T)$ for all $T>t$ and $z=\pi,x,i,\epsilon$). We have:
\begin{eqnarray*}
E(\pi)&=&E\left(\lambda x_t+ \hat E_t \sum_{T\ge t}(\xi\beta)^{T-t}\left(\xi \beta \lambda x_{T+1}+(1-\xi)\beta \pi_{T+1}\right)\right),\\
\implies E(x)&=&\frac{1-\beta}{\lambda} E(\pi),
\end{eqnarray*}
and 
\begin{eqnarray*}
\pi_t&=&\lambda x_t+ \sum_{T\ge t}(\xi\beta)^{T-t}\left(\xi \beta \lambda E(x)+(1-\xi)\beta E(\pi)\right)=\lambda x_t+ \beta E(\pi),\\
\implies x_t&=&\lambda^{-1}(\pi_t-\beta E(\pi)).
\end{eqnarray*}
Substituting for $x_t$ and also for expectations in (\ref{eq:IHAD}) gives an expression for RPE inflation:
\begin{eqnarray*}
x_t&=&\lambda^{-1}(\pi_t-\beta E(\pi))\\
&=&-\sigma i_t+ \epsilon_t+\sum_{T\ge t}\beta^{T-t}\left((1-\beta)E(x)+\sigma E(\pi)-\sigma\beta E(i)+\beta E(\epsilon)\right), \\
\implies \pi_t&=&-\lambda\sigma i_t+\lambda\epsilon_t+(1+\frac{\lambda\sigma}{1-\beta})E(\pi)-\frac{\beta\lambda\sigma}{1-\beta}E(i)+\frac{\lambda\beta}{1-\beta}E(\epsilon).
\end{eqnarray*}
Let $\hat z:=(z_1,z_2)'$ denote the vector of state-contingent RPE values of $z$ for any variable, $z$. Note that $E(z)=\bar q z_2+(1-\bar q)z_1$. Then the infinite horizon RPE solution for inflation satisfies:
\begin{eqnarray*}
\hat \pi=\left(1+\frac{\lambda\sigma}{1-\beta}\right)\tilde K \hat \pi-\lambda\sigma\left(I-\beta \tilde K\right)^{-1}\hat i+\lambda\left(I-\beta \tilde K\right)^{-1}\hat \epsilon,
\end{eqnarray*}
where $I$ is the identity matrix and 
\begin{eqnarray*}
\tilde K:=\begin{pmatrix}1-\bar q & \bar q\\
1-\bar q & \bar q\end{pmatrix}.
\end{eqnarray*}
Premultiplying both sides of the last equation by $\left(I-\beta \tilde K\right)$ and rearranging yields
\begin{eqnarray}
\left(I-(1+\lambda\sigma)\tilde K\right)\hat \pi=-\lambda\sigma\hat i+\lambda\hat \epsilon \label{eq: RPE IH fp}.
\end{eqnarray}
From the proof of Proposition \ref{prop1} and \ref{prop: RPE existence}, it can be seen that any solution of (\ref{eq: RPE IH fp}) is also a RPE of (\ref{eq:IS})-(\ref{eq:MP}). Hence, the infinite horizon model (\ref{eq:IHAD})-(\ref{eq:MP2}) admits the same RPE as (\ref{eq:IS})-(\ref{eq:MP}), and therefore an incoherent model can admit RPE under infinite horizon learning under some conditions. The result is summarized in the following proposition.

\begin{proposition}\label{prop:IH coherence2}
Consider (\ref{eq:IHAD})-(\ref{eq:MP2}) and suppose $\epsilon_2\ge0$. Then:
\begin{enumerate}
    \item[i.] A restricted perceptions equilibrium (RPE) exists if and only if $\epsilon_1\ge\bar{\epsilon}_{RPE}$, where $\bar{\epsilon}_{RPE}$ depends on the model's parameters, see Equation \eqref{eq: RPE cutoff} in the Appendix \ref{appe sec: prop3}, and satisfies $\bar{\epsilon}_{RPE}=-\infty$ if $q=1$.
    \item[ii.] $\bar{\epsilon}_{REE}\ge\bar{\epsilon}_{RPE}$ if and only if $p+q\ge1$.
\end{enumerate}
\end{proposition}

\subsection{Endogenous Bounded Rationality} \label{appe sec: EndoBR}

Following \cite{Moberly}, this section models bounded rationality as an optimal choice by agents who face a cost of paying attention in the spirit of \cite{gabaix2020}. To that end, consider the modified version of (\ref{eq:IS})-(\ref{eq:MP})
\begin{eqnarray}
x_t&=&M_{\epsilon_t} E_tx_{t+1}-\sigma(i_t-E_t\pi_{t+1})+\epsilon_t \label{ISe},\\
\pi_t&=&\lambda x_t +M_{f,\epsilon_t}\beta E_t\pi_{t+1},\\
i_t&=&\max\{\psi\pi_t,-\mu\} \label{TRe},
\end{eqnarray}
where $\epsilon_t,p,q$, etc., are defined in the main text, and $p<1$, $q=1$, $\epsilon_1=\epsilon<0$ and $\epsilon_2=0$. In line with section 3.1 of \cite{Moberly} we assume that $0\le M_{f,\epsilon_t}\le 1$ and $0\le M_{\epsilon_t}\le 1$ can switch values when $\epsilon_t$ changes: $(M_{f,\epsilon_t},M_{\epsilon_t})=(M_{f,i},M_{i})$ if and only if $\epsilon_t=\epsilon_i$, for $i=1,2$. 

We will shortly describe how $M_{f,\epsilon_t}$ and $M_{\epsilon_t}$ arise endogenously following the approach of \cite{Moberly}. First, we establish the set of MSV solutions. For a given $M_{f,\epsilon_t},M_{\epsilon_t}$, there are four possible types of MSV solutions: PP, ZP, PZ and ZZ solutions. Let $l\in\{PP,ZP,PZ,ZZ\}$ denote a MSV solution. Then in the ZP and PP equilibria, the ``high state" ($\epsilon_t=\epsilon_2$) equilibrium outcomes are given by:
\begin{eqnarray*}
    x^l_2=\pi^l_2=i^l_2=r^l_2=mc^l_2=0,
\end{eqnarray*}
where $r^l_2=i^l_2-\pi^l_2$ is the ex ante real interest rate and $mc^l:=(\phi+\sigma)x^l_2$ is marginal cost.\footnote{See \cite{Moberly} for the microfoundations.} For the PZ and ZZ solutions, the ``high state" outcomes are given by:
\begin{eqnarray*}
    x^l_2&=&\left(1-M_2-\frac{\sigma \lambda}{1-\beta M_{f,2}}\right)^{-1}\sigma\mu,\\
    \pi^l_2&=&\frac{\lambda}{1-\beta M_{f,2}}x^l_2,\\
    mc^l_2&=&(\phi+\sigma)x^l_2,\\
    r^l_2&=&-\mu-\pi^l_2.
\end{eqnarray*}
The low state PP equilibrium is characterized by:
\begin{eqnarray*}
    x^{PP}_1&=&\left(1-pM_1+\frac{(\psi-p)\sigma \lambda}{1-p\beta M_{f,1}}\right)^{-1}\epsilon_1,\\
    \pi^{PP}_1&=&\frac{\lambda}{1-p\beta M_{f,1}}x^{PP}_1,\\
    mc^{PP}_1&=&(\phi+\sigma)x^{PP}_1,\\
    r^{PP}_1&=&\psi \pi^{PP}_1-p\pi^{PP}_1.
\end{eqnarray*}
Similarly, the low state ZP equilibrium is characterized by:
\begin{eqnarray*}
    x^{ZP}_1&=&\left(1-pM_1-\frac{p\sigma \lambda}{1-p\beta M_{f,1}}\right)^{-1}\left(\sigma\mu+\epsilon_1\right),\\
    \pi^{ZP}_1&=&\frac{\lambda}{1-p\beta M_{f,1}}x^{ZP}_1,\\
    mc^{ZP}_1&=&(\phi+\sigma)x^{ZP}_1,\\
    r^{ZP}_1&=&-\mu-p\pi^{ZP}_1.
\end{eqnarray*}
The low state PZ equilibrium is given by:
\begin{eqnarray*}
    x^{PZ}_1&=&\left(1-pM_1+\frac{(\psi-p)\sigma \lambda}{1-p\beta M_{f,1}}\right)^{-1}\left((1-p)\left(M_1x^{PZ}_2+\left(\frac{(p-\psi)\sigma M_{f,1}\beta}{1-p\beta M_{f,1}}+\sigma\right)\pi^{PZ}_2\right) +\epsilon_1\right),\\
    \pi^{PZ}_1&=&\frac{\lambda x^{PZ}_1+M_{f,1}\beta (1-p)\pi^{PZ}_2}{1-p\beta M_{f,1}},\\
    mc^{PZ}_1&=&(\phi+\sigma)x^{PZ}_1,\\
    r^{PZ}_1&=&\psi \pi^{PZ}_1-p\pi^{PZ}_1-(1-p)\pi^{PZ}_2.
\end{eqnarray*}
Finally, low state ZZ equilibrium is given by:
\begin{eqnarray*}
    x^{ZZ}_1&=&\left(1-pM_1-\frac{p\sigma \lambda}{1-p\beta M_{f,1}}\right)^{-1}\left((1-p)\left(M_1x^{ZZ}_2+\left(\frac{p\sigma M_{f,1}\beta}{1-p\beta M_{f,1}}+\sigma\right)\pi^{ZZ}_2\right) +\epsilon_1+\sigma\mu\right),\\
    \pi^{ZZ}_1&=&\frac{\lambda x^{ZZ}_1+M_{f,1}\beta (1-p)\pi^{ZZ}_2}{1-p\beta M_{f,1}},\\
    mc^{ZZ}_1&=&(\phi+\sigma)x^{ZZ}_1,\\
    r^{ZZ}_1&=&-\mu-p\pi^{ZZ}_1-(1-p)\pi^{ZZ}_2.
\end{eqnarray*}

Following \cite{Moberly} we model the choice of attention of each household and firm. In our framework, households and firms need to balance the loss of not paying attention with the cost of paying attention. First, consider the household problem. The household consumption function in the high state is given by.\footnote{For convenience we set $\sigma=1$.}
\begin{eqnarray*}
c^l_t&=&\sum_{h\ge 0}(\beta m_2)^h ((1-\beta)x^l_{2}-\beta r^l_{2})\\
&=& \frac{X_2}{1-\beta m_2},
\end{eqnarray*}
where $M_2=m_2$ in equilibrium and $X_2:=(1-\beta)x^l_{2}-\beta r^l_{2}$. Following, \cite{Moberly} we can show that the attention decision in the high state \textit{when $X_2\neq 0$} (i.e. the optimal $m_2$) depends chiefly on the following quantity:
\begin{eqnarray*}
E\left[\left(\frac{\partial c^l_t}{\partial m_{2}}\right)^2\right]=\frac{(\beta X_2)^2}{(1-\beta m_{d,2})^4},
\end{eqnarray*}
where the derivative is evaluated at some default level of attention $m_{d,2}\in[0,1]$. The households' optimal attention parameter in the high state of solution $l$, $\bar m^l_2$, for a given $M_1$, $M_{f,1}$, $M_2$ and $M_{f,2}$, is 
\begin{eqnarray*}
    \bar m^l_2= \max\left(m_{d,2},1-\frac{\xi_c^2}{E\left[\left(\frac{\partial c^l_t}{\partial m_{2}}\right)^2\right]} \right),
\end{eqnarray*}
where the parameter $\xi_c$ is the scale-free cost of attention. Having pinned down $M_2=\bar m^l_2$, we can characterize optimal $m_1$. In the low state, consumption is given by:
\begin{eqnarray*}
c^l_t&=&E_t\left\{\sum_{h\ge 0}\beta^h \Pi_{k=1}^h m_{t+k}((1-\beta)x_{t+h}-\beta(r_{t+h}-\epsilon_{t+h}))\right\}\\
&=&E_t\left\{\sum_{h\ge 0}\beta^h \Pi_{k=1}^h m_{t+k}X_{t+h}\right\},
\end{eqnarray*}
where $m_{t+k}=m_1$ and $X_{t+k}=X_1:=(1-\beta)x_{1}-\beta(r_{1}-\epsilon_{1})$ if $\epsilon_{t+k}=\epsilon_1$, otherwise $m_{t+k}=m_2$ and $X_{t+k}=X_2:=(1-\beta)x_{2}-\beta r_{2}$. For $h\ge 1$ we
have
\begin{eqnarray*}
E_t\beta^h \Pi_{k=1}^h m_{t+k}X_{t+h}&=&
 \beta^h\left((m_1p)^hX_1+\left((m_1p)^{h-1}m_2+(m_1p)^{h-2}(m_2)^{2}+\hdots+m_2^h\right)(1-p)X_2\right)\\
 &=& \beta^h\left((m_1p)^h X_1+(1-p)\frac{m_2^{h+1}-m_2(m_1p)^h}{m_2-m_1p}X_2\right).
\end{eqnarray*}
Substituting the last equation into the equation for $c_t$ and rearranging yields:
\begin{eqnarray*}
    c^l_t&=&\frac{1}{1-\beta m_1 p}X_1+\frac{(1-p)m_2}{m_2-m_1p}\left(\frac{\beta m_2}{1-\beta m_2}-\frac{\beta m_1 p}{1-\beta m_1 p}\right) X_2.
\end{eqnarray*}
Note that
\begin{eqnarray*}
    \frac{(1-p)m_2}{m_2-m_1p}\left(\frac{\beta m_2}{1-\beta m_2}-\frac{\beta m_1 p}{1-\beta m_1 p}\right) = \frac{(1-p)\beta m_2}{(1-\beta m_2)(1-p \beta m_1)}.
\end{eqnarray*}
Hence, the equation for $c_t$ can be expressed as
\begin{eqnarray*}
    c^l_t&=&\frac{1}{1-\beta m_1 p}X_1+\frac{(1-p)\beta m_2}{(1-\beta m_2)(1-p \beta m_1)}X_2\\
    &=& \frac{1}{1-\beta m_1 p}\left(X_1+\frac{(1-p)\beta m_2}{1-\beta m_2}X_2\right).
\end{eqnarray*}

In equilibrium, $M_1=m_1$ and $M_2=m_2$. \cite{Moberly} shows that the attention decision in the low state (i.e. the optimal $m_1$) depends chiefly on the following quantity:
\begin{eqnarray*}
E\left[\left(\frac{\partial c^l_t}{\partial m_{1}}\right)^2\right]=\frac{(\beta p)^2\left(X_1(1-\bar m^l_2\beta)+\bar m^l_2(1-p)\beta X_2\right)^2}{(1-\beta p m_{d,1})^4(1-\bar m^l_2\beta)^2},
\end{eqnarray*}
where the derivative is evaluated at some default level of attention $m_{d,1}\in[0,1]$ and $m_2=\bar m^l_2$ is assumed. The households' optimal attention parameter in solution $l$, $\bar m^l_1$, for given $M_1$, $M_2$ is 
\begin{eqnarray*}
    \bar m^l_1= \max\left(m_{d,1},1-\frac{\xi_c^2}{E\left[\left(\frac{\partial c^l_t}{\partial m_{1}}\right)^2\right]} \right),
\end{eqnarray*}
where again the parameter $\xi_c$ is the scale-free cost of attention.

Now consider the firm problem.\footnote{We assume that firms have their own cognitive discount factor (see \cite{gabaix2020}, footnote 13). We arrive at qualitatively similar results if we assume that households and firms have the same congitive discount factor.} The firm pricing function in the high state of solution $l$ is given by

\begin{eqnarray*}
q^l_t&=&(1-\beta\theta)E_t\left\{\sum_{h\ge 0} (\beta\theta m_{f,2})^h\left(\pi_{t+1}+\hdots \pi_{t+h}+mc_{t+h}\right)\right\}\\
&=&(1-\beta\theta)\sum_{h\ge 0} (\beta\theta m_{f,2})^h mc^l_{2}+(1-\beta\theta)\sum_{h\ge 1} (\beta\theta m_{f,2})^h h\pi^l_{2}\\
&=& \frac{(1-\beta\theta)}{1-\beta \theta m_{f,2}}mc^l_2+\frac{(1-\beta\theta)\beta \theta m_{f,2}}{(1-\beta \theta m_{f,2})^2}  \pi^l_2,
\end{eqnarray*}
where $q$ is the relative price, $1-\theta$ is the probability that a firm can reset its price in a given period (see \cite{gabaix2020}) and $M_{f,2}=m_{f,2}\left(\theta+(1-\theta)\frac{1-\beta\theta}{1-\beta \theta m_{f,2}}\right)$ in equilibrium. The relevant quantity for attention choice is $E\left[\left(\frac{\partial q^l_t}{\partial m_{f,2}}\right)^2\right]$ where 
\begin{eqnarray*}
\frac{\partial q^l_t}{\partial m_{f,2}}=\frac{\beta\theta(1-\beta\theta)(mc^l_2(1-\beta\theta m_{f,2})+\pi^l_2(1+m_{f,2}\beta\theta))}{(1-\beta\theta m_{f,2})^3},
\end{eqnarray*}
and the derivative is evaluated at some default level of attention $m_{d,f,2}\in[0,1]$. The firm's optimal attention parameter in solution $l$, $\bar m^l_{f,1}$, for given $M_{f,2}$ is 
\begin{eqnarray*}
    \bar m^l_{f,2}= \max\left(m_{d,f,2},1-\frac{\xi_f^2}{E\left[\left(\frac{\partial q^l_t}{\partial m_{f,2}}\right)^2\right]} \right),
\end{eqnarray*}
where the parameter $\xi_f$ is the scale-free cost of attention. Now consider the low state pricing function:
\begin{eqnarray*}
q^l_t&=&(1-\beta\theta)E_t\left\{\sum_{h\ge 0} (\beta\theta)^h \Pi_{k=1}^h m_{f,t+k}\left(\pi_{t+1}+\hdots \pi_{t+h}+mc_{t+h}\right)\right\}\\
&=&(1-\beta\theta)E_t\left\{\sum_{h\ge 0} (\beta\theta)^h \Pi_{k=1}^h m_{f,t+k} mc_{t+h}\right\}+\\&&(1-\beta\theta)E_t\left\{\sum_{h\ge 1} (\beta\theta)^h \Pi_{k=1}^h m_{f,t+k} \left(\pi_{t+1}+\hdots+\pi_{t+h}\right)\right\}\\
&=& \frac{(1-\beta\theta)}{1-p \beta \theta m_{f,1}}mc_1+\frac{(1-\beta\theta)(1-p)\beta \theta m_{f,2}}{(1-\beta \theta m_{f,2})(1-p \beta \theta m_{f,1})}  mc_2\\
&+&(1-\beta\theta)E_t\left\{\sum_{h\ge 1} (\beta\theta)^h \Pi_{k=1}^h m_{f,t+k} \left(\pi_{t+1}+\hdots+\pi_{t+h}\right)\right\},
\end{eqnarray*}
where
\begin{eqnarray*}
    E_t\sum_{h\ge 0} (\beta\theta)^h \Pi_{k=1}^h m_{f,t+k} mc_{t+h}=\frac{1}{1-p\beta \theta m_{f,1}}mc_1+\frac{(1-p)\beta \theta m_{f,2}}{(1-\beta \theta m_{f,2})(1-p \beta \theta m_{f,1})}  mc_2,
\end{eqnarray*} 
can be derived by following steps used to derive the consumption function. Now focus on the term involving expected future inflation. 
\begin{eqnarray*}
&&E_t\left\{\sum_{h\ge 1} (\beta\theta)^h \Pi_{k=1}^h m_{f,t+k} \left(\pi_{t+1}+\hdots+\pi_{t+h}\right)\right\}\\
&=&(1-p)\sum_{h\ge 1} (\beta\theta m_{f,2})^h h \pi_2\\
&+&p(1-p)\left(\beta\theta m_{f,1}\pi_1+\beta\theta m_{f,1} \sum_{h\ge 1} (\beta\theta m_{f,2})^h \left(\pi_1+h\pi_2\right)\right)\\
&+&p^2(1-p)\left(\beta\theta m_{f,1}\pi_1+(\beta\theta m_{f,1})^2 2\pi_1+(\beta\theta m_{f,1})^2 \sum_{h\ge 1} (\beta\theta m_{f,2})^h \left(2\pi_1+h\pi_2\right)\right)\\
&+&\hdots\\
&=&(1-p)\sum_{i\ge0}p^i \left(\sum_{k=1}^{i}(\beta\theta m_{f,1})^k k \pi_1 + (\beta\theta m_{f,1})^i\sum_{h\ge 1}(\beta\theta m_{f,2})^h\left(i \pi_1+h \pi_2\right)\right).
\end{eqnarray*}
First, consider the term $\sum_{i\ge0}p^i\sum_{k=1}^{i}(\beta\theta m_{f,1})^k k \pi_1 $:
\begin{eqnarray*}
    \sum_{i\ge0}p^i \sum_{k=1}^{i}(\beta\theta m_{f,1})^k k \pi_1 &=&p\beta\theta m_{f,1}(1+p+p^2+\hdots)\pi_1\\&+&(p\beta\theta m_{f,1})^2(1+p+p^2+\hdots)2\pi_1\\
    &+&(p\beta\theta m_{f,1})^3(1+p+p^2+\hdots)3\pi_1\\
    &+&\hdots\\
    &=& \frac{p\beta\theta m_{f,1}}{(1-p)(1-p \beta\theta m_{f,1})^2}\pi_1,
\end{eqnarray*}
where the last equality uses the fact that $\sum_{h\ge 1}a^h h=a/(1-a)^2$ if $|a|<1$. Now consider the term:
\begin{eqnarray*}
&&\sum_{i\ge0}(p\beta\theta m_{f,1})^i\sum_{h\ge 1}(\beta\theta m_{f,2})^h i \pi_1=\frac{\beta\theta m_{f,2}}{1-\beta\theta m_{f,2}}\sum_{i\ge0}(p\beta\theta m_{f,1})^i i \pi_1\\&=&\frac{\beta\theta m_{f,2}}{1-\beta\theta m_{f,2}}\sum_{i\ge1}(p\beta\theta m_{f,1})^i i \pi_1=\frac{p(\beta\theta)^2 m_{f,2} m_{f,1}}{(1-p\beta\theta m_{f,1})^2(1-\beta\theta m_{f,2})}\pi_1.
\end{eqnarray*}
Finally, consider the term:
\begin{eqnarray*}
\sum_{i\ge0} (p\beta\theta m_{f,1})^i\sum_{h\ge 1}(\beta\theta m_{f,2})^h h \pi_2= \sum_{i\ge0} (p\beta\theta m_{f,1})^i\frac{\beta\theta m_{f,2}}{(1-\beta\theta m_{f,2})^2} \pi_2=\frac{\beta\theta m_{f,2}}{(1-p\beta\theta m_{f,1})(1-\beta\theta m_{f,2})^2} \pi_2
\end{eqnarray*}
Therefore:
\begin{eqnarray*}
&&E_t\left\{\sum_{h\ge 1} (\beta\theta \Pi_{k=1}^h m_{f,t+k})^h \left(\pi_{t+1}+\hdots+\pi_{t+h}\right)\right\}\\
&=&(1-p)\left(\frac{p\beta\theta m_{f,1}}{(1-p)(1-p\beta\theta m_{f,1})^2}+\frac{p (\beta\theta)^2m_{f,1}m_{f,2}}{(1-p\beta\theta m_{f,1})^2(1-\beta\theta m_{f,2})}\right)\pi_1\\&+&\frac{(1-p)\beta\theta m_{f,2}}{(1-p \beta\theta m_{f,1})(1-\beta\theta m_{f,2})^2}\pi_2.
\end{eqnarray*}
Putting these things together: 
\begin{eqnarray*}
    q_t&=& \frac{(1-\beta\theta)}{1-\beta \theta m_{f,1} p}mc_1+ \frac{(1-\beta\theta)(1-p)\beta \theta m_{f,2}}{(1-\beta \theta m_{f,2})(1-p \beta \theta m_{f,1})}  mc_2\\
    &+& \left(\frac{(1-\beta\theta)p\beta\theta m_{f,1}}{(1-p\beta\theta m_{f,1})^2}+\frac{(1-\beta\theta)(1-p)p (\beta\theta)^2m_{f,1}m_{f,2}}{(1-p\beta\theta m_{f,1})^2(1-\beta\theta m_{f,2})}\right)\pi_1\\&+&\frac{(1-\beta\theta)(1-p)\beta\theta m_{f,2}}{(1-p \beta\theta m_{f,1})(1-\beta\theta m_{f,2})^2}\pi_2.
\end{eqnarray*}
The relevant quantity for attention choice is $E\left[\left(\frac{\partial q^l_t}{\partial m_{f,1}}\right)^2\right]$ where the derivative is evaluated at some default level of attention $m_{d,f,1}\in[0,1]$. For brevity, the expression for $E\left[\left(\frac{\partial q^l_t}{\partial m_{f,1}}\right)^2\right]$ 
 is omitted from these notes. The firm's optimal attention parameter in solution $l$, $\bar m^l_{f,1}$, given $\bar m^l_{f,2}$ is 
\begin{eqnarray*}
    \bar m^l_{f,1}= \max\left(m_{d,f,1},1-\frac{\xi_f^2}{E\left[\left(\frac{\partial q^l_t}{\partial m_{f,1}}\right)^2\right]} \right).
\end{eqnarray*}

We define an endogenous BRE as follows.

\begin{definition} An \textbf{endogenous bounded rationality equilibrium (BRE)} is a tuple, $\mathbf{m^*}=(m^*_1,m^*_{f,1},m^*_2,m^*_{f,2})$, and a vector of endogenous variables, $\mathbf{x^*}=(x_1,x_2,\pi_1,\pi_2,i_1,i_2)$, such that 
\begin{enumerate}
    \item $m^*_1=\bar m_1$, $m^*_{f,1}=\bar m_{f,1}$, $m^*_2=\bar m_2$, and $m^*_{f,2}=\bar m_{f,2}$ taking $M_1=m^*_{1}$, $M_2=m^*_{2}$, $M_{f,1}=m^*_{f,1} \left(\theta+(1-\theta)\frac{1-\beta\theta}{1-\beta \theta m^*_{f,1}}\right)$, $M_{f,2}=m^*_{f,2} \left(\theta+(1-\theta)\frac{1-\beta\theta}{1-\beta \theta m^*_{f,2}}\right)$, and $\mathbf{x^*}$ as given.
    \item $\mathbf{x^*}$ solves (\ref{ISe})-(\ref{TRe}) given $M_1=m^*_{1}$, $M_2=m^*_{2}$, $M_{f,1}=m^*_{f,1} \left(\theta+(1-\theta)\frac{1-\beta\theta}{1-\beta \theta m^*_{f,1}}\right)$ and $M_{f,2}=m^*_{f,2} \left(\theta+(1-\theta)\frac{1-\beta\theta}{1-\beta \theta m^*_{f,2}}\right)$.
\end{enumerate}
 \end{definition}

An endogenous BRE is a BRE in which the agents' discount factors are optimally chosen taking the economy-wide discount factors as given.

We solve for endogenous BRE numerically using the calibration: $\beta=0.99$, $\sigma=1$, $\psi=2$, $\lambda=0.02$, $p=0.9$, $q=1$, $\xi_c=\xi_f=0.01$ and $m_{d,1}=m_{d,2}=m_{d,f,1}=m_{d,f,2}=0.7$. The scale-free attention cost parameters are set to 0.01 which means that households/firms pay attention to variables that make a 1\% difference for decisions on average. The relatively low value of the default discount factor means that agents initially discount the future heavily. For each calibration of $\epsilon_1$ we solve for endogenous BRE.

For the ZP and PP cases, we note that since $x^l_2=\pi^l_2=r^l_1=mc^l_2=0$ for $l=ZP,PP$, the optimal discount factors, $\bar m^l_2$ and $\bar m^l_{f,2}$ are undetermined in the ZP and PP solutions. We therefore assume that agents set $ m_2=\bar m^l_1$ and $m_{f,2}=\bar m^l_{f,1}$ in the ZP and PP equilibria. In other words, agents are assumed to choose the same discount factor in both states to solve the low state optimization problem. Characterizing the endogenous BRE in these cases boils down to solving for $\bar m^l_1$ and $\bar m^l_{f,1}$. We find that neither the ZP nor the PP solution exists if $\epsilon_1<-0.014$. Intuitively, a large (negative) shock necessitates greater endogenous attention, which in turn implies high endogenous BRE values of the discount parameters and hence no solution. For this calibration, no REE exists for sufficiently negative values of $\epsilon_1$, but a RPE exists for any value of $\epsilon_1$.

In general, a sufficiently negative value of $\epsilon_1$ implies no solution; support restrictions on the shock are needed to generate endogenous BRE.\\

\noindent \textbf{Remark:} \textit{For any standard calibration of the model, the ZP or PP solution only exists if $\epsilon_1>\bar \epsilon_{EBRE}$ for some $\bar \epsilon_{EBRE}<0$.}\\

Intuitively, $\epsilon_1=-\infty$ implies $E\left[\left(\frac{\partial c^l_t}{\partial m_{1}}\right)^2\right]=\infty$ for $l=ZP,PP$ and for any $m_1$. Therefore $M_1=m^*_1=1$ is the only candidate endogenous BRE value of $m^*_1$. However, if $M_1=1$ then the model is incoherent for high values of $|\epsilon_1|$.

For the PZ and ZZ cases, we find: $\bar m^l_{2}=m^*_2=0.8977$ and $\bar m^l_{f,2}=m^*_{f,2}=0.9808$ which implies $M_{f,2}=0.9677$ for $l=PZ,ZZ$. We then solve for the remaining low state discount factors for different values of $\epsilon_1$. As in the ZP and PP cases, we find that the PZ and ZZ solutions may not exist for $\epsilon_1<-0.14$. Therefore, the support of the shock must be restricted for an endogenous BRE to exist.

\subsection{Forward Guidance Puzzle} \label{appe sec: FG}

Central banks have relied heavily on forward guidance (FG) in recent decades. A large literature established that promises to keep interest rates lower for longer at the ZLB can have implausibly large effects on inflation and output in standard New Keynesian environments. Moreover, a promise to cut a future interest rate has larger effects on today’s inflation than the same cut in the current rate, and the effects become \textit{unbounded} as the timing of the rate cut is pushed into the infinite future. These counterintuitive predictions are sometimes referred to as the ``forward guidance puzzle" (see \cite{GibbsMcClung}). 
 In order to focus squarely on the implausible effects of anticipated \textit{future} rate cuts, we consider the following canonical thought experiment: suppose at time $t=0$, the central bank promises to (a) peg the interest rate at steady state until $t=T-1>0$, then (b) peg the interest rate below steady state at $t=T$, and finally (c) set interest rates according to a policy rule (e.g. an active Taylor rule) for $t>T$. Formally we have:
\begin{eqnarray}
x_t&=&M E_tx_{t+1}-\sigma(i_t-NE_t\pi_{t+1}) \label{FGAbs1},\\
\pi_t&=&\lambda x_t+\beta M_f E_t\pi_{t+1},\\
i_t&=&\begin{cases}
0 \text{   for } t=0,\hdots, T-1\\
\bar i<0 \text{   for } t=T\\
\psi \pi_t \text{   for }  t> T.
\end{cases} \label{FGAbs3}
\end{eqnarray}
To fix things, we define the forward guidance puzzle following \cite{Diba}.\\

\noindent \textit{\textbf{Definition} (Forward Guidance Puzzle). When the policy rate is pegged ($i_t=0$) for $t=0,\hdots,T-1$, the time-$0$ response to inflation and output to an expected policy rate cut at time-$T$ ($i_T=\bar i<0$) goes to infinity with $T$ (i.e. $lim_{T\rightarrow\infty} \partial \pi_0/\partial i_T=lim_{T\rightarrow\infty} \partial x_0/\partial i_T=-\infty$).}\\

Intuitively, the forward guidance puzzle emerges if the time-$0$ response of inflation or output to a promise to cut interest rates at time $T>0$ is \textit{strictly increasing} in $T$. Using terminology from \cite{FarhiWerning}, this ``anti-horizon" effect of monetary policy implies that a 100-basis point cut in the current policy rate causes a smaller rise inflation today than a promise to cut by 100-basis points 10 years from now, which has a smaller effect today than a promise to cut 1000 years from now, and so on. 

The model can be solved recursively through the method of undetermined coefficients combined with backward induction given agents' expectations about the economy after forward guidance ends. First, $\psi>1$ implies the unique equilibrium, $Y_t=0$ for $t>T$. This determines $E_T Y_{T+1}=0$, which implies $Y_T=\Gamma^{br}$ where $\Gamma^{br}$ is a function of $\bar i$ and the other model parameters. Therefore:\begin{eqnarray*}
Y_{T-1}&=&A_{br,z}\Gamma^{br},\\
Y_{T-2}&=&A_{br,z}^2\Gamma^{br},\\
\vdots\\
Y_{0}&=&A_{br,z}^T\Gamma^{br},
\end{eqnarray*}
 where
\begin{eqnarray*}
A_{br,z}:=\left(
\begin{array}{cc}
 M & \sigma N \\
 M \lambda & M_f\beta+\lambda\sigma N
\end{array},
\right)
\qquad
\Gamma^{br}:=\left(
\begin{array}{cc}
 -\sigma \bar i  \\
 - \lambda \sigma \bar i 
\end{array}.
\right)
\end{eqnarray*}
Alternatively, we can represent the solution as a VAR(1) process. Define $j:=T-t$ and $Y_t=a_{T-j}$. Then $a_{j}$ is given by
\begin{eqnarray*}
a_0&=&\Gamma^{br}, \\
a_j&=&A_{br,z}a_{j-1} \text{  for } j>0.
\end{eqnarray*}
The roots of $A_{br,z}$ are inside the unit circle if and only if $(M-1)(1-M_f\beta)+\lambda\sigma N<0$. Hence, if $(M-1)(1-M_f\beta)+\lambda\sigma N<0$ then $\lim_{j\rightarrow\infty} a_j=0$ and therefore $\lim_{T\rightarrow\infty}\partial \pi_0/\partial i_T=0$. If $(M-1)(1-M_f\beta)+\lambda\sigma N>0$ then the roots of $A_{br,z}$ are outside the unit circle and one can show that $\lim_{T\rightarrow\infty}\partial \pi_0/\partial i_T=-\infty$. We summarize this result, which is nearly a restatement of Proposition 4 of \cite{gabaix2020}, as a theorem. 
\begin{theorem}\label{Theo FG}
Consider the forward guidance model (\ref{FGAbs1})-(\ref{FGAbs3}). 
\begin{enumerate}
    \item The model does not exhibit the forward guidance puzzle if $(M-1)(1-M_f\beta)+\lambda\sigma N<0$.
    \item The model exhibits the forward guidance puzzle under RE ($M=M_f=N=1$).
    \end{enumerate}
\end{theorem}
The theorem demonstrates that the same condition ensuring coherence/completeness in the occasionally-binding constraint framework rules out the forward guidance puzzle. The model with full-information RE is susceptible to the puzzle.

\subsubsection{Adaptive Learning and Forward Guidance Puzzle}
Now we consider the effects of forward guidance when agents adaptively forecast inflation and output. We study two models of adaptive learning. The first model is given by the following system of equations:
\begin{eqnarray}
x_t&=&\hat E_tx_{t+1}-\sigma(i_t-\hat E_t\pi_{t+1}) \label{FGAas1},\\
\pi_t&=&\lambda x_t+\beta \hat E_t\pi_{t+1},\\
\hat E_tx_{t+1}&=&\gamma_{x,t}x_{t-1}+(1-\gamma_{x,t})\hat E_{t-1}x_t,\\
\hat E_t\pi_{t+1}&=&\gamma_{\pi,t}\pi_{t-1}+(1-\gamma_{\pi,t})\hat E_{t-1}\pi_t,\\
i_t&=&\begin{cases}
0 \text{   for } t=0,\hdots, T-1\\
\bar i<0 \text{   for } t=T\\
\psi \pi_t \text{   for }  t> T.
\end{cases} \label{FGAas3}
\end{eqnarray}
It is trivial to show that $\partial \pi_t/\partial i_T=\partial x_t/\partial i_T=0$ for all $t<T$ in this framework with learning. Expectations are backward-looking and predetermined in each period $t<T$ and hence the ``anticipated" interest rate cut has no effect on inflation and output until the shock hits the economy at $t=T$. 

\begin{proposition}
The adaptive learning model (\ref{FGAas1})-(\ref{FGAas3}) does not exhibit the forward guidance puzzle and forward guidance announcements have no contemporaneous impact on the economy ($\partial \pi_0/\partial i_T=\partial x_0/\partial i_T=0$ for all $T$). \label{prop:FGP1}\end{proposition}

The last proposition is not entirely robust to the type of decision rules that learning agents are assumed to have. To see this, consider a second model that features infinite horizon decision rules and adaptive learning:
\begin{eqnarray}
x_t&=&-\sigma i_t+ \hat E_t \sum_{k\ge t}\beta^{k-t}\left((1-\beta)x_{k+1}+\sigma \pi_{k+1}-\sigma\beta i_{k+1}\right) \label{FGAaas1}, \\
\pi_t&=&\lambda x_t+ \hat E_t \sum_{k\ge t}(\xi\beta)^{k-t}\left(\xi \beta \lambda x_{k+1}+(1-\xi)\beta \pi_{k+1}\right) , \\
\hat E_tx_{k+1}&=&\hat E_tx_{t+1}=\gamma_{x,t}x_{t-1}+(1-\gamma_{x,t})\hat E_{t-1}x_t,\\
\hat E_t\pi_{k+1}&=&\hat E_t\pi_{t+1}=\gamma_{\pi,t}\pi_{t-1}+(1-\gamma_{\pi,t})\hat E_{t-1}\pi_t,\\
i_t&=&\begin{cases}
0 \text{   for } t=0,\hdots, T-1\\
\bar i<0 \text{   for } t=T\\
\psi \pi_t \text{   for }  t> T,
\end{cases} \label{FGAaas3}
\end{eqnarray}
where $\lambda:=(1-\xi\beta)(1-\xi)/\xi$. Under infinite horizon learning, agents need to forecast the path of the nominal interest rate in addition to the paths of inflation and output. The following assumption about interest rate forecasts encodes the belief that the forward guidance announcement is credible:
\begin{eqnarray}
\hat E_0i_{k+1}&=&\begin{cases}
0 \text{   for } k=0,\hdots,T-2\\
\bar i \text{   for } k=T-1\\
\gamma_{i,0}i_{-1}+(1-\gamma_{i,0})\hat E_{-1}i_0 \text{   for } k\ge T.
\end{cases}\label{credible}
\end{eqnarray}
In other words, (\ref{credible}) shows how learning agents might form expectations if the forward guidance announcement (\ref{FGAaas3}) is perceived as credible. On the other hand, interest rate expectations at $t=0$ are given by  
\begin{eqnarray}
\hat E_0i_{k+1}=\gamma_{i,0}i_{-1}+(1-\gamma_{i,0})\hat E_{-1}i_0,\label{notcredible}
\end{eqnarray}
when the announcement is \textit{not} credible. In either case, the forward guidance puzzle is absent. 

\begin{proposition}
Consider the infinite-horizon adaptive learning model (\ref{FGAaas1})-(\ref{FGAaas3}). \label{prop:FGP2}
\begin{enumerate}
\item[i.] If the announcement is credible (interest rate expectations are given by (\ref{credible})) then there is no forward guidance puzzle and $\partial x_0/\partial i_T=-\sigma\beta^T$ and $\partial \pi_0/\partial i_T=-\lambda \sigma\beta^T$.
\item[ii.] If the announcement is not credible (interest rate expectations are given by (\ref{notcredible})) then there is no forward guidance puzzle and $\partial x_0/\partial i_T=\partial \pi_0/\partial i_T=0$ for any $T$.
\end{enumerate}
\end{proposition}

We refer interested readers to \cite{EGP2021} and \cite{ColeJMCB} for more on forward guidance under infinite-horizon learning.

\subsection{Learning REE: Alternatives Forecasting Models} \label{appe sec: prop8}

Proposition \ref{prop: REE E-stable} assumes that agents believe that output and inflation follow a two-state process, consistent with REE. However, the REE law of motion can be represented in a variety of different ways. For instance, consider the following perceived laws of motion for inflation and output:
\begin{eqnarray}
Y^e_t&=&a_{\epsilon_{t-k}} \label{altPLM1},\\
Y^e_t&=&a_{\epsilon_{t-k}}+b\epsilon_{t-k}, \label{altPLM2}\\
Y^e_t&=&a+b_{\epsilon_{t-k}}\epsilon_{t-k}, \label{altPLM3}\\
Y^e_t&=&a_{\epsilon_{t-k}}+b_{\epsilon_{t-k}}\epsilon_{t-k}, \label{altPLM4}\\
Y^e_t&=&a+b\epsilon_{t-k}, \label{altPLM5}\\
z^e_{t} &=& a_z+b_z z_{t-1} \label{altPLM6}
\end{eqnarray}
where $z\in\{\pi,x\}$, $k=0,1$ and $a_{\epsilon_{t-k}}$, $b_{\epsilon_{t-k}}$ may assume different values depending on $\epsilon_{t-k}$. Again, $Y^e_t$ denotes the subjective forecast of $Y_t$ implied by the forecasting model.

If learning agents instead had one of the PLMs (\ref{altPLM1})-(\ref{altPLM6}) and estimated the parameters of those models recursively, e.g. using least squares, would they eventually have self-confirming views about inflation and output? In other words, would the data confirm their belief that $Y_t$ follows one of the processes (\ref{altPLM1})-(\ref{altPLM6})? If agents observe $\epsilon_t$ and $Y_t$ when forecasting at time $t$, then beliefs formed under PLMs of the form (\ref{altPLM1})-(\ref{altPLM6}) can only become self-confirming if a REE exists. Hence, we refer to (\ref{altPLM1})-(\ref{altPLM6}) as ``REE-consistent beliefs''.

\begin{proposition} \label{prop: REE PLMs}
Suppose agents condition time-$t$ forecasts on current (time-$t$) variables. Then REE-consistent beliefs (\ref{altPLM1})-(\ref{altPLM6}) can only be self-confirming if a REE exists.
\end{proposition}

Proposition \ref{prop: REE PLMs} makes it apparent that agents including the demand shock, $\epsilon_t$, in their (piecewise) linear forecasting model (or $Y_t$ in the case of (\ref{altPLM6})) cannot develop self-confirming views about the economy if a REE does not exist (incoherence). This result has implications for how we should think about learning and equilibrium in the case of incoherence. Since none of the above ``REE-consistent beliefs'', i.e., PLMs consistent with a REE following a two-state process, could converge to a self-confirming equilibrium whenever the REE does not exist, it means that we should look at different PLMs in case of incoherence, such as a RPE.

\subsubsection{Proof of Proposition \ref{prop: REE PLMs}} 

Consider (\ref{altPLM1})-(\ref{altPLM6}), let $Y^e_t$ denote the subjective forecast of $Y_t$ implied by a given forecasting model, and assume that agents observe $\epsilon_t$ and $Y_t$ when forecasting at time $t$. 
Furthermore, to deal with possible multiplicity of time-$t$ temporary equilibria, i.e. a time-$t$ solution of (\ref{eq:IS})-(\ref{eq:MP}) given forecasts and $\epsilon_t$ with binding ZLB ($s_t=0$) and a solution with slack ZLB constraint ($s_t=1$), we simply assume that $\epsilon_t$ determines $s_t$. E.g. if $\epsilon_t=\epsilon_j$ and $s_{k}=0$ for some $k<t$ such that $\epsilon_k=\epsilon_j$, then we impose $s_{t}=0$.

(i) First consider (\ref{altPLM1})-(\ref{altPLM5}).

\paragraph{Case $k=0$.} 

If $k=0$ and expectations are formed under PLMs (\ref{altPLM1})-(\ref{altPLM5}) then $Y^e_t$ follows a two-state process: $Y^e_t=Y^e_j$ if $\epsilon_t=\epsilon_j$. Further, $\hat E_t Y_{t+1}=Pr(\epsilon_{t+1}=\epsilon_1|\epsilon_t)Y^e_1+(1-Pr(\epsilon_{t+1}=\epsilon_1|\epsilon_t))Y^e_2$ is a two-state process. 
Therefore, if $k=0$ then $Y^e_j=Y_j$ is necessary and sufficient for the agents to have self-confirming beliefs under the PLMs (\ref{altPLM1})-(\ref{altPLM5}). These self-confirming beliefs imply: $\hat E_t Y_{t+1}=Pr(\epsilon_{t+1}=\epsilon_1|\epsilon_t)Y_1+(1-Pr(\epsilon_{t+1}=\epsilon_1|\epsilon_t))Y_2$. Substituting $\hat E_t Y_{t+1}$ into the model and solving for $Y_1$ and $Y_2$ straightforwardly implies that $Y_1$, $Y_2$ is a REE. Hence, beliefs formed under (\ref{altPLM1})-(\ref{altPLM5}) with $k=0$ are only self-confirming if a REE exists. 

\paragraph{Case $k=1$.}

 Beliefs are only self-confirming under the PLMs (\ref{altPLM1})-(\ref{altPLM5}) with $k=1$ if $Y^e_j=E(Y_t|\epsilon_{t-1}=\epsilon_j)$ for $j=1,2$ where $E$ denotes the true mathematical expectation operator. Further, $\hat E_t Y_{t+1}$ formed under (\ref{altPLM1})-(\ref{altPLM5}) follows a two-state process and therefore temporary equilibrium $Y_t$ follows a two-state process: $Y_j$, where $Y_j$ is the actual equilibrium value of $Y$ given $Y^e_j$ and $\epsilon_t=\epsilon_j$ for $j=1,2$. It follows that beliefs are self-confirming if and only if $E(Y_t|\epsilon_{t-1}=\epsilon_1)=p Y_1+(1-p)Y_2$ and $E(Y_t|\epsilon_{t-1}=\epsilon_2)=(1-q) Y_1+q Y_2$. Therefore, if agents have self-confirming beliefs under PLMs (\ref{altPLM1})-(\ref{altPLM5}) with $k=1$ then $\hat E_t Y_{t+1}=Y^e_{t+1}=p Y_1+(1-p)Y_2$ if $\epsilon_t=\epsilon_1$ and $\hat E_t Y_{t+1}=Y^e_{t+1}=(1-q) Y_1+qY_2$ otherwise. Substituting  $\hat E_t Y_{t+1}$ into the model reveals that $Y_1$, $Y_2$ is a REE.

(ii) Now consider (\ref{altPLM6}). If agents observe time$-t$ information when forming time$-t$ expectations then
\begin{eqnarray}
\hat E_t z_{t+1} &=& a_z+b_z z_t, \label{CEE2}
\end{eqnarray}
where $z\in\{\pi,x\}$. We say that (\ref{altPLM6}) yields self-confirming beliefs if agents correctly understand the mean and serial correlation of $x$ and $\pi$, i.e., $a_z=(1-b_z)E(z_t)$, $b_z=(E(z_tz_{t-1})-a_zE(z_t))/E(z_{t-1}^2)$. Given fixed $a_z$, $b_z$ and expectations (\ref{CEE2}), $Y_t$ is a two-state process: $Y_j$, where $Y_j$ is the actual value of $Y_t$ given expectations and $\epsilon_t=\epsilon_j$. This implies $E(z_tz_{t-1})=q \bar q z_2^2+((1-q)\bar q+(1-p)(1-\bar q))z_1z_2+p(1-\bar q)z_1^2$, $E(z_t^2)=\bar q z_2^2+(1-\bar q)z_1^2$, $E(z_t)=\bar q z_2+(1-\bar q)z_1$. Solving for $a_z$ and $b_z$ and substituting these values into (\ref{CEE2}) yields:
\begin{eqnarray*}
\hat E_t (z_{t+1}|\epsilon_t=\epsilon_1)&=& p z_1+(1-p)z_2, \\
\hat E_t (z_{t+1}|\epsilon_t=\epsilon_2)&=& q z_2+(1-q)z_1.
\end{eqnarray*}
Substituting expectations into the model and solving for $z_1$, $z_2$ straightforwardly reveals that $z_1$ and $z_2$ must be a REE. Therefore, (\ref{altPLM6}) is not consistent with a non-rational equilibrium of an incoherent model if agents have current information.\footnote{Note that our result is related to \cite{EvansMcGough2018JME}, who study E-stability of REE in linear models when agents cannot observe exogenous shocks.}

We conclude that if beliefs formed under PLMs (\ref{altPLM1})-(\ref{altPLM6}) are self-confirming then a REE exists. Consequently, (\ref{altPLM1})-(\ref{altPLM6}) are not consistent with any non-rational equilibrium of an incoherent model.

\subsection{E-stability of BR-RPE} \label{appe sec: estab-BR-RPE}

Analogous to the RPE case considered in section \ref{sec: learning}, there is a unique E-stable bounded rationality restricted perceptions equilibrium (BR-RPE).

 \begin{proposition} \label{prop: BRRPE E-stable}
Consider (\ref{eq:IS})-(\ref{eq:MP}) and assume $\epsilon_2\ge0$. If $\epsilon_1>\bar{\epsilon}_{BR,RPE}$, then:
\begin{enumerate}
    \item[i.] There is a unique E-stable bounded rationality restricted perceptions equilibrium (BR-RPE).
    \item[ii.] The E-stable BR-RPE is either the PP BR-RPE or the ZP BR-RPE if $(M-1)(1-M_f\beta)+\lambda\sigma N\ge0$.
    \item[iii.] The E-stable BR-RPE is the unique BR-RPE if $(M-1)(1-M_f\beta)+\lambda\sigma N<0$.
\end{enumerate}
        \end{proposition}

\noindent \textbf{Proof.} To assess E-stability of each BR-RPE, we express the BR-RPE unconditional mean of inflation and output as a function of agents' expectations, $Y^e$:
\begin{eqnarray*}
\bar Y^{PP}(Y^e)&:=&\hat A_PY^e+\bar \Gamma^{PP}, \label{TPP}\\
\bar Y^{ZP}(Y^e)&:=&\left(\bar q \hat A_P+(1-\bar q)\hat A_Z\right)Y^e+\bar \Gamma^{ZP}, \label{TZP}\\
\bar Y^{PZ}(Y^e)&:=&\left((1-\bar q)\hat A_P+\bar q \hat A_Z\right)Y^e+\bar \Gamma^{PZ}, \label{TPZ}\\
\bar Y^{ZZ}(Y^e)&:=&\hat A_ZY^e+\bar \Gamma^{ZZ}, \label{TZZ}
\end{eqnarray*}
where $\bar \Gamma^i$ collect terms that do not depend on beliefs, $Y^e$, and
\begin{eqnarray*}
\hat A_P&:=& 
\begin{pmatrix}
 \frac{M}{1+\lambda \sigma \psi} & \frac{N \sigma-M_f \beta \sigma \psi}{1+\lambda \sigma \psi} \\
 \frac{M\lambda}{1+\lambda \sigma \psi} & \frac{M_f\beta+N \lambda\sigma}{1+\lambda \sigma \psi}
\end{pmatrix}, \hspace{1cm}
\hat A_Z:= 
\begin{pmatrix}
 M & N \sigma \\
 M\lambda & M_f\beta+N \lambda\sigma
\end{pmatrix}.
\end{eqnarray*}
It immediately follows that 
\begin{eqnarray*}
DT_{\bar Y^{PP}}&=&\hat A_P-I, \label{TPP2}\\
DT_{\bar Y^{ZP}}&=&\bar q \hat A_P+(1-\bar q)\hat A_Z-I, \label{TZP2}\\
DT_{\bar Y^{PZ}}&=&(1-\bar q)\hat A_P+\bar q \hat A_Z-I, \label{TPZ2}\\
DT_{\bar Y^{ZZ}}&=&\hat A_Z-I \label{TZZ2}.
\end{eqnarray*}

\paragraph{Case $\delta=(M-1)(1-M_f\beta)+N\sigma\lambda<0$.} It is straightforward to show that the real parts of the eigenvalues of $DT_{\bar Y^{PP}}$ and $DT_{\bar Y^{ZZ}}$ are negative if $\delta<0$. Therefore, the PP BR-RPE and ZZ BR-RPE are E-stable if they exist.
The ZP RPE is E-stable if and only if
\begin{eqnarray*}
tr(DT_{\bar Y^{ZP}})&=& \delta+M M_f\beta-1-\bar q\frac{\lambda\sigma(M+M_f\beta+N\lambda\sigma)\psi}{1+\lambda\sigma\psi}<0,\\
det(DT_{\bar Y^{ZP}}))&=&\frac{-\delta(1+\lambda\sigma\psi)}{1+\lambda\sigma\psi}+\frac{\lambda\sigma\psi \bar q(\delta+1)},{1+\lambda\sigma\psi}>0
\end{eqnarray*}
where $tr(B)$ denotes the trace of matrix $B$. Because $-1<\delta$, the ZP BR-RPE is E-stable in the case $\delta<0$. Further, this holds for any $\bar q$, and therefore the PZ BR-RPE is E-stable if it exists, as $tr(DT_{\bar Y^{PZ}}),
det(DT_{\bar Y^{PZ}})$ have the same form as $tr(DT_{\bar Y^{ZP}}),
det(DT_{\bar Y^{ZP}})$ with $\bar q$ replaced by $1-\bar q$.

By the proof of Proposition \ref{prop: BR completeness} (setting $q=\bar q$ and $p=1-\bar q$) there is a unique BR-RPE if $\delta<0$. Therefore, there is a unique E-stable BR-RPE if $\delta<0$.\\

\paragraph{Case $\delta\ge 0$.}It is straightforward to show that the real parts of the eigenvalues of $DT_{\bar Y^{PP}}$ are negative and the real part of an eigenvalue of $DT_{\bar Y^{ZZ}}$ is non-negative if $\delta\ge0$. Therefore, the PP BR-RPE is E-stable and the ZZ BR-RPE is not E-stable in the case $\delta\ge0.$

The ZP RPE is E-stable if and only if $tr(DT_{\bar Y^{ZP}})<0<det(DT_{\bar Y^{ZP}})$, which holds if and only if $\eta_{ZP}=\bar q(\lambda\sigma (M+M_f\beta(1-M)+N \lambda \sigma )\psi)-(M-1+M_f\beta(1-M)+N\lambda\sigma)(1+\lambda\sigma\psi)>0$. From the proofs of Propositions \ref{prop:BR coherence} and \ref{prop: BRRPE existence}:
\begin{eqnarray*}
&&\epsilon^{PP,BR,RPE}-\epsilon^{ZP,BR,RPE}_2=v_b \eta_{ZP},
\end{eqnarray*}
where $v_b:=\frac{(\lambda \epsilon_2 + (\lambda\sigma - \delta\psi^{-1}) \mu)\psi}{((1 - \bar q) (M+ M_f \beta(1 - M) + N \lambda \sigma) \psi ((1 - M_f \beta) (1 - 
         M) + \lambda \sigma (\psi - N \bar q) + (1 - 
         \bar q) (M_f \beta + M (1 - M_f \beta))))}>0$. Therefore, if the ZP RPE is E-stable then $\epsilon^{PP,BR,RPE}>\epsilon^{ZP,BR,RPE}_2$ and the condition for PP existence becomes $\epsilon_1>\epsilon^{PP,BR,RPE}$ and the condition for ZP existence becomes $\epsilon^{PP,BR,RPE}\ge\epsilon_1>\epsilon^{ZP,BR,RPE}_2$ as demonstrated in the proofs of Propositions \ref{prop:BR coherence} and \ref{prop: BRRPE existence}.\footnote{If $\bar q=1$, the PP exists and is E-stable if and only if $\epsilon_1>\epsilon^{PP,BR,RPE}$ and the ZP exists and is E-stable if and only if $\epsilon_1\le\epsilon^{PP,BR,RPE}$. The ZZ and PZ solutions cannot be E-stable if $\delta\ge0.$ } Hence, if the ZP RPE exists and is E-stable then the PP solution does not exist.  

Next consider the PZ solution. The PZ solution is E-stable if and only if 
\begin{eqnarray*}
tr(DT_{\bar Y^{PZ}})&=& \frac{-2+M+M_f\beta +N\lambda\sigma-2\lambda\sigma\psi}{1+\lambda\sigma\psi}-\bar q\frac{\lambda\sigma(M+M_f\beta+N\lambda\sigma)\psi)}{1+\lambda\sigma\psi}<0,\\
det(DT_{\bar Y^{PZ}}))&=&\frac{1-M_f\beta+M(M_f\beta-1)+(\psi-N)\lambda\sigma}{1+\lambda\sigma\psi}-\bar q\frac{\lambda\sigma\psi(\delta+1)}{1+\lambda\sigma\psi}>0,
\end{eqnarray*}
which holds if and only if $0<den^{PZ,BR,RPE}$ where $den^{PZ,BR,RPE}$ is equal to $den^{PZ,BR}$ defined in the Proposition \ref{prop:BR coherence} proof when $q=\bar q$ and $p=1-\bar q$. From the proof of Proposition \ref{prop: BRRPE existence}, the PZ RPE only exists in the case $\delta\ge0$ if $den^{PZ,BR,RPE}<0$. Hence the PZ BR-RPE is never E-stable if $\delta\ge0$.

Therefore, the PP BR-RPE is the only E-stable BR-RPE solution when $\epsilon_{1}>\epsilon^{PP,BR,RPE}$, and the ZP BR-RPE is the only E-stable BR-RPE solution when $\epsilon^{PP,BR,RPE}\ge\epsilon_1>\epsilon^{ZP,BR,RPE}_2$. It follows that a unique E-stable BR-RPE solution exists when $\epsilon_{1}>\bar{\epsilon}_{BR,RPE}$.

\subsection{Is the RPE reasonable?}\label{appe sec: RPEreasonable}
In a RPE, agents have badly misspecified beliefs. Agents forecast the means of inflation and output as if they believe those variables are constant or mean-plus-noise, despite the fact that these variables would obviously follow a persistent two-state Markov chain in a RPE. Why would we consider RPE reasonable? Should agents be expected to detect their mis-specification over time simply by looking at time series data? Several comments are in order.

\begin{figure}[htp]
 \caption{Region of Coherence of the REE and of the RPE}
 \centering
    \begin{subfigure}{2in}
        \includegraphics[scale=.33]{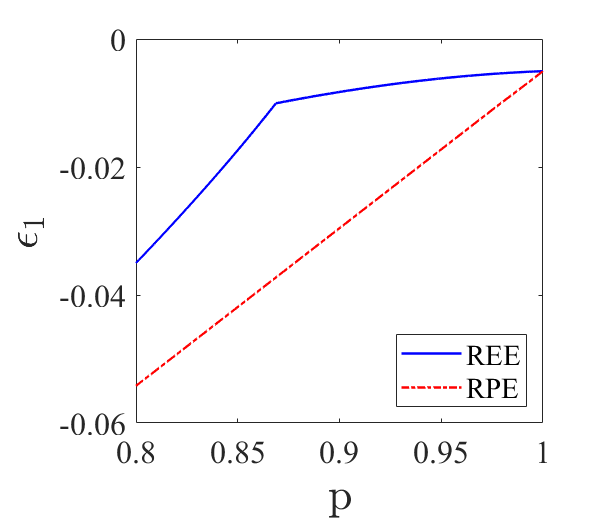}
		\caption*{(a)}
        \label{fig: coherence0.98}
    \end{subfigure}
    \begin{subfigure}{2in}
      \includegraphics[scale=.33]{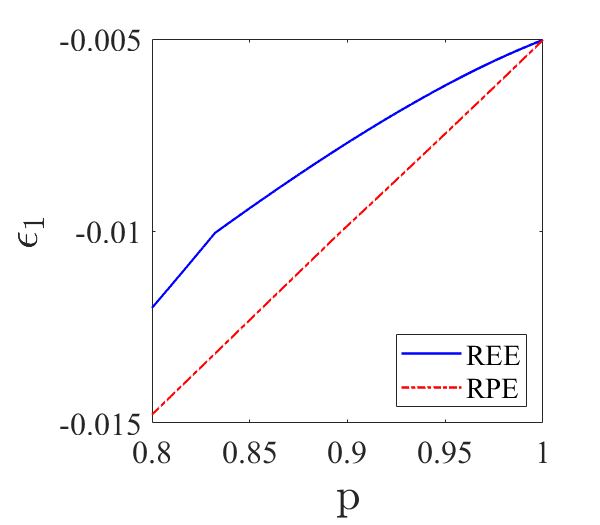}
		\caption*{(b) $q=0.9$}
        \label{fig: coherence0.9}
    \end{subfigure}
     \begin{subfigure}{2in}
       \includegraphics[scale=.33]{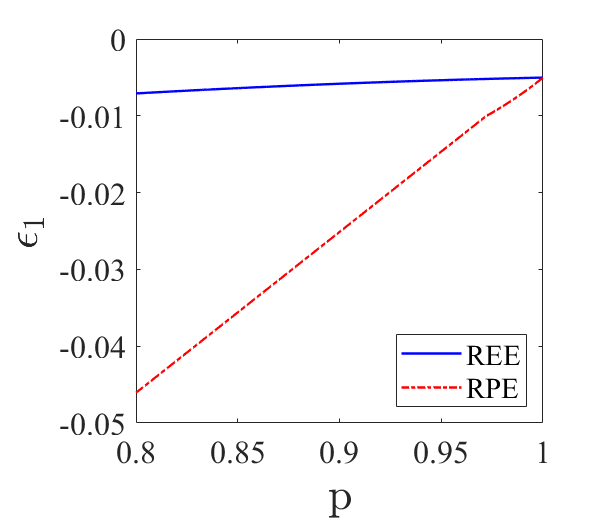}
		\caption*{(c) $\lambda=0.2$}
         \label{fig: coherence lambda}
     \end{subfigure}
     \caption*{\footnotesize
        Note: The area above the blue (red) curve depicts values of $\epsilon_1$ and $p$ for which at least one REE (RPE) exists. Other parameter values: $\beta=0.99$, $\sigma=1$, $\lambda=0.02$, $q=0.98$, $\epsilon_2=0.$}
\label{fig: cohe regions }
\end{figure}
 
First, if a REE exists, then we could argue these RPE are implausible. In this case, agents could learn to do better, because there would likely be a learnable REE. But incoherence precludes REE, and as shown in Appendix \ref{appe sec: prop8}, it implies that agents fail to form self-confirming expectations using a variety of different forecasting models that condition on the demand shock or even a lag of the endogenous variables. In the case of incoherence of REE, the RPE is thus a potentially reasonable alternative, because it relaxes the condition for the existence of self-confirming equilibria. Figure \ref{fig: cohe regions } visualizes the difference between the combination of values of the negative shock, $\epsilon_1,$ and of its persistence, $p$, that yields coherence in the REE and in the RPE cases. The area above the blue line and the red line defines the set of pairs $(\epsilon_1,p)$ so that at least one REE and RPE exist, respectively. Panel (a) shows that the difference between the region of the parameter space for which there is coherence in the two cases is substantial. In particular, unless the persistence, $p,$ of the negative demand shock falls below 0.87, RE admits an equilibrium only for very small negative shocks. Panel (b) shows that both regions are quite sensitive---they shrink by around a quarter---to the value of the persistence of the other state where $\epsilon_2=0.$ Finally, panel (c) shows that the region of coherence of REE shrinks quite substantially as prices becomes more flexible, while this is not the case for the RPE. The curse of flexibility is therefore a much more pronounced problem for REE than for RPE, just as Figure \ref{fig: BR cohe regions } (c) shows, which is very intuitive because the curse hinges on the rationality and forward-lookingness of the agents.

\begin{figure}[htp]
 \caption{Maximum Expected ZLB Duration in a ZP Solution}
 \centering
    \begin{subfigure}{2in}
        \includegraphics[scale=.35]{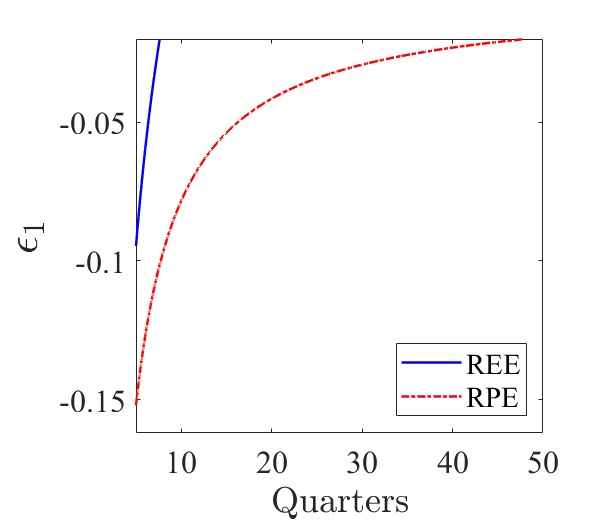}
		\caption*{(a)}
    \end{subfigure}
    \begin{subfigure}{2in}
      \includegraphics[scale=.35]{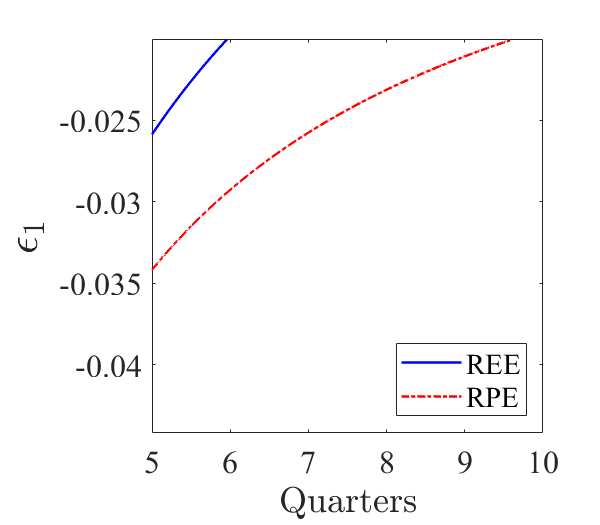}
		\caption*{(b) $q=0.9$}
    \end{subfigure}
     \begin{subfigure}{2in}
       \includegraphics[scale=.35]{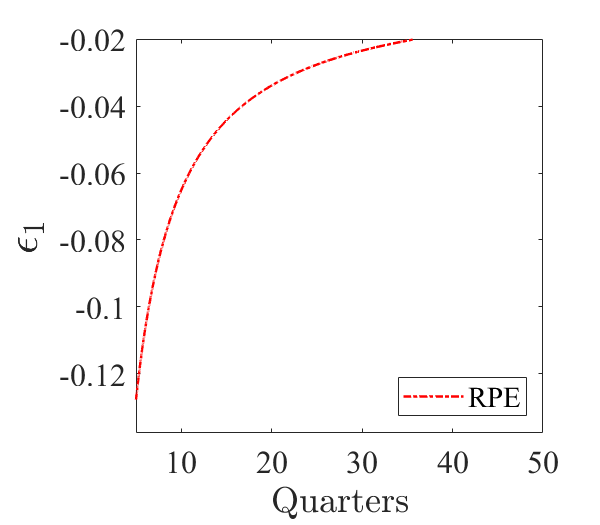}
		\caption*{(c) $\lambda=0.2$}
     \end{subfigure}
     \caption*{\footnotesize
        Note: The blue (red) curve depicts the maximum expected duration ZLB ($(1-p)^{-1}$) we can generate for given $\epsilon_1$ in a REE (RPE) ZP solution. The figure only depicts values of $\epsilon_1$ for which a REE ZP or RPE ZP solution exists. Other parameter values: $\beta=0.99$, $\sigma=1$, $\lambda=0.02$, $q=0.98$, $\epsilon_2=0.01$.}
\label{fig: LT duration }
\end{figure}

The Figure \ref{fig: cohe regions } results suggest that a fundamentals-driven RE liquidity trap must be relatively short-lived in the case of a REE compared to the duration of actual liquidity trap events experienced by Japan, the Euro Area and the U.S. In contrast, a fundamentals-driven RPE liquidity trap can be more persistent. Figure \ref{fig: LT duration } depicts the maximum expected duration of the liquidity trap (equal to $(1-p)^{-1}$) that we can generate in a ZP REE or ZP RPE for different combinations of demand shock, $\epsilon_1$. It can be seen that liquidity traps cannot be very persistent in a REE, whereas the RPE liquidity traps can be highly persistent, particularly if $q$ is relatively large as in panel (a).\footnote{Note that $p=0.965$ produces an expected liquidity trap duration of around 28 quarters, which is the length of the 2008-2015 ZLB episode in the U.S.} Panel (c) again shows that the curse of flexibility is a more pronounced problem for the REE. The BRE results are not depicted in Figure \ref{fig: LT duration }, but Proposition \ref{prop: BR completeness} implies that we can generate permanent ZLB events in a BRE for very negative shocks.

\begin{figure}[htp]
 \caption{Simulations when REE does not exist and a RPE exists}
 \centering
    \begin{subfigure}{3.2in}
        \includegraphics[scale=.385]{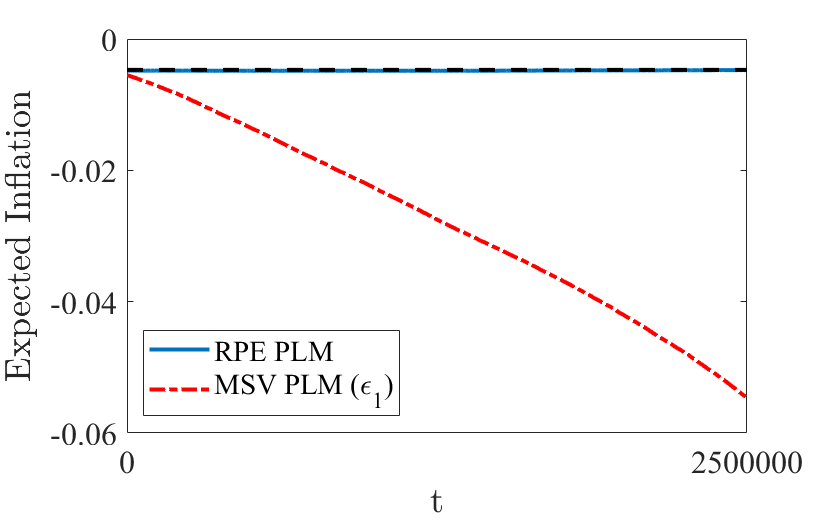}
        \caption*{(a) RPE vs MSV beliefs}
        \label{fig: RPEvsMSV}
    \end{subfigure}
    \begin{subfigure}{3.2in}
       \includegraphics[scale=.4]{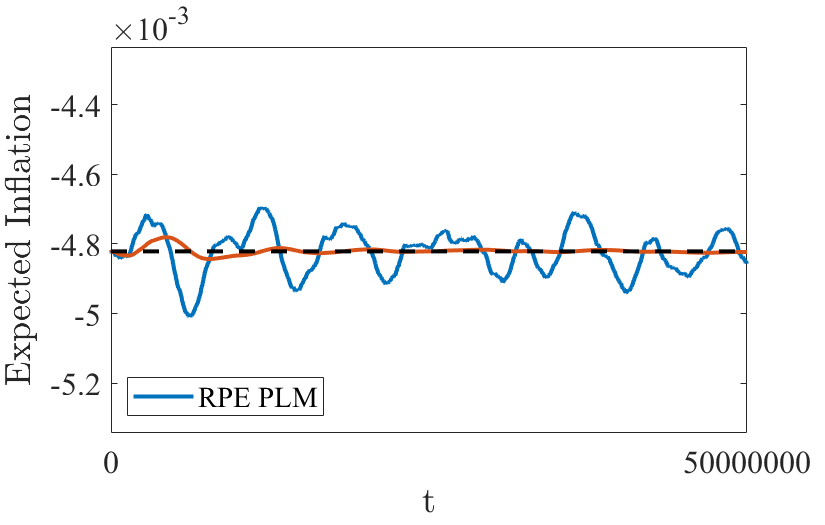}
					\caption*{(b) RPE beliefs}
        \label{fig: RPEsimu}
    \end{subfigure}
    
     \caption*{\footnotesize
        Note: The model is calibrated so that an E-stable RPE ZP solution exists, but no MSV REE exists. The constant gain is small and set to $g_t=0.00001$ for all $t$. $\beta=0.99$, $\sigma=1$, $\lambda=0.02$, $p=0.85$, $q=0.98$, $\epsilon_1=-0.04$, $\epsilon_2=0$.}
\label{fig: simulations}
\end{figure}

Second, suppose the model is incoherent under RE, but an E-stable RPE exists and the economy is in it. One could argue that agents inhabiting the RPE would notice that RPE inflation and output follow a two-state process. Hence, agents would then stop setting one-period ahead inflation and output expectations equal to the long run average of those variables, and start to estimate a two-state forecasting model in their attempt to learn these dynamics. Our previous propositions already suggest this might be a bad idea (see Proposition \ref{prop: REE E-stable} and Appendix \ref{appe sec: prop8}). Can they reach another---not self-confirming---equilibrium? Figure \ref{fig: simulations} (a) depicts the results from simulating the learning dynamics for the case of MSV-consistent beliefs and also for the case of RPE-consistent beliefs, assuming a small constant gain.\footnote{For MSV learning simulation, we initialize the forecast, $Y^e_{j,1}$ to match the state-contingent mean of inflation/output in the RPE when $\epsilon_t=\epsilon_j$. In other words, we assume that agents observe actual endogenous variables in the RPE switching with $\epsilon_t$ during periods $t<1$ and then they decide to make their forecasts consistent with the switching at $t=1$. We use the same initialization for RPE beliefs. Learning agents are assumed to have lagged information as defined in Section \ref{sec: RPE}.} It clearly shows that MSV-consistent beliefs are explosive even with very small gain parameter, while, on the contrary, the RPE-consistent beliefs are not. Panel (b) in Figure \ref{fig: simulations} displays the dynamics of expected inflation (and its cumulative average in red) from which it is evident that RPE expectations remain in some neighborhood around their RPE values.\footnote{Moreover, simulations---not reported---also show that RPE-consistent beliefs tend to revert to RPE values even with decreasing-gain and when initial beliefs are a small distance from RPE values. Intuitively, the RPE-consistent beliefs could also be explosive (into deflationary spirals) whenever the gain parameter is too large or initial beliefs are very far from the RPE value.}
Numerical simulation therefore suggests another reason why the RPE might be a good alternative. If a RPE exists---and a REE does not---and if agents try to learn using the REE PLM, then the economy will derail into deflationary spirals. On the contrary, if agents try to learn the RPE, then expectations remain stable and ``centered" on the correct RPE values---provided that the gain parameter is small and initial inflation and output expectations are not too far away from the average inflation and output rate in the RPE. 

Third, it is important to recall from Proposition \ref{prop: CC temp RPE} that the assumption of learning by itself ensures coherence and completeness, provided that agents have lagged information. Thus, while tight model restrictions are needed to characterize self-confirming equilibrium, the economy can \textit{always} be in a market-clearing temporary equilibrium.

Of course there could be other non-rational equilibria such as the consistent expectations equilibrium (CEE) considered by \cite{JorgensenLansing2021}, the stochastic consistent expectations equilibria (SCEE) of \cite{HommesZhu} or \cite{AiraudoHajdini}.  Our numerical analysis indicates that these more sophisticated non-rational equilibria may not exist for some plausible calibrations of the model.\footnote{In a SCEE, agents' forecasts introduce a lag of inflation and output into the model, which prevents us from analytically examining the existence of SCEE in our model with an occasionally binding constraint. See, e.g., \citetalias{AscariMavroeidis}} Thus, the RPE may even be the best alternative among non-rational equilibria of our model with $M=M_f=N=1$, but CEE or SCEE existence remains an open question. However, whether or not these alternative non-rational equilibria exist is not relevant for the main result of this paper: rationally incoherent models can be non-rationally coherent, i.e., admit non-rational equilibria.

\subsection{RPE and Continuous Shocks}\label{appe sec: RPEcontinuous}
To get closed-form solutions for both REE and RPE, we must assume that $\epsilon_t$ follows a discrete-valued Markov chain. To the best of our knowledge, no paper provides conditions for existence and uniqueness of RE equilibrium which can be applied to a model similar to our model under the assumption that $\epsilon_t$ is both persistent and continuously distributed.\footnote{See \cite{Mendes} for analytical existence results under the assumption that $\epsilon_t$ is a mean-zero, i.i.d process.} However, while it is hard to characterize REE in a model with continuous shocks and an occasionally binding constraint, it is relatively easy to derive RPE. 

To illustrate, consider the model (\ref{eq:IS})-(\ref{eq:MP}) and suppose instead that $\epsilon_t=\rho \epsilon_{t-1}+v_t$ where $\rho\in[0,1)$ and $v_t \sim \mathcal{N}(0,\,\sigma_v^{2})$. In a RPE of this economy, agents' forecasts are given by $\hat E_t\pi_{t+1}=a_{\pi}$, $\hat E_tx_{t+1}=\frac{1-\beta}{\lambda}a_{\pi}$ consistent with the RPE studied in the previous sections. Substituting these expectations into the model gives the following RPE law of motion for inflation:
\begin{align}
    \pi_t&=\begin{cases}
    (1+\lambda\sigma)a_{\pi}+\lambda\sigma \mu+\lambda \epsilon_t \hspace{.2cm}\text{ if } s_t=0,\\
    \frac{1+\lambda\sigma}{1+\lambda\sigma\psi}a_{\pi}+\frac{\lambda}{1+\lambda\sigma\psi}\epsilon_t \hspace{1.2cm}\text{ if } s_t=1.
    \end{cases}
\end{align}
Let $h(a_{\pi})$ denote $E(\pi_t)$ as a function of $a_{\pi}$, and let $\sigma_{\epsilon}:=\sqrt{\frac{\sigma_v^2}{1-\rho^2}}$. Then:
\begin{align}
h(a_{\pi})&= Pr\left(s_t=0\right)E(\pi_t|s_t=0)+\left(1-Pr\left(s_t=0\right)\right)E\left(\pi_t|s_t=1\right), 
\end{align}
where $s_t=0$ indicates that the ZLB is binding.
To compute RPE, we need to compute $Pr(s_t=0)$, $E(\pi_t|s_t=0)$ and $E(\pi_t|s_t=1)$ as functions of $a_{\pi}$. Let $\Phi$ and $\phi$ denote the standard normal probability distribution function and standard normal probability density function, respectively. Further, define:
\begin{align}
    L(a_{\pi}):=(\sigma_{\epsilon}\lambda)^{-1}\left(-\mu/\psi-(1+\lambda\sigma)a_{\pi}-\lambda\sigma\mu\right). 
\end{align}
It follows that:
\begin{align*}
Pr(s_t=0)&=\Phi(L(a_{\pi})), \nonumber \\
E(\pi_t|s_t=0)&=(1+\lambda\sigma)a_{\pi}+\lambda\sigma \mu -\frac{\lambda\sigma_{\epsilon}\phi(L(a_{\pi}))}{\Phi(L(a_{\pi}))}, \\
E(\pi_t|s_t=1)&=\frac{1+\lambda\sigma}{1+\lambda\sigma\psi}a_{\pi} +\frac{\lambda\sigma_{\epsilon}\phi(L(a_{\pi}))}{(1+\lambda\sigma\psi)(1-\Phi(L(a_{\pi})))}.
\end{align*}
Therefore, we have :
\begin{align}
    h(a_{\pi})&=\frac{1+\lambda\sigma}{1+\lambda\sigma\psi}a_{\pi}+\Phi(L(a_{\pi}))\left(\frac{(1+\lambda\sigma)\lambda\sigma\psi}{1+\lambda\sigma\psi}a_{\pi}+\lambda\sigma\mu\right) -\frac{\phi(L(a_{\pi}))\lambda^2 \sigma_{\epsilon}\sigma\psi}{1+\lambda\sigma\psi}. \label{eq:h}
\end{align}

There is a RPE if and only if there exists $\bar a_{\pi}\in \mathbb{R}$ such that $h(\bar a_{\pi})=\bar a_{\pi}$. One can show there exists a unique maximum of $h( a_{\pi})-  a_{\pi}$, denoted $a^*_{\pi}$, and consequently there is either no RPE solution or there are exactly two RPE solutions.\footnote{To see this, note that $\Phi$ is strictly decreasing in $a_{\pi}$ and $\Phi$ and $L$ are injective functions and that $h'(a_{\pi})-1=\frac{\lambda\sigma(1-\psi)}{1+\lambda\sigma\psi}+\frac{\Phi(L(a_{\pi}))\lambda\sigma\psi(1+\lambda\sigma)}{1+\lambda\sigma\psi}$. Then under the Taylor Principle ($\psi>1$), there exists a unique maximum, $a^*_{\pi}$, such that $h'(a^*_{\pi})-1=0$ and $h'(a_{\pi})-1>0$ ($h'(a_{\pi})-1<0$) for all $a_{\pi}<a^*_{\pi}$ ($a_{\pi}>a^*_{\pi}$). For brevity, we abstract from the special case in which $h(a^*_{\pi})=a^*_{\pi}$.}  A necessary and sufficient condition for existence of the RPE is $h(a^*_{\pi})-a^*_{\pi}\ge0$. We summarize the result as a proposition.

\begin{proposition} \label{prop: cont shock}
Consider (\ref{eq:IS})-(\ref{eq:MP}) and suppose that $\epsilon_t=\rho \epsilon_{t-1}+v_t$ where $v_t \sim \mathcal{N}(0,\,\sigma_v^{2})$. Then:
\begin{enumerate}
    \item[i.] Two restricted perceptions equilibria (RPE) exist if and only if $h(a^*_{\pi})>a^*_{\pi}$
    where $a^*_{\pi}$ is given by
    \begin{align*}
        a^*_{\pi}&=L^{-1}\left(\Phi^{-1}\left(\frac{\psi-1}{(1+\lambda\sigma)\psi}\right)\right).
    \end{align*}
    \item[ii.] A RPE does not exist if and only if $h(a^*_{\pi})<a^*_{\pi}$.
\end{enumerate}
\end{proposition}

\begin{figure}[htp!]
\caption{Existence and Multiplicity of RPE with Continuous Shocks}
\begin{center}
	\begin{minipage}{\textwidth}
		\centering
		\includegraphics[scale=.5]{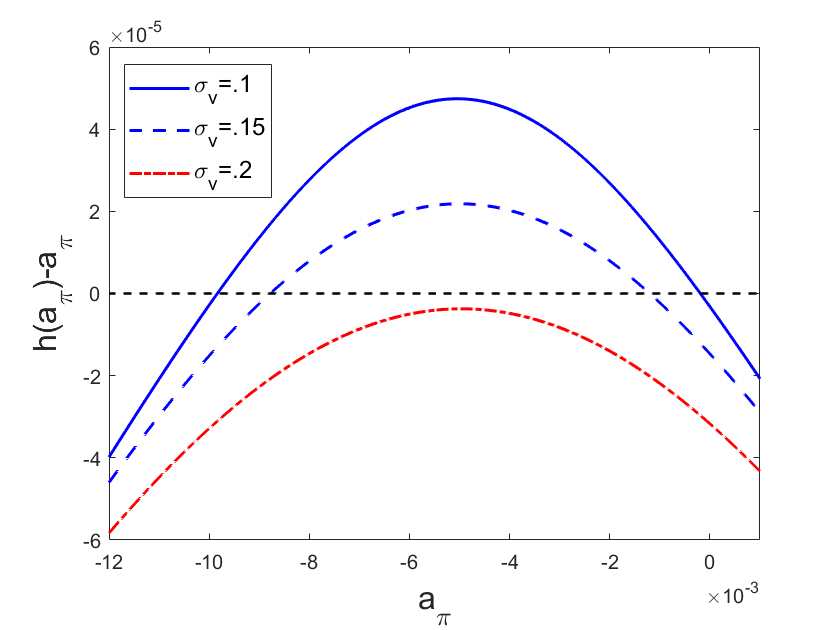}
	\end{minipage}
\end{center}\label{fig: RPE conshock}
\end{figure}

\begin{figure}[htp!]
			\caption{RPE Existence}
			\begin{center}
				\begin{minipage}{.5\textwidth}
					\centering
					\includegraphics[scale=.33]{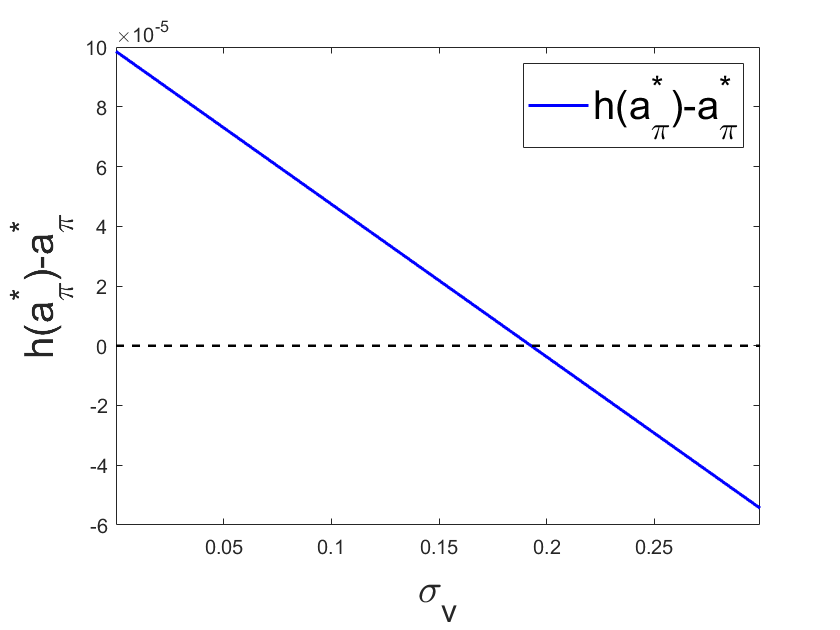}
					\caption*{(a) Shock Variance}
				\end{minipage}\hfill
				\begin{minipage}{.5\textwidth}
					\centering
					\includegraphics[scale=.33]{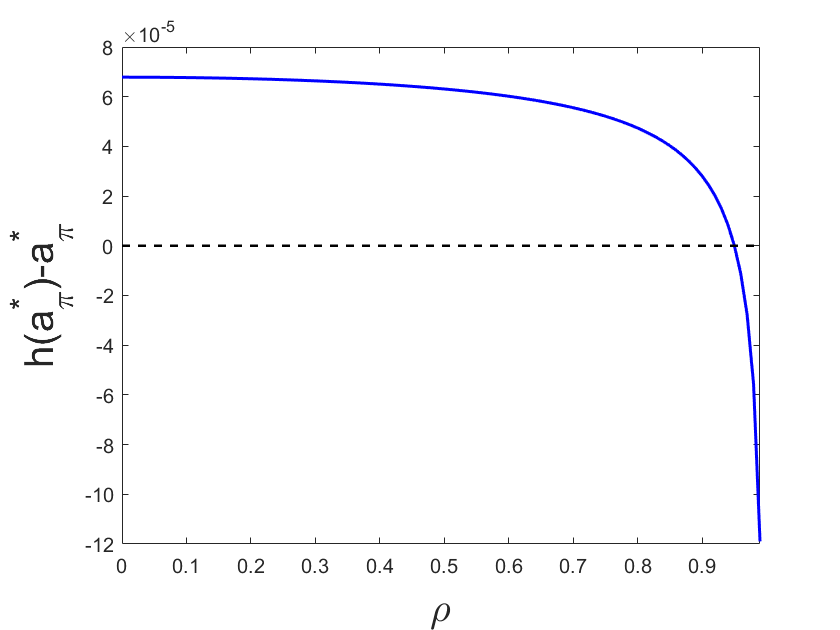}
					\caption*{(b) Shock Persistence}
				\end{minipage}
				\begin{minipage}{.5\textwidth}
					\centering
					\includegraphics[scale=.33]{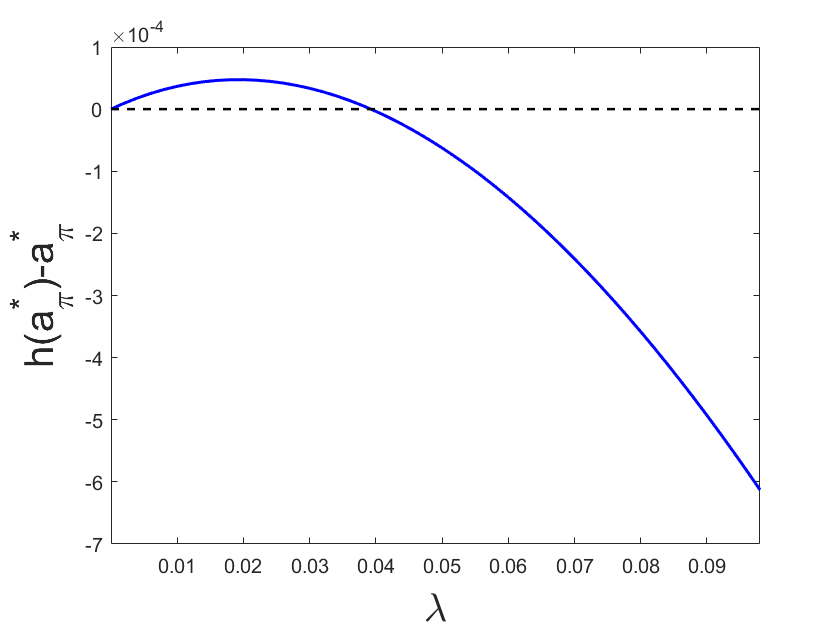}
					\caption*{(c) Price Flexibility}
				\end{minipage}\hfill
				\begin{minipage}{.5\textwidth}
					\centering
					\includegraphics[scale=.33]{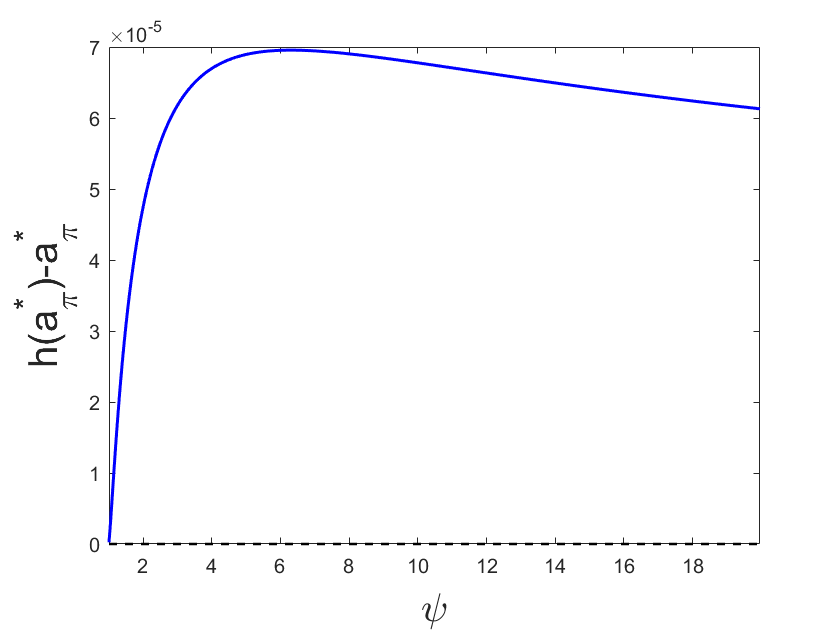}
					\caption*{(d) Activeness of Policy}
				\end{minipage}
			\end{center}\label{fig: RPE conshock_multi}
		\end{figure}

By inspecting (\ref{eq:h}), one can see that increasing the variance and persistence of the shocks (i.e. increasing $\sigma_{v}$ and $\rho$) or decreasing price rigidity (i.e. increasing $\lambda$) reduces $h(a_{\pi})-a_{\pi}$, which must be positive for an (actually two) RPE to exist. Consequently, sufficiently high values of $\sigma_{v}$, $\rho$ or $\lambda$ preclude existence of RPE in the model with continuous, persistent shocks. Figure \ref{fig: RPE conshock} plots $h(a_{\pi})-a_{\pi}$ for three different values of $\sigma_v$, assuming $\rho=0.8$. It is evident that larger values of $\sigma_v$ shifts $h(a_{\pi})-a_{\pi}$ down.\footnote{Figures \ref{fig: RPE conshock} and \ref{fig: RPE conshock_multi} plot $h(a^*_{\pi})-a^*_{\pi}$ for different calibrations of key parameters. In both figures we use the following benchmark calibration unless otherwise noted: $\beta=0.99$, $\sigma=1$, $\psi=2$, $\lambda=0.02$, $\rho=0.8$, $\sigma=0.1$.} Notice in the figure that the RPE levels of inflation are always less than the zero inflation steady state level, and hence the numerical RPE we consider display a deflationary bias akin to the deflationary bias studied under RE in \cite{NakataSchmidt2019} or \cite{BianchiMelosiRottner}. Figure \ref{fig: RPE conshock_multi} plots $h(a^*_{\pi})-a^*_{\pi}$ for different values of other key parameters in calibrated models. To interpret the panels in the Figure recall that $h(a^*_{\pi})-a^*_{\pi}>0$ for the RPE to exist. The figure shows that the RPE is less likely to exist if the shock variance or persistence is high, or if prices are more flexible. Hence, the same insights from the simple two-state process example carry over to the case of continuous shocks (see Figure \ref{fig: cohe regions }).

\subsection{Variation on a theme: REE with lagged expectations}\label{appe sec: lagged}

This section briefly looks at the possibility of the existence of other equilibria, in which agents have imperfect information in the sense that they do not observe the contemporaneous shock. Throughout this paper we stuck to the standard assumption that ``rational'' agents observe the demand shock contemporaneously (i.e. $\epsilon_t$ is included in agents' time-$t$ information set). This would be a natural assumption if for example $\epsilon_t$ is a shock to the households' preferences as in \cite{Eggertsson}. However, the assumption that agents observe $\epsilon_t$ with a lag (so that $\epsilon_{t-1}$, but not $\epsilon_t$, is included in agents' time-$t$ information set) permits the study of some additional non-rational equilibria which may exist in rationally incoherent models.
 
To illustrate existence of these additional ``lagged expectations equilibria" (LEE), consider the model (\ref{eq:IS})-(\ref{eq:MP}) and suppose $q=1$, $\epsilon_2=0$. Further suppose that agents believe inflation and output follows the same persistent two-state Markov chain as the shock (just like rational agents) but instead agents do not know $\epsilon_t$ and hence agents attach $p^2$ probability to the prospect that $\epsilon_{t+1}=\epsilon_1$ when forecasting at time $t$ in the temporary state, instead of attaching $p$ probability to this event as agents with full-information RE would do. Under this assumption about agents' time-$t$ information set, the economy either returns to the steady state with zero inflation or the steady state with zero interest rates after $\epsilon_t=\epsilon_2$. The ``temporary state" value of output when $\epsilon_t=\epsilon_1$ (assuming for simplicity that we go back to the zero inflation steady state) is given by:
\begin{eqnarray}
x_t&=&\nu(p^2)\hat E_tx_{t+1}-\sigma \max\{\frac{\psi \lambda}{1-\beta p^2}x_t,-\mu\}+\epsilon_1, \\
\text{where} \hspace{3em} \nu(p^2)&:=& \left(1+\frac{\lambda \sigma}{1-\beta p^2}\right)>1, \nonumber
\end{eqnarray}
which we obtain by substituting the Phillips curve and Taylor rule into (\ref{eq:IS}). From this equation, it is apparent that for any $p$, sufficiently low values of $\epsilon_1$ preclude unconstrained interest rates, just as in the case of full information RE. Thus, for a sufficiently large demand shock, output will be given by;
\begin{eqnarray}
x_t=\frac{1}{1-p^2\nu(p^2)}(\sigma \mu+\epsilon_1),
\end{eqnarray}
if a solution of the model exists at all.  We call this solution a lagged expectation equilibrium (LEE). It is a self-confirming equilibrium because agents correctly forecast the conditional mean of output and inflation (e.g. $E(x_t|\epsilon_{t}=\epsilon_1)=\frac{1}{1-p^2\nu(p^2)}(\sigma \mu+\epsilon_1)$ and $E(x_t|\epsilon_{t}=\epsilon_2)=0$).\footnote{In the first period such that $\epsilon_t=\epsilon_2$, we have $x_t\neq0$. However, $E(x_t|\epsilon_{t}=\epsilon_2)=E(\pi_t|\epsilon_{t}=\epsilon_2)=0$ because state $2$ is an absorbing state. Thus, the LEE is a non-rational equilibrium in which agents have self-confirming beliefs about the state-contingent conditional means of endogenous variables.} Note that $p^2\nu(p^2)<p\nu(p)$, and therefore if $p^2\nu(p^2)<1<p\nu(p)$ we will have a LEE given any $\epsilon_1$, but only a REE if $\epsilon_1$ is sufficiently close to zero. REE existence always implies existence of LEE, but the opposite is not true. This simple exercise reveals that there can be additional deviations from RE, beyond the scope of this paper, which are useful for understanding an incoherent model.

\end{subappendices}

\end{document}